\documentstyle[multicol,aps,rmp,amsfonts,graphics,epsfig]{revtex}            
\begin{document}                       
\title{Superconductivity in Fullerides}
\author{O. Gunnarsson}
\address{      
Max-Planck-Institut f\"ur Festk\"orperforschung, D-70506 Stuttgart,
Germany}
\maketitle
\begin{abstract}
Experimental studies of superconductivity properties of 
fullerides are
briefly reviewed. Theoretical calculations of the electron-phonon coupling,
in particular for the intramolecular phonons, are discussed extensively. 
The calculations are compared with coupling constants deduced from a number of 
different experimental techniques. It is discussed why the A$_3$C$_{60}$   
are not Mott-Hubbard insulators, in spite of the large Coulomb
interaction. Estimates of the Coulomb 
pseudopotential $\mu^{\ast}$, describing the effect of the Coulomb 
repulsion on the superconductivity, as well as possible electronic
mechanisms for the superconductivity are reviewed. The calculation of 
various properties within the Migdal-Eliashberg theory and attempts to
go beyond this theory are described.
\end{abstract}
%TCIMACRO{\TexButton{twocolumn}{\begin{multicols{2}}}
%BeginExpansion
\begin{multicols}{2}
%EndExpansion
\tableofcontents
\section{Introduction}\label{introduction}
\begin{figure}[h]
\unitlength1cm
\begin{minipage}[t]{8.5cm}
\centerline{
\resizebox{2in}{!}{\includegraphics{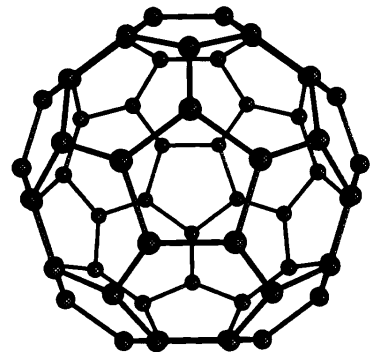}}
}
\caption[]{\label{fig0}The C$_{60}$ molecule.} 
\end{minipage}
\hfill
\end{figure}

The fullerenes (C$_{60}$, C$_{70}$,...) have attracted much 
 interest since their discovery by Kroto {\it et al.} (1985), not least
 because of their appealing, symmetric shape (see Fig. \ref{fig0}).
This interest increased dramatically when Kr\"atschmer {\it et al.} (1990)
discovered how to produce C$_{60}$ in large enough quantities to be able
 to make solids of a size which allowed traditional solid state experiments.
Very soon Haddon {\it et al.} (1991) found that intercalation of alkali metal
 atoms in solid C$_{60}$ leads to metallic behavior. 
Shortly afterwards it was found that some of these alkali-doped C$_{60}$ 
compounds are superconducting with a transition temperature $T_c$ which
is only surpassed by the cuprates (Hebard {\it et al.}, 1991; Rosseinsky
{\it et al.}, 1991; Holczer {\it et al.}, 1991a; Tanigaki {\it et al.},
1991; Fleming {\it et al.}, 1991). Thus $T_c$ is 33 K for RbCs$_2$C$_{60}$
 (Tanigaki
{\it et al}, 1991) and for Cs$_3$C$_{60}$ under pressure $T_c=40$ K has
 been reported (Palstra {\it et al}, 1995).
The great interest in the superconductivity of the alkali-doped C$_{60}$
 compounds is in particular due to these systems being a completely new 
class of superconductors, the large value of $T_c$ and the question whether 
or not such a large value of $T_c$ can be caused by the coupling to phonons
alone.
There has therefore been a great effort over the last five years to 
characterize and understand both the normal state and superconducting
properties of fullerides.       
The present review  deals only with C$_{60}$ compounds, since no other 
superconducting fullerides are known at present.         
\end{multicols}

\begin{figure}[h]
\unitlength1cm
\begin{minipage}[t]{17.cm}
\centerline{
\rotatebox{-1}{\resizebox{3in}{!}{\includegraphics{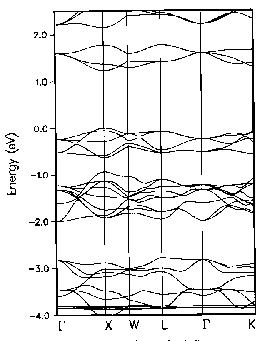}}}
}
\caption[]{\label{fig1} 
Some of the subbands around the Fermi energy for solid C$_{60}$ in the 
Fm${\bar 3}$ structure.
The bands at about -0.5 eV are the $h_u$ bands, which are occupied in solid C$_{60}$,
and the bands at about 1.5 eV are the $t_{1u}$ bands, which become populated in 
A$_n$C$_{60}$. The bands around -1.5 eV results from the overlapping $h_g$ and
$g_g$ bands. (From Erwin (1993).)} 
\end{minipage}
\hfill
\end{figure}

\begin{multicols}{2}
C$_{60}$ is the most symmetric
molecule in the sense that its point group (icosahedral)
with 120 symmetry operations is the largest point group
of the known molecules. It has the shape of a soccer ball. The 60 carbon atoms are
all equivalent and form 12 pentagons and 20 hexagons.

The C$_{60}$ molecules condense to a solid of weakly bound molecules. 
While the shortest separation between two atoms on the same molecule is about
1.4 \AA, the shortest separation between two atoms on different molecules
is about 3.1 \AA. The fullerites are therefore molecular solids, where
many of the molecular properties essentially survive in the solid. 
The discrete levels of a free C$_{60}$ molecule are only weakly 
broadened in the solid, leading to a set of essentially nonoverlapping
bands with a width of about 1/2 eV as is illustrated in Fig. \ref{fig1}.
The system therefore has two very different energy scales, the 
intramolecular ($E_I\sim 30$ eV) and the intermolecular ($W\sim 1/2$ eV)
 energy scales.  For an undoped C$_{60}$ solid, the $h_u$ band is full
and the $t_{1u}$ band is empty, and this system is  therefore a band insulator.
When solid C$_{60}$ is doped by alkali atoms,
  the alkali atoms donate about one
electron each to the $t_{1u}$ band
 (Erwin and Pederson, 1991; Saito and Oshiyama, 1991b; 
 Martins and Troullier, 1992; 
Satpathy {\it et al.}, 1992). Since the $t_{1u}$ band can take six electrons,
it is half-full for A$_3$C$_{60}$ (A=K, Rb), which is a metal
(Haddon {\it et al.}, 1991). A$_4$C$_{60}$
is, however, an insulator (Murphy {\sl et al.}, 1992; Kiefl
 {\sl et al.}, 1992; Benning {\it et al.}, 1993)
 although the $t_{1u}$ band is only partly filled and should be a 
metal according to band theory (Erwin, 1993; Erwin and Bruder, 1994).
 A$_6$C$_{60}$ is a band
insulator. Recently, AC$_{60}$ has attracted much interest because of its
rich and interesting phase diagram and because of its possibly 
quasi-one dimensional character (Stephens {\it et al.}, 1994; 
Chauvet {\it et al.}, 1994, 
Erwin {\it et al.}, 1995).
TDAE-C$_{60}$ has a ferromagnetic-like transition at $T=16$ K
(Allemand {\it et al.}, 1991). If the fullerides are considered as
organic compounds, they  have the highest superconducting and
(ferro)magnetic transition temperatures for organic compounds.

In A$_3$C$_{60}$ (A=K, Rb) the C$_{60}$ molecules are located on a
fcc lattice.
At low temperatures, the C$_{60}$ molecules take one of two likely
orientations in an essentially random way (Stephens {\it et al.}, 1991).
This orientational disorder 
has a substantial effect on some of the electronic properties
(Gelfand and Lu, 1992ab; 1993; Mele and Erwin, 1994),
 but has been neglected in most calculations.
 A certain short-range correlation 
of the orientations is expected on theoretical grounds (Gunnarsson
{\it et al.}, 1991; Mazin {\it et al.}, 1993a,
Erwin and Mele, 1994) and has been observed experimentally
(Teslic {\it et al.}, 1995).
The alkali atoms are located in the tetrahedral and octahedral holes.
In Rb NMR of Rb$_3$C$_{60}$ 
one would then expect to observe two lines. At low temperatures,
however, three lines are seen     (Walstedt {\it et al.}, 1993;
Zimmer {\it et al.}, 1993), raising some
 questions about our understanding of the precise structure.

The effective Coulomb interaction between two electrons on a C$_{60}$
molecule in a solid is  about 1-1${1\over 2}$ eV                          
(Lof {\it et al.}, 1992; Br\"uhwiler {\it et al.}, 1992;
Pederson and Quong, 1992; Antropov {\it et al.},
1992). The long-ranged Coulomb interaction in  
the A$_3$C$_{60}$ compounds leads to  a charge carrier plasmon,
due to the oscillations of the $t_{1u}$ electrons. This plasmon has the energy
$\omega_{pl}\sim 1/2$ eV,    an intermediate coupling constant
$(g/\omega_{pl})^2\sim 1$ (Antropov {\it et al.}, 1993b; Knupfer {\it et al.},
1993) and unusal dispersion (Gunnarsson {\it et al.}, 1996) and broadening
(Liechtenstein {\it et al.}, 1996). 

In systems where the superconductivity is driven by the electron-phonon
interaction, the transition temperature $T_c$ can be calculated 
in the Eliashberg (1960) theory.  A simple estimate of $T_c$ can 
be obtained from the McMillan (1968) formula 
\begin{equation}\label{eq:1.1}
T_c={\omega_{ln}\over 1.2  }{\rm exp}\lbrack - {1.04(1+ \lambda) \over
\lambda-\mu^{\ast}(1+0.62\lambda)} \rbrack,
\end{equation}
where $\omega_{ln}$ is a typical phonon frequency
(logarithmic average), $\lambda$ is the
 electron-phonon coupling, and $\mu^{\ast}$ is the Coulomb pseudopotential,
which describes the effects of the repulsive Coulomb interaction. 
To estimate $T_c$ we then in particular need to consider $\lambda$
and $\mu^{\ast}$.

C$_{60}$ has intramolecular vibrations (phonons) with energies up to
 $\omega_{ph}\sim 0.2$ eV. Only phonons with A$_g$ or H$_g$ symmetry couple 
to the $t_{1u}$ electrons. It was very early proposed that these phonons
drive the superconductivity (Varma {\it et al.}, 1991; Schluter {\it et
al.}, 1992ab; Mazin {\it et al.}, 1992). Estimates of the electron-phonon
interaction $\lambda\sim {1\over 2}-1$ fall in the right range to explain
the experimental values of $T_c$. It has therefore become widely, but
not universally, accepted that the intramolecular H$_g$ phonons drive
the superconductivity. Other phonon modes, such as librations,
 C$_{60}$-C$_{60}$ and alkali-C$_{60}$ vibrations are believed to 
play a small role for the superconductivity.
It has, however, also been argued that an electronic mechanism may drive
the superconductivity (Chakravarty {\it et al.}, 1991; Baskaran and Tosatti,
1991; Friedberg {\it et al.}, 1992).

The attractive interaction between two electrons induced by the
electron-phonon interaction is small ($\sim 1/10$ eV) compared with
the Coulomb repulsion between two electrons on the same C$_{60}$
molecule ($U\sim 1-1{1\over 2}$ eV). 
Nevertheless, for conventional superconductors it is argued that the effects 
of the Coulomb interaction are drastically reduced by retardation effects,
due to the very different energy scales for phonons and electrons.
The dimensionless Coulomb pseudopotential $\mu^{\ast}$                 
is therefore believed to be drastically renormalized for such
systems. For doped C$_{60}$ solids it has been controversial whether
the intramolecular ($E_I$) or intermolecular ($W$) (Anderson, 1991) 
energy scale is relevant for the 
retardation effects. In the latter case one may expect the retardation 
effects to be small, since $\omega_{ph} \sim W$. It has therefore 
been asserted that the phonons alone cannot explain the superconductivity
in A$_3$C$_{60}$ (Anderson, 1991).
Later work has provided support for the intermolecular energy scale
being the relevant one, but nevertheless found that the electron-phonon
mechanism may be sufficient if the Coulomb interaction is screened
as efficiently as predicted by RPA (Gunnarsson and Zwicknagl, 1992).

In the A$_3$C$_{60}$ (A=K, Rb) compounds many energy scales 
are similar $\omega_{ph} \sim \tilde E_{JT}\lesssim \omega_{pl}
 \sim W \lesssim U$,
where $\tilde E_{JT}$ is the Jahn-Teller energy of  C$_{60}^{-3}$, which may 
be of the order of several tenths of an eV. 
Furthermore, certain coupling constants are intermediate ($\lambda \sim 1$), 
$(g/\omega_{pl})^2\sim 1$). This is a situation where we may expect
 particularly interesting physics, but also a situation which is particularly
difficult to treat theoretically. $W<U$ suggests very strong
correlation effects (Lof {\it et al.}, 1992), which have, however, 
 not been considered  in most treatment of   
fullerides. The similar magnitude of 
$\omega_{ph}$ and  $W$ raises doubt about 
the strength of the retardation effects, as discussed above,  and it 
implies that Migdal's theorem is not valid. Corrections to Migdal's theorem
have been considered in a few cases (see, e.g., Pietronero {\it al.}, 
 1995, Takada 1993).
Jahn-Teller-like effects, beyond what is included in the Migdal theory,
 may also play a role (Auerbach {\it et al.}, 1994; Manani {\it et al.}, 1994).
The fact that $\tilde E_{JT}$ may not be much smaller than $W$ could also be 
important for the insulating character of A$_4$C$_{60}$ (Fabrizio and 
Tosatti, 1996).
In view of this, it is not surprising that
many of the normal state properties are unusual (see, e.g., 
Sawatzky (1995)) and several interesting, unresolved 
questions remain.   

There have been a number of reviews of the fullerenes. Several articles
appeared in J. Chem. Phys. Solid Vol. {\bf 53}, No. 12 (Fischer and Cox, 1992),  
in Vol. 30 No. 8 of Carbon (1992), in Vol. 48
 of Solid State Physics (1994), e.g., by Pickett (1994) and by Lieber and Zhang
(1994), and in a book edited by Billups and Ciufolini (1993). 
Early reviews were given by Weltner {\it et al.} \ (1989), by Kroto {\it et al.} \ (1991)
by Wilson {\it et al.} \ (1992) and by Loktev (1992). 
A number of conference proceedings have appeared (Hammond and Kuck, 1992;
Kumar {\it et al.}, 1993;
Fischer {\it et al.}, 1993; Prassides {\it et al.}, 1994, 
Kuzmany {\it et al.}, 1993; 1994a; 1995a). 
The field has further been reviewed 
by Holczer and Whetten (1992) and by Ramirez (1994) from an experimental 
and by Gelfand (1994) from a theoretical point of view.
A general review with the emphasis on the superconductivity was given by 
Hebard (1992).
Specific topics have recently been review by Golden {\it et al.}
(1995) (high energy spectroscopies), Kuzmany {\it et al.} (1995b)
(infrared spectroscopies), Pintschovius (1996) (neutron scattering)
Kuzmany {\it et al.} (1994b) (Raman scattering) and Buntar and Weber
(1996) (magnetic properties).
The theory of superconductivity in general has been presented in several 
text-books, e.g., Schrieffer (1964), Parks (1969) and Tinkham (1975),
and in reviews, e.g., Allen and Mitrovic (1982) and Carbotte (1990),
and it will not be addressed here.
Popular descriptions of the fullerenes have been given by Baggot (1994)
and Aldersey-Williams (1995).

\noindent
\begin{minipage}{3.375in}
\begin{table}[h]
\caption[]{Structure,    valence  and band involved for certain classes
of fullerene superconductors. (AE = Alkaline Earth, RE = Rare Earth)}
\begin{tabular}{lll}
Class &   Structure   & Valence  \\
\tableline
A$_3$C$_{60}$    &   Fm${\bar 3}$m   &  $n=3$ ($t_{1u}$)   \\
Na$_2$A$_x$C$_{60}$  &  Pa3   &   $n\le 3$ ($t_{1u}$)   \\
A$_{3-x}$Ba$_x$C$_{60}$  &   Fm${\bar 3}$m &  $n\ge 3$ ($t_{1u}$)  \\
(AE)$_x$C$_{60}$     &   Im${\bar 3}$, sc, orthorhomb &  $n >  6$ ($t_{1g}$) \\
(RE)$_x$C$_{60}$  &      orthorhomb     &     \\
\end{tabular}
\label{table0}
\end{table}
\end{minipage}

\section{Experimental results}
\subsection{Transition temperature}

The superconducting fullerides may be divided in five  classes; 
1)                 A$_3$C$_{60}$       (A = K, Rb,   Cs or some 
combination of these elements, 2)                  Na$_2$A$_x$C$_{60}$
($x\le 1$) ( A = K, Rb, Cs or a combination of these, 3)    
 A$_{3-x}$Ba$_x$C$_{60}$ (A=K, Rb, Cs),    4)     
Ca$_x$C$_{60}$ ($x \sim 5$) (Kortan {\it et al.},  1992a)
and Ba$_6$C$_{60}$ (Kortan {\it et al.},  1992b) or Ba$_4$C$_{60}$ (Baenitz
{\it et al.}, 1995) and Sr$_6$C$_{60}$ (Kortan {\it et al.}, 1994) 
and 5) Yb$_{2.75}$C$_{60}$ and Sm$_x$C$_{60}$. A summary is given in Table
\ref{table0}.

These classes differ in particular in  their structure and the C$_{60}$ 
valence. In the first class and in the second class with $x=1$
 three electrons are doped
into the $t_{1u}$ band, while in the second class with $x<1$ the doping 
is less than three electrons.
 In the third class the filling of the
$t_{1u}$ band is larger than three and in the fourth class the $t_{1u}$
band is full and the $t_{1g}$ band is 
populated (Chen {\it et al.}, 1992;  Wertheim {\it et al.},
1992; Erwin and Pederson, 1993; Saito  and Oshiyama, 1993;
 Knupfer {\it et al.}, 1994). 
 The $t_{1g}$ states furthermore  strongly hybridize with alkaline 
earth states
(Chen {\it et al.}, 1992; Erwin and Pederson, 1993; 
Saito and Oshiyama, 1993), in contrast to the very weak
hybridization of the $t_{1u}$ states with the alkali states in
A$_3$C$_{60}$.

In the first three classes, the C$_{60}$ molecules are located on a fcc
lattice. 
 The first (Stephens {\it et al.}, 1991) and third (Yildirim, 1996b) class have
 the space group  Fm${\bar 3}$m,
 and  the molecules more or less randomly take one of two preferential 
orientations. In the second class the space group is Pa${\bar 3}$
(Kniaz {\it et al.}, 1993; Prassides {\it et al.}, 1994), possibly with
 some distortions for  Na$_2$KC$_{60}$
 and Na$_2$RbC$_{60}$ (Maniwa {\it et al.}, 1995). For these systems
there are four C$_{60}$ molecules per unit cell, 
all having different orientations.
In the fourth group
 Ca$_5$C$_{60}$ has a simple cubic structure with the molecules at the
fcc lattice positions (Kortan {\it et al.}, 1992a), Ba$_4$C$_{60}$ has
 an orthorhombic structure (Baenitz {\it et al.}, 1995), 
Ba$_6$C$_{60}$ (Kortan {\it et al.}, 1992b) and Sr$_6$C$_{60}$ (Kortan {\it et al.}, 1994)
have a bcc structure (Im${\bar 3}$).

\begin{figure}[h]
\unitlength1cm
\begin{minipage}[t]{8.5cm}
\centerline{
\rotatebox{270}{\resizebox{3in}{!}{\includegraphics{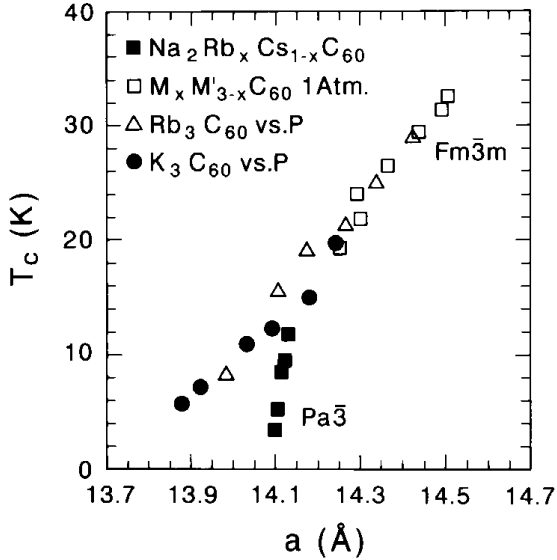}}}
}
\caption[]{\label{fig2}
 $T_c$ as a function of the lattice parameter $a$ for
the Fm${\bar 3}$m (triangles, circles and open squares)
 and Pa${\bar 3}$ (filled squares) superconductors. For K$_3$C$_{60}$
 and
Rb$_3$C$_{60}$ the lattice parameter was varied by changing the pressure,
while for M$_x$M$^{'}_{3-x}$C$_{60}$ (with M, M$^{'}$=K, Rb, Cs) the
lattice parameter was varied by changing the composition.
(After Yildirim et al. 1994).}
\end{minipage}
\hfill
\end{figure}

The $T_c$ as a function of the lattice parameter $a$ has been determined
by a large number of groups for both the first (Tanigaki {\it et al.},  1991;
Sparn {\it et al.}, 1991;
Fleming {\it et al.},  1991; Zhou {\it et al.},  1992) and second (Tanigaki {\it et al.},  1992;
Yildirim {\it et al.},  1994; Mizuki {\it et al.}, 1994, Yildirim {\it et al.}, 1995)  
(for $x=1$) class.
 In Fig. \ref{fig2} these results are summarized. 
The figure illustrates that in the first class, $T_c$ varies smoothly with
$a$, relatively independently of the alkali atoms in the compound. Thus       
if pressure is applied to Rb$_3$C$_{60}$ to give the same $a$ as for
K$_3$C$_{60}$ at normal pressure,
 the corresponding $T_c$'s are approximately equal (Zhou, 1992).
It is therefore the lattice parameter, not the alkali atom, which 
determines $T_c$.            This seems to exclude the alkali vibrations
 as the main
driving force for the superconductivity, since the vibration frequency 
and thereby $T_c$ would then be expected to depend on the mass of the alkali
atoms. Recent work has,
however,  questioned these results, finding a different
pressure-volume relation for Rb$_3$C$_{60}$ 
than the earlier work and thereby a different $T_c-a$ dependence 
(Diederichs {\it et al.}, 1996). A certain deviation between Rb$_2$CsC$_{60}$
under pressure and Rb$_3$C$_{60}$ was also observed by Schirber {\it et al.}
(1993). 
This would imply that changing
the lattice parameter by physical or chemical pressure does not
lead to identical results, although the deviations are not dramatic. 
 A smooth variation with lattice parameter is observed
for the second group ($x=1$) (see, Fig. \ref{fig2}). It is, however,
interesting that the slopes of the two classes are
very different.

(NH$_3$)$_4$Na$_2$CsC$_{60}$ probably also belongs to the first class
(with the likely structure Fm${\bar 3}$ or Fm${\bar 3}$m) and its 
$T_c=29.6$ K is essentially what one would expect
from the large lattice parameter $a=14.473$ \AA \ (Zhou {\it et al.}, 1993). 
Cs$_3$C$_{60}$ has a different (noncubic) structure and a different pressure
dependence than the other A$_3$C$_{60}$ compounds 
and it should not be considered as a memeber of the first class. 
The maximum $T_c$ found for Cs$_3$C$_{60}$
is 40 K (Palstra {\it et al.}, 1995).
Another interesting system is NH$_3$K$_3$C$_{60}$ which like K$_3$C$_{60}$ 
has three electrons in the $t_{1u}$ orbital but an orthorhombic
distortion of the fcc lattice (Rosseinsky {\it et al.}, 1993; Zhou {\it et al.}, 1995).
 At ambient pressure
 this system shows no superconductivity above 4 K and it is possibly
an insulator. Under pressure, however,  superconductivity occurs
with the maximum observed $T_c$ being 28 K.
It had been suggested that only cubic fullerides can be superconductors,
  but Cs$_3$C$_{60}$ and NH$_3$K$_3$C$_{60}$ under
 pressure now provide examples
of  noncubic superconducting fullerides. These compounds nevertheless
raise interesting questions about the role
of the deviation from cubic symmetry, since the application of pressure
reduces this symmetry lowering and increases $T_c$ (Palstra {\it et al.}, 1995;
Zhou {\it et al.}, 1995). Alternatively, the closeness to a Mott transition
may be important for the pressure dependence (Palstra {\it et al.}, 1995;
 Zhou {\it et al.}, 1995).

\begin{figure}[h]
\unitlength1cm
\begin{minipage}[t]{8.5cm}
\centerline{
\rotatebox{270}{\resizebox{3in}{!}{\includegraphics{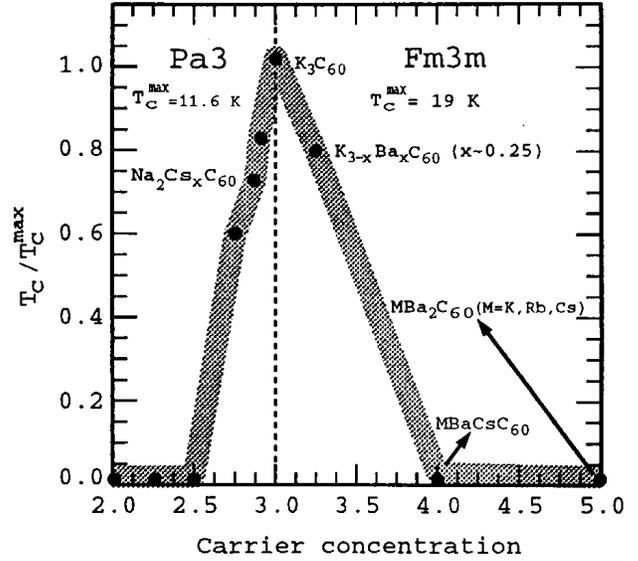}}}
}
\caption[] {\label{fig2a}$T_c$ as a function of C$_{60}$ valence $n$ for the
Na$_2$Cs$_x$C$_{60}$ ($x\le 1$, $n\le 3$) and K$_{3-x}$Ba$_x$C$_{60}$ ($n\ge
3$)
compounds. $T_c$ is scaled by the $T_c$ ($T_c^{max}$) of the
end members Na$_2$CsC$_{60}$ and K$_3$C$_{60}$.
 The heavy line is a guide for the eye. Observe the
different lattice structures. (After Yildirim {\it et al.}, 1996a).}
\end{minipage}
\hfill
\end{figure}

Yildirim {\it et al.} \ (1996a) have studied Na$_2$Cs$_x$C$_{60}$ for $x=0.25$,
0.50 and 0.75. The values of $T_c$ found for these compounds are
shown in Fig. \ref{fig2a} as a function of the expected valence $n=2+x$.
It is found that $T_c$ drops quickly as the valence is reduced. 
For $n=2.5$ no superconductivity was observed above $T=0.5$ K,
while the systems are believed to be metallic down to $n$=2.25 
(Yildirim {\it et al.}, 1996a).

Yildirim {\it et al.} \ (1995; 1996ab) studied Rb$_{3-x}$Ba$_x$C$_{60}$ for
 $x$=0.25, 1 and 2.
Based on Raman data and theoretical considerations, they concluded
that the C$_{60}$ valence is $n=3+x$ and        
that Ba is essentially completely ionized.                               
The results for $T_c$ are shown in Fig. \ref{fig2a}.
$T_c$ drops rapidly with $x$ and for $x=1$ $(n=4)$ and $x=2$ ($n=5$) 
no superconductivity
was observed above $T=0.5$ K. These compounds show a very small
Pauli susceptibility, and it was tentatively concluded that they are
 weakly metallic (Yildirim {\it et al.}, 1996b).
Fig. \ref{fig2a} suggests a rapid drop of $T_c$ when the C$_{60}$ valence
deviates from 3. The reduction for $n>3$, but not for $n<3$ (see Sec. VIIA.2),
   may be explainable by the reduction of the 
density of states with increasing filling, taking the merohedral disorder
into account (Yildirim {\it et al.}, 1995). 
The reduction of $T_c$ as $n$ is reduced below 3 seems not be understood.
Fig. \ref{fig2a} furthermore appears to 
be inconsistent with the conclusion that stoichiometric A$_3$C$_{60}$
is an insulator (Lof {\it et al.}, 1992). In such a case one might expect $T_c$
to increase when the occupancy of the $t_{1u}$ orbital
moves away    from integer values, in analogy with the high $T_c$
 superconductors, while Fig. \ref{fig2a} shows the opposite trend.

In the fourth group the measured transition temperatures are  8.4 K 
(Ca$_x$C$_{60}$), 4 K (Sr$_x$C$_{60}$) and 7 K (Ba$_x$C$_{60}$).

Recently, superconductivity has been found in RE$_x$C$_{60}$, where RE
is a rare earth atom. \"Ozdas {\it et al.} \ (1995) discovered that
 Yb$_{2.75}$C$_{60}$ is superconducting with  $T_c=6$ K.
This system has a slight orthorhombic distortion 
of   the Pa${\bar 3}$ structure  and the structure is Pcab.
Based on EXAFS measurements it was concluded that Yb is divalent,
which would imply the C$_{60}$ valence 5.5 (\"Ozdas {\it et al.}, 1995). 
Chen and Roth (1995) found superconductivity in Sm$_x$C$_{60}$ ($x<3$)
with $T_c=8$ K.

It is also interesting to observe that Li$_2$AC$_{60}$ is not       
superconducting (Hirosawa {\it et al.}, 1994). This system has a different
structure than Na$_2A$C$_{60}$ and A$_3$C$_{60}$. It has the space group
Fm${\bar 3}$m, but with an orientational state of C$_{60}$
which is best  modeled as a quasi-spherical unit
(Hirosawa {\it et al.}, 1994).

\subsection{Superconductivity gap}

The superconductivity gap $\Delta$ is of great interest, since a value of
the reduced gap $2\Delta/T_c$ which is substantially larger than the BCS
value (3.53) indicates that strong-coupling effects are important.
For intermediate coupling, it is found that (Mitrovic {\it et al.}, 1984)      
(see, also Carbotte (1990) for a discussion)
\begin{equation}\label{eq:2.1}
{2\Delta\over T_c}=3.53\lbrack 1+12.5({T_c\over \omega_{ln}})^2
{\rm ln}({\omega_{ln}\over 2T_c})\rbrack,
\end{equation}
where $\omega_{ln}$ is the logarithmic average phonon frequency.              
This formula says that the deviation from the BCS result is small if 
$T_c/\omega_{ln}<<1$.                            
An early measurement of $\Delta$ was performed using point contact tunneling
in STM, giving $2\Delta/T_c=5.3$ for Rb$_3$C$_{60}$         (Zhang et al., 1991b).   
Later STM measurements on single crystals have given reduced gaps 
$2\Delta/T_c=5.4$ (Jess {\it et al.}, 1994) and $2\Delta/T_c=2.0-4.0$
(Jess {\it et al.}, 1996). The variations in $2\Delta/T_c$ 
were ascribed to variations in the local stoichiometry at the tip position.
From  measurements using NMR the values 3.0 for K$_3$C$_{60}$ and 4.1
for Rb$_3$C$_{60}$ (Tycko et al., 1992), 4.3 for K$_3$C$_{60}$
 (Sasaki et al., 1994) and $3.4\pm 0.2$ K$_3$C$_{60}$ 
(Auban-Senzier {\it et al.}, 1993)
were deduced.
Recent NMR measurements were found to be in good agreement with the 
BCS value for the gap (Stenger {\it et al.}, 1995).
 Using muon spin relaxation, Kiefl et al. (1993) obtained 
the   value 3.6 for Rb$_3$C$_{60}$. Recent optical measurements have given
the values 3.44 for K$_3$C$_{60}$ and 3.45 for Rb$_3$C$_{60}$ (Degiorgi et al., 1994,
Degiorgi, 1995). Finally, the reduced gap was      measured in photoemission,
giving the value 4.1 for Rb$_3$C$_{60}$ (Gu et al., 1994). 
We note that the data obtained from different experiments show a substantial
variation. Most recent results, however, tend to scatter around the 
weak-coupling limit and are  
consistent with this limit,  
if a reasonable uncertainty is assumed for these  experiments.

\subsection{Isotope effect}

\begin{figure}[h]
\unitlength1cm
\begin{minipage}[t]{8.5cm}
\centerline{
\rotatebox{269}{\resizebox{!}{3in}{\includegraphics{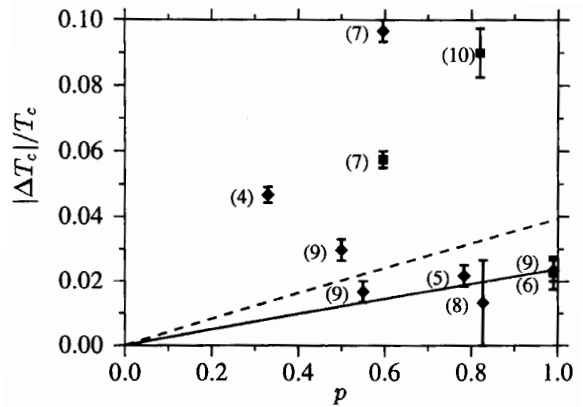}}}
}
\caption[]{\label{fig2aa}The isotope shift $\Delta T_c$ of $T_c$
 for K$_3$C$_{60}$ (squares)
and Rb$_3$C$_{60}$ (diamonds) as a function of the fraction $p$ of the
substitution ($p=1$ completely ${}^{13}C$). The dashed line shows
the BCS result $\alpha=0.5$ and the full line $\alpha=0.3$.
The results are obtained from (4) Ebbesen {\sl et al.} (1992),
(5) Ramirez {\sl et al.}  (1992a), (6) Chen and Liber (1992),
(7) Zakhidov {\sl et al.} (1992), (8) Bethune {\sl et al.} (1992),
(9) Chen and Lieber (1993), (10) Auban-Senzier {\sl et al.} (1993).
(After Deaven and Rokhsar, 1993).}
\end{minipage}
\hfill
\end{figure}

The isotope effect may provide interesting information about the
mechanism for superconductivity. In the BCS theory 
for a system with only one type of ions with the mass $M$, the transition  
temperature behaves as $T_c\sim M^{-\alpha}$, where $\alpha=0.5$.   
For the A$_3$C$_{60}$ compounds, the C$_{60}$ phonons are expected
to be the important ones, and one is therefore interested in the
isotope effect when ${}^{12}C$ is substituted by  ${}^{13}C$. For the 
case of complete (99$\%$) substitution, it was found that 
$\alpha=0.30\pm0.06$ for K$_3$C$_{60}$ (Chen and Lieber, 1992) and $\alpha=
0.30\pm0.05$ for Rb$_3$C$_{60}$ (Chen and Lieber, 1993). Interesting and 
contradictory results have been obtained for incomplete substitution.
Chen and Lieber (1993) found that for a single peak mass distribution of
C$_{60}$ molecules (Rb$_3$(${}^{13}$C$_{0.55}{}^{12}$C$_{0.45})_{60}$),
the value of $\alpha$ was consistent with the result for complete
substitution. On the other hand, they found that a two peak mass distribution
(Rb$_3({}^{13}$C$_{60})_{0.5}({}^{12}$C$_{60})_{0.5}$) gave a much
larger value of $\alpha$ ($\sim 0.8$). A value of $\alpha$ which is larger
than 0.5     is unusual, but it was also observed by three other
groups. Ebbesen {\it et al.} (1992a) obtained $\alpha=1.4\pm 0.5$ for 
Rb$_3$C$_{60}$ (33$\%$ substitution), 
Zakhidov {\it et al.} (1992) obtained $\alpha=(1.2-1.43)\pm0.2$
for K$_3$C$_{60}$ and $\alpha=(2-2.25)\pm 0.25$ for Rb$_3$C$_{60}$
(60$\%$ substitution) and Auban-Senzier {\it et al.} (1993) obtained 
$\alpha=1.45\pm 0.3$ for 82 $\%$ substitution.
 However, these results
were observed for distributions of C$_{60}$ masses with essentially only
one peak, while the $\alpha>0.5$ was obtained by Chen and Lieber (1993)
for a two-peak distribution. Finally Ramirez {\it et al.} (1992a) obtained     
$\alpha=0.37\pm 0.05$ for mass distribution of 15 $\%$ ${}^{12}$C$_{60}$
and the rest (${}^{13}$C$_{0.9}{}^{12}$C$_{0.1})_{60}$.  
For the partially substituted compounds, the experimental situation 
thus appears to be unclear.
The results are summarized in Fig. \ref{fig2aa}.
The theoretical considerations have so far, however, been made for 
completely substituted systems, where $\alpha=0.3$ should be
the best available experimental result. The theoretical treatment 
of partially substituted compounds remains a challenging problem.

The isotope effect with respect to the alkali dopants have also been studied.
It was found that within the experimental uncertainty there is no
isotope effect (Ebbesen {\it et al.} 1992b; Burk {\it et al.} 1994ab). This is 
consistent with the experimental result that applying pressure to Rb$_3$C$_{60}$ 
so that the lattice parameter is reduced to that of K$_3$C$_{60}$ leads to
about the same $T_c$ for both systems (Fleming {\it et al.} 1991). This strongly
suggests that the alkali ions have a very weak influence on $T_c$ except
for the indirect influence via the lattice parameter.

\subsection{Other properties}

A$_3$C$_{60}$ is a type II superconductor. The lower critical field       
 $H_{c1}$ has been measured to be 130 Oe (Holczer {\sl et al.}, 1991b)
for K$_3$C$_{60}$ and 120 Oe (Sparn   {\sl et al}, 1992) and 60$\pm5$ Oe
(Ramirez {\sl et al.}, 1994) for Rb$_3$C$_{60}$.
To estimate the penetration depth $\lambda(0)$, one can now use       
the relationship with    $H_{c1}$                               
$\lambda(0)$ (Tinkham, 1975)
\begin{equation}\label{eq:2.2}
H_{c1}={\phi_0\over 4\pi \lambda(0)^2}{\rm ln}({\lambda(0) \over \xi_{GL}}),
\end{equation}
where $\phi_0=hc/e$ and $\xi_{GL}$ is the Ginzburg-Landau coherence 
length. Putting $\xi_{GL}=30$ \AA,
which is reasonable in view of the results below,
 one obtains the values $\lambda(0)=$ 
2500 \AA \ for K$_3$C$_{60}$ and 2600 and 3700 \AA \ for Rb$_3$C$_{60}$.         
The penetration depth has also been estimated from $\mu$SR with the results
4800 \AA \ for K$_3$C$_{60}$ and 4200 \AA \ for Rb$_3$C$_{60}$ (Uemura {\it et al.},
 1991; 1993),
and from the optical conductivity with the result 8000 \AA \ for   
both K$_3$C$_{60}$ and Rb$_3$C$_{60}$ (Degiorgi {\it et al.}, 1992) and from
NMR data giving 6000 \AA \ for K$_3$C$_{60}$ and 4600 \AA \ for Rb$_3$C$_{60}$
(Tycko {\it et al.}, 1992).
Uemura {\sl et al.} (1991) also                             
deduced the temperature dependence of penetration depth
$\lambda(T)$. They could fit these results with the formula
$\lbrack 1-(T/T_c)^{\alpha}\rbrack$, with $\alpha=3.2$. From this
they concluded that K$_3$C$_{60}$ probably is an isotropic
s-wave superconductor, as expected.
Based on values for $\lambda(0)$ and $\xi$, Uemura {\it et al.} (1991)
also estimated an ``effective'' Fermi temperature $T_F^{eff}$.
Combining their small value (470 K) for $T_F^{eff}$ with $T_c$ in
a $T_c$ vs. $T_F^{eff}$ plot, they found that the A$_3$C$_{60}$ belongs
to a group of ``exotic'' superconductors, including the High $T_c$ cuprates
and the organic BEDT systems.

The upper critical field $H_{c2}(0)$ is of interest, since it allows
an estimate of the coherence length via (Tinkham, 1975)
\begin{equation}\label{eq:2.3}
H_{c2}(0)={\phi_0\over 2 \pi \xi_{GL}}.
\end{equation}
Several groups have measured $H_{c2}^{'}\equiv dH_{c2}/dT$ and then obtained
the value of $H_{c2}(0)$ from the Wertheimer-Helfand-Hohenberg 
(Wertheimer {\it et al.}, 1966) 
theory.
Based on this, $\xi_{GL}$ was deduced to be 26 \AA \ for K$_3$C$_{60}$ (Holczer
{\it et al.}, 1991b) and 20 \AA \ for Rb$_3$C$_{60}$ (Sparn {\it et al.}, 1992)
from dc-magnetization. The high-field susceptibility  gave the values
 29 \AA \ to 33 \AA \ (Boebinger {\it et al.}, 1992)
and 35 \AA \ (Holczer and Whetten, 1992) for K$_3$C$_{60}$ and 30 \AA \ 
(Holczer and Whetten, 1992) for Rb$_3$C$_{60}$. 
From the ratio of $\xi_{GL}/\lambda$ it is clear that $A_3$C$_{60}$
is a type II superconductor.
The magnetic properties of the C$_{60}$ superconductors have been reviewed
by Buntar and Weber (1996).

The jump in the specific heat at $T_c$ was estimated by Ramirez {\it et al.} \ (1992)
to be $\Delta C=68\pm13$ mJ/mole-K$^2$ for K$_3$C$_{60}$
and by Burkhart and Meingast (1996) to be 64$\pm$14 and 75$\pm$14
 mJ/mole-K$^2$ for K$_3$C$_{60}$ and Rb$_3$C$_{60}$, respectively.
The Hebel-Slichter peak was not seen in early NMR measurements 
(Tycko {\it et al.}, (1992), but          
has been seen in $\mu$SR (Kiefl {\it et al.}, (1993))
and more recent NMR (Stenger {\it et al.}, 1995) measurements.
The vertex glass and vertex fluid phase diagram of A$_3$C$_{60}$
has been studied by Lin {\it et al.} (1994).

In Table \ref{table00} we have summarized some superconducting 
parameters. The table has mainly been extracted from Ramirez (1994),
but it has been partly updated in the view of newer experiments.

\noindent
\begin{minipage}{3.375in}
\begin{table}[t]
\caption[]{Superconducting parameters for K$_3$C$_{60}$ and Rb$_3$C$_{60}$.
(Mainly
after Ramirez (1994)).}
\begin{tabular}{lll}
Property   & K$_3$C$_{60}$  & Rb$_3$C$_{60}$  \\
\tableline
$T_c$     & 19.5 K    & 29.5 K    \\
$\lambda$ ($\mu$SR) &   $4800\pm 200$ \AA &  $4200$ \AA          \\
$\lambda$ (optical) &   $8000\pm 500$ \AA &  $8000\pm 500$ \AA   \\
$\lambda$ (NMR)     &   $6000$ \AA        &  4600 \AA            \\
$\xi_0$ ($\chi_{ac}$) & 29-35 \AA         &  30 \AA              \\
$\Delta C/T_c$        & $64\pm14$ mJ/mole-K$^2$ &  $75\pm14$ mJ/mole-K$^2$ \\
$2\Delta/k_bT_c$ (NMR)  & 3.0             & 4.1                  \\
$2\Delta/k_bT_c$ (optical)  & 3.44            & 3.45                 \\
$2\Delta/k_bT_c$ ($\mu$SR)  &                 & $3.6\pm0.3$          \\
$2\Delta/k_bT_c$ (tunneling) & $5.3\pm0.2$  &  $5.2\pm0.3$   \\
$2\Delta/k_bT_c$ (tunneling) &              &  $5.4, 2.0-4.0$ \\
\end{tabular}

\label{table00}
\end{table}
\end{minipage}

\section{Phonons and electron-phonon coupling}
The phonons and the electron-phonon interaction are very important
for the properties of alkali-fullerides. The electron-phonon
interaction is usually assumed to cause the superconductivity.
It should, however, also be important for the transport properties
in general, and may also play a substantial role for other 
electronic properties.

\begin{figure}[h]
\unitlength1cm
\begin{minipage}[t]{8.5cm}
\centerline{
\rotatebox{269}{\resizebox{!}{3in}{\includegraphics{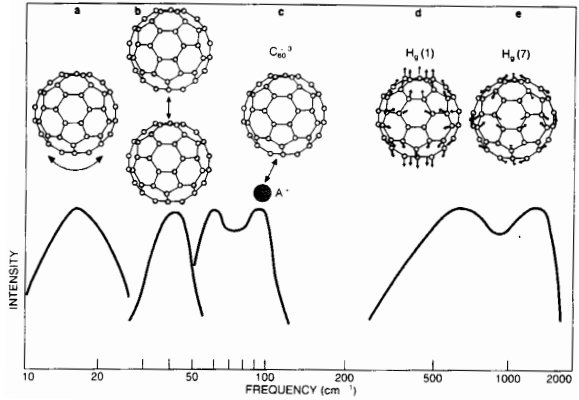}}}
}
\caption[]{\label{figphonon}Schematic representation of
various phonons in A$_3$C$_{60}$ compounds. The figure shows (a) librations,
(b) optic C$_{60}$-C$_{60}$ phonons, (c) A-C$_{60}$ phonons and (d)-(e)
 intramolecular H$_{g}$ modes. The figure indicates the radial and
 tangential character of the low-lying and and high-lying H$_g$ modes,
respectively. (After Hebard, 1992).}

\end{minipage}
\hfill
\end{figure}

The phonons of alkali-fullerides can be divided in subgroups, which       
reflect the molecular nature of the solid. The highest modes are due to
 intramolecular vibrations, with energies in the range 273-1575 cm$^{-1}$
(0.034-0.195 eV) for an undoped C$_{60}$
solid (Bethune {\it et al.},  1991). The intermolecular phonons
occur at substantially lower energies in the range up to about 
140 cm$^{-1}$ (17 meV) (Prassides {\it et al.},  1991;
Pintschovius, 1996). These modes contain both 
C$_{60}$-C$_{60}$  and A-C$_{60}$ (A=K, Rb) vibrations.
Finally, there are librations, which occur at an energy of about
4-5  meV (Christides {\it et al.},  1992; Schober {\it et al.}, 1994).
These various phonons are schematically illustrated in Fig. \ref{figphonon}.
It is believed that the intramolecular modes play the main role for
the superconductivity. We therefore focus on these modes, but also the 
other modes are discussed.

\subsection{Intramolecular modes}

For the intramolecular phonons, the effects of the interactions between
the C$_{60}$ molecules are small.     These phonons can therefore
be approximately classified by the icosahedral point group.
In alkali-fullerides, the $t_{1u}$ band is (partly) populated. The
 coupling of the
phonons to this band is therefore of particular interest. From 
symmetry arguments, it follows that only the two phonons with A$_g$
and the eight phonons with H$_g$ symmetry can couple (Varma {\it et al.},
1991; Lannoo {\it et al.}, 1991).
We therefore focus our interest on these modes.

\subsubsection{Calculations of phonon frequencies}

There have been a number of calculations of the phonon frequencies.
In Table \ref{tableI} we show the results of  calculations
using the semi-empirical MNDO method (Varma {\it et al.},  1991), an
empirical force-constant model (Jishi {\it et al.}, 1992)
and the {\sl ab initio} density functional method (Hohenberg and Kohn, 1964;
Kohn and Sham, 1965) in the local density approximation (LDA)
but using different numerical methods ( 
Antropov {\it et al.},  1993b; Quong {\it et al.},  1993; 
Faulhaber {\it et al.},  1993). 
These calculations were all for a free C$_{60}$ molecule, except the
one by Antropov {\it et al.},  (1993b), which was for K$_3$C$_{60}$ in
the so-called uni-directional structure, where all C$_{60}$ molecules
have the same orientation (where a cubic axis and a two-fold molecular
axis coincide).
These calculations can be considered as representative for the
large number of calculations that have been performed (see also, e.g., 
Negri {\it et al.},  1988; Adams {\it et al.}, 1991; 
Kohanoff {\it et al.},  1992; Jones {\it et al.}, 
 1992; Onida and Benedek, 1992; Bohnen {\it et al.}, 1995).
 These results are compared with experimental
results obtained from Raman scattering (Bethune {\it et al.},  1991).
We can see that the best empirical and {\sl ab initio} calculations have
errors of the order one per cent.
The small errors in, for instance, the calculations of  
Quong {\it et al.}, (1993) suggests that the LDA errors for the frequencies
are small.

Bohnen {\it et al.} (1995) have calculated the changes in the 
phonon frequencies when going from solid C$_{60}$ to K$_3$C$_{60}$,
and found a softening of the modes, in particular the high-frequency
ones.

\begin{minipage}{3.375in}
\begin{table}
\caption[]{
Phonon frequencies (in cm$^{-1}$) for the H$_g$ and A$_g$ modes.
Different theoretical calculations are compared with
experimental results for undoped solid C$_{60}$.}
\begin{tabular}{ccccccc}
\multicolumn{1}{c}{Mode}&
\multicolumn{6}{c}{Energy} \\
   & Varma\tablenotemark[1]  &  Jishi\tablenotemark[2]
 & Antropov\tablenotemark[3] & Quong\tablenotemark[4] &
 Faulhaber\tablenotemark[5]
& Exp.\tablenotemark[6] \\
\tableline
H$_g$(8)&1721&1575&1462&1576&1567&1575 \\
H$_g$(7)&1596&1401&1387&1443&1425&1428 \\
H$_g$(6)&1406&1217&1290&1244&1200&1250 \\
H$_g$(5)&1260&1102&1091&1093&1067&1099 \\
H$_g$(4)& 924& 788& 785& 767& 750&774   \\
H$_g$(3)& 711& 708& 753& 727& 640&710  \\
H$_g$(2)& 447& 439& 454& 439& 421&437   \\
H$_g$(1)& 263& 269& 281& 258& 249&273   \\
A$_g$(2)&    &1468&1463&1499&1493&1470    \\
A$_g$(1)&    & 492& 458& 478& 459&496      \\
\end{tabular}
\label{tableI}
\tablenotetext[1] {Varma {\it et al.}, 1991}
\tablenotetext[2] {Jishi {\it et al.}, 1992}
\tablenotetext[3] {Antropov {\it et al.}, 1993}
\tablenotetext[4] {Quong {\it et al.}, 1993}
\tablenotetext[5] {Faulhaber {\it et al.}, 1993}
\tablenotetext[6] {Bethune {\it et al.}, 1991}
\end{table}

\end{minipage}

 The four highest modes are primarily
made up of vibrations tangentially to the C$_{60}$ surface, while
the three lowest modes have mainly radial character (Schluter {\it et al.}, 1992ab;
Antropov {\it et al.},  1993b), as indicated in Fig. \ref{figphonon}.
 The reason is that the molecule is much stiffer
against tangential than radial distortions, since the former require larger
bond length changes.
The phonon frequencies are high, extending up to about 0.2 eV.
This is due to both the light mass of carbon and to the stiffness of
the C-C bond.

\subsubsection{Calculations of electron-phonon coupling}
The electron-phonon coupling plays a crucial role in the theory of
superconductivity. We introduce the dimensionless coupling
to the phonon mode $\nu$ (Rainer, 1986)
\begin{eqnarray}\label{eq:3.1}
\lambda_{\nu}={2\over N(0)}\sum_{\bf q}{1\over \omega_{\nu {\bf q}}}
\sum_{n,m,{\bf k}}|g_{n{\bf k},m({\bf k+q})}(\nu)|^2
\nonumber \\
\times\delta(\varepsilon_{n{\bf k}})\delta(\varepsilon_{m({\bf k+q})}-
\varepsilon_{n{\bf k}}-\omega_{\nu {\bf q}}),
\end{eqnarray}
where $N(0)$ is the electron density of states per spin at the Fermi energy
and $\omega_{\nu{\bf q}}$ is the energy of the $\nu$th phonon with
the wave vector ${\bf q}$. The energy of the $n$th electronic band
with the wave vector ${\bf k}$ is $\varepsilon_{n{\bf k}}$.
Finally, $g_{n{\bf k},m({\bf k+q})}(\nu)$ is the matrix element 
between the two electronic state $n{\bf k}$ and $m({\bf k+q})$
 of the potential created when a phonon $\nu{\bf q}$ is excited. 
Thus it follows
\begin{equation}\label{eq:3.2}
g_{n{\bf k},m({\bf k+q})}(\nu)=\langle n{\bf k}|\Delta V_{\nu{\bf q}}|
m({\bf k+q})\rangle {1\over \sqrt{2M\omega_{\nu {\bf q}}}},
\end{equation}
where $\Delta V_{\nu{\bf k}}$ is the induced potential per unit displacement
along the phonon normal coordinate,    $\sqrt{2M\omega_{\nu{\bf q}}}$
is the phonon amplitude and $M$ is the reduced mass.

The calculation of $\lambda$ requires a double sum over a Brillouin zone 
of a quantity ($g$) which is very time consuming to calculate.
In general, an accurate calculation of $\lambda$ is therefore very
complicated. For the intramolecular modes in the fullerides,
important simplifications follow from the large difference between
the intramolecular ($E_I$) and intermolecular ($W$) energy scales
(Lannoo {\it et al.}, 1991). One can then assume that the phonon eigenvalues and 
eigenvectors are independent of ${\bf q}$. If in addition we
assume that $W<<E_I$ in the calculation of $g_{n{\bf k},m({\bf k+q})}(\nu)$,
but keep the full structure of $|n{\bf k}>$,                      
 the complicated Brillouin zone sums can be performed
analytically, and the problem is essentially reduced to a molecular
problem (Lannoo {\it et al.},  1991).
The accuracy of these assumptions has been tested for a simple 
tight-binding model, where it was possible to perform the Brillouin
zone summations (Antropov {\it et al.}, 1993b). It was found that the simplified     
calculation agreed fairly well with the full calculation, justifying
the assumptions of Lannoo {\it et al.} \ (1991). 

 Thus it is sufficient to calculate
the shift $\Delta \varepsilon_{\nu \alpha}$  of the $t_{1u}$ level $\alpha$
for a free C$_{60}$ molecule per unit displacement of the $\nu$th 
phonon coordinate. One then finds that 
\begin{equation}\label{eq:3.3}
\lambda \sim N(0)\sum_{\nu\alpha}{\Delta \varepsilon_{\nu\alpha}^2
\over \omega_{\nu}^2}.
\end{equation}
 This gives a molecular specific quantity 
which must then be multiplied by the density of states $N(0)$, which is 
determined by the intermolecular hopping. The molecular calculation
therefore gives a value for $\lambda_{\nu}/N(0)$, and it is appropriate
to compare this quantity from different calculations. We observe that
the quantity $\lambda_{\nu}/N(0)$  grows with the intramolecular
energy scale, while $N(0)$ is inversely proportional to the
intermolecular hopping. The large difference between these two energy
scales therefore favors a large value of $\lambda$ (Schluter {\it et al.},
1992a). 
It is also clear that as the lattice parameter increases, the overlap
between the C$_{60}$ molecules is decreased, reducing the band width.
This leads to a larger density of states $N(0)$ and a larger $\lambda$,
explaining the increase in $T_c$ with lattice parameter (Rosseinsky 
{\it et al.}, 1991; Flemming {\it et al.}, 1991; Tanigaki {\it et al.},
1993; Ramirez, 1994)
(see, also Fig. \ref{fig2}).

We notice that the results are quite sensitive to the phonon eigenvectors,
and small changes in the eigenvectors can lead to large shifts of
the coupling strength between different phonons.
One can show, however,
 that $\sum_{\nu\alpha} (\Delta \varepsilon_{\nu\alpha})^2$
is independent of the phonon eigenvectors (Antropov {\it et al.},  1993b).
 A shift of coupling strength
between phonons with similar energies therefore has a small effect
on the total value of $\lambda$, while a shift between phonons of 
very different energies can dramatically change the total $\lambda$
due to the energy denominator in Eq. (\ref{eq:3.3}). 
For instance, in the calculation of Antropov {\it et al.} \ (1993b), 
mixing 5 $\%$ of the weight ($\sqrt{0.05}$) of the seventh eigenvector
into the eighth and vice versa changes the couplings 0.020 and 0.022
in Table \ref{tableII} to 0.010 and 0.030, i.e. a small change 
in the total coupling. The same mixing between the second and the 
eighth eigenvectors changes the couplings 0.006 and 0.022 to 0.033
and 0.019, i.e. a large change of the total coupling.

\noindent
\begin{minipage}{3.375in}
\begin{table}
\caption[]{The partial electron-phonon coupling
 constants $\lambda_{\nu}/N(0)$
 (in eV) according to different theoretical calculations.
The energies (in cm$^{-1}$) of
the modes for the undoped system are shown.}
\begin{tabular}{cccccc}
\multicolumn{1}{c}{Mode}&
\multicolumn{1}{c}{Energy}&
\multicolumn{4}{c}{$\lambda_{\nu}/N(0)$}  \\
 &    & Varma\tablenotemark[1] &  Schluter\tablenotemark[2] &
 Antropov\tablenotemark[3] & Faulhaber\tablenotemark[4] \\
\tableline
H$_g$(8)&1575&.011 & .009 & .022&  .009 \\
H$_g$(7)&1428&.034 & .013 & .020 & .015      \\
H$_g$(6)&1250&.000 & .003 & .008 & .002      \\
H$_g$(5)&1099&.006 & .001 & .003 & .002      \\
H$_g$(4)& 774&.000 & .007 & .003 & .010      \\
H$_g$(3)& 710&.001 & .004 & .003 & .001      \\
H$_g$(2)& 437&.001 & .007 & .006 & .010      \\
H$_g$(1)& 273&.003 & .008 & .003 & .001      \\
$\sum$ H$_g$ &  &.056 & .052 & .068 & .049    \\
\end{tabular}
\label{tableII}
\tablenotetext[1] {Varma {\it et al.}, 1991}
\tablenotetext[2] {Schluter {\it et al.}, 1992b}
\tablenotetext[3] {Antropov {\it et al.}, 1993b}
\tablenotetext[4] {Faulhaber {\it et al.}, 1993}
\end{table}
\end{minipage}

In Table \ref{tableII} we show some typical results of calculations    
of $\lambda/N(0)$. The calculation of Varma {\it et al.} \ (1991) was based
 on the
semi-empirical MNDO method. The calculation of Schluter {\it et al.} \ (1992b)
used an empirical bond charge model  for the phonons and a LDA calculation
for the electronic part. The work of Antropov {\it et al.} \ (1993b) and
Faulhaber {\it et al.} \ (1993) used {\sl ab initio} LDA calculations both
to generate the phonons and to calculate their interactions with
the electrons. There are several other calculations. Mazin {\it et al.} \ (1992)
performed an {\sl ab initio} LDA calculation in the ASA-LMTO formalism
and used      an empirical phonon model to obtain $\lambda/N(0)=
0.037$ eV. Coulon {\it et al.} \ (1992) performed an {\sl ab
initio} LDA calculation using a pseudopotential Gaussian method 
and found $\lambda/N(0)\ge 0.04$ eV.  Asai and Kawaguchi (1992) 
obtained $\lambda/N(0)=0.070$ eV.
The coupling to the A$_g(2)$ mode for a free molecule was estimated by
 Stollhoff (1991) to be
$\lambda_{A_g(2)}/N(0)=0.0025$ eV and by Scherrer and Stollhoff (1993)
to be 
$\lambda_{A_g(2)}/N(0)=0.01$ eV. Relatively
small values for the coupling to the A$_g$(2) mode were also deduced by
Schl\"uter {\it et al.} \ (1992a) ($\lambda_{A_g(2)}/N(0)=0.005$ eV),
Antropov {\it et al.} \ (1993b) ($\lambda_{A_g(2)}/N(0)=0.009$ eV) and
Pickett {\it et al.} \ (1994)  ($\lambda_{A_g(2)}/N(0)=0.002$ eV).
These calculations do not include the screening effects discussed below.
The electron-phonon interaction has also been calculated by
Jishi and Dresselhaus (1992) and by Chen and Goddard (1993). 

Table \ref{tableII} illustrates
 that there are substantial differences between
the different calculations in terms of the distribution of coupling
strength between the different modes. Thus Varma {\it et al.} \ (1991) found that
almost all the coupling strength is to the two highest modes. Antropov
{\it et al.} \ (1993b) found more coupling strength to the lower phonons and this
was still more true for the calculations of Schluter {\it et al.} \ (1992b) and
Faulhaber {\it et al.} \ (1993). Nevertheless, all four calculations find
the strongest coupling to one of the uppermost phonons.
The large variation in the distribution of coupling between the different
phonons in Table \ref{tableII}
may to a substantial extent be due to the sensitivity to the
phonon eigenvectors discussed below Eq. (\ref{eq:3.3}).
These variations suggest that there are substantial uncertainties in the 
calculations.

In estimates of the total $\lambda$,  the contributions from the
A$_g$ modes are usually   not
included. The reason is that the A$_g$ contribution is expected to be
screened out in A$_3$C$_{60}$ (A=K, Rb) (Schluter {\it et al.},  1992b).
This is illustrated by Raman scattering results for A$_3$C$_{60}$
(A=K, Rb) (Duclos {\it et al.}, 1991, Pichler {\it et al.}, 1992).
As discussed in Sec. III.3, the additional broadening of the A$_g$ modes
in A$_3$C$_{60}$ compared with C$_{60}$ should be a measure of the
electron-phonon coupling. Experiments, however, show little or no 
extra broadening of the A$_g$(2) in the metallic phase. For instance,
Pichler {\it et al.} (1992) estimated the          
additional broadening to be  only about 3 cm$^{-1}$,
implying $\lambda_{A_g(2)}/N(0) \sim 0.0002$, i.e., an extremely 
small coupling.

The corresponding additional broadening for many of the H$_g$ modes is,
on the other hand, large, indicating a much stronger coupling for these 
modes.
The symmetric A$_g$ mode leads
to the same displacement of all the $t_{1u}$ levels on a given molecule,
and this displacement can largely be compensated by a transfer of charge
to this molecule. This screening is much less efficient for the H$_g$
modes, since these do not shift the center of gravity of the $t_{1u}$
levels. Simple estimates within RPA and assuming static screening, suggest
that the screening could reduce the coupling to the A$_g$ mode by 
one to two orders of magnitude (Antropov {\it et al.}, 1993b).
It would be interesting to investigate nonadiabatic effects,
going beyond static screening, since the 
A$_g$(2) phonon energy is comparable to the energy scale of the 
screening.

\noindent
\begin{minipage}{3.375in}
\begin{table}
\caption[]{The density of states $N(0)$ (per eV and spin) for K$_3$C$_{60}$
 and Rb$_3$C$_{60}$ according to different band structure   calculations.
Some of the results have been interpolated by Gelfand (1994).}
\begin{tabular}{cc}
$N(0)$ (K$_3$C$_{60}$)  &  $N(0)$ (Rb$_3$C$_{60}$)/$N(0)$ (K$_3$C$_{60}$)  \\
\tableline
  6.6\tablenotemark[1]   &   1.26\tablenotemark[1]   \\
  7.0\tablenotemark[2]   &          \\
  8.6\tablenotemark[3]   &   1.18\tablenotemark[3]    \\
  8.6\tablenotemark[4]    &   1.21\tablenotemark[4]    \\
  9.0\tablenotemark[5]    &   1.14\tablenotemark[5]     \\
  9.8\tablenotemark[6]    &  1.27\tablenotemark[6]     \\
\end{tabular}
\label{tableIIIa}
\tablenotetext[1] {Erwin and Pickett (1991; 1992)}
\tablenotetext[2] {Antropov (1992)}
\tablenotetext[3] {Huang {\it et al.} \ (1992)}
\tablenotetext[4] {Satpathy {\it et al.} \ (1992)}
\tablenotetext[5] {Novikov {\it et al.} \ (1992)}
\tablenotetext[6] {Troullier and Martins (1992)}
\end{table}

\end{minipage}
To obtain values of $\lambda$, we also need estimates of the density 
of states at the Fermi energy $N(0)$. There have been several 
theoretical calculations of $N(0)$ for
 K$_3$C$_{60}$. Full potential, linear combination of Gaussian 
orbitals gave $N(0)=6.6$ states/eV-spin for the  
socalled uni-directional structure, where all the C$_{60}$ molecules 
have the same orientation and there is 
one C$_{60}$ molecule per unit cell (Erwin and Pickett, 1991; 1992).
An ASA-LMTO calculation gave $N(0)=8.6$ states/eV-spin
 (Satpathy {\it et al.},  1992) and a full potential LMTO gave 7.0 state/eV-spin
(Antropov, 1992).  These and some other results are collected in
 Table \ref{tableIIIa}.
A calculation for the bi-directional structure, where the molecules have
two different directions, gave $N(0)=7.5$ states/eV-spin (Antropov,
1992). The most relevant calculation is probably, however, for the 
orientationally disordered structure. This has only been performed
within a tight-binding approach (Gelfand and Lu, 1992ab; 1993). 
All the structures in $N(\varepsilon)$ are then smeared out, and 
$N(0)$ is slightly larger than $3/W$, where $W$ is the band width. 
If we assume $W\sim 0.6$ eV, as many band structure calculations 
find (Erwin and Pickett, 1991; 1992; Antropov, 1992), this gives 
$N(0)$ of the order 5-6 states/eV-spin.

Experimentally, the density of states has been estimated using a 
number of different    techniques. Many estimates are based on
 the NMR relaxation  rate. It has often been assumed that the
Fermi contact interaction dominates the relaxation, and large values of
$N(0)$ were then obtained. It has, however, been shown that the relaxation 
is dominated by the spin-dipolar interaction, due to the $p_z$
character ($\sim$ 95 $\%$) of the states close to the Fermi energy 
(Antropov {\it et al.}, 1993a).
These results were supported by Tycko {\it et al.} \ (1993), who found that the 
spin-dipolar coupling dominates for RbC$_{60}$ and CsC$_{60}$. 
Antropov {\it et al.} \ (1993) deduced that $N(0)=7.2$ and 8.1 
states/eV-spin for
K$_3$C$_{60}$ and Rb$_3$C$_{60}$, respectively. These values are 
used in the following. 

$N(0)$ has also been estimated from the specific heat jump $\Delta C$
at the superconductivity transition
(Ramirez {\it et al.}, 1992a; Burkhart and Meingast, 1996). Using the 
formula
\begin{equation}\label{eq:3.3a}
{\Delta C \over T_c}\sim N(0)(1+\lambda)(1.43+O(\lambda^2)),
\end{equation}
and $\lambda=0.068N(0)$ (Antropov {\it et al.}, 1993b) Burkhart and Meingast (1996) 
obtained $N(0)=6.6$ and 7.4 states/eV-spin
for K$_3$C$_{60}$ and Rb$_3$C$_{60}$, respectively. Using instead $\lambda=0.147N(0)$
(Gunnarsson {\it et al.}, 1995), they found $N(0)=5.3$ and 5.9  states/eV-spin
for K$_3$C$_{60}$ and Rb$_3$C$_{60}$, respectively. 

The density of states has furthermore been estimated from Pauli contribution to
the  susceptibility, obtained from SQUID and EPR measurements, as 
 in Table \ref{tableIIIb}.
The SQUID results were corrected for the diamagnetic contribution from 
the cores. The result of Wong {\it et al.} \ (1992) was  also corrected for
the Landau orbital diamagnetism, while Ramirez {\it et al.} \ (1992b) argued
that this contribution is negligible in the present systems.
The EPR measurements contain no diamagnetic contributions. 
The susceptibility may have an appreciable Stoner enhancement
(Ramirez {\it et al.}, 1992b). 
Based on LDA calculations, this enhancement factor was estimated
to be 1.34 and 1.42 for K$_3$C$_{60}$ and Rb$_3$C$_{60}$, respectively (Antropov {\it et al.},
 1993a). The accuracy of the LDA in this context is, however, not
known. This Stoner enhancement has {\sl not} been taken into account 
in Table \ref{tableIIIb}, and the numbers in the table  should be divided
by the appropriate Stoner enhancement.
The susceptibility has recently been calculated for a three-fold
degenerate Hubbard model, describing the $t_{1u}$ band and the 
on-site Coulomb interaction $U$.\cite{Aryasetiawan}
 A quantum Monte Carlo lattice method  
was used to obtain the enhancement of the Pauli susceptibility due
to the Coulomb interaction (Aryasetiawan {\it et al.}, 1996). 
For typical values of $U\sim 1.2-1.3$ eV,
an enhancement of the order of a factor three was obtain. This would reduce
the values of $N(0)$ estimated from the susceptibility and roughly
bring them in line with other estimates.

We observe that it is hard to estimate $N(0)$ from photoemission
or yield spectroscopy (De Seta and Evangelisti, 1995),
due to a large broadening of the band, which is probably caused 
by phonon and plasmon satellites (Knupfer {\it et al.}, 1993)
or other many-body effects.
Various    experimental estimates of $N(0)$ have been discussed by 
Gelfand (1994).

Table \ref{tableIII} shows the  
 contribution from the H$_g$ modes to 
$\lambda$ using the   values of $N(0)$ deduced from NMR, i.e., $N(0)=$    
7.2 and 8.1 states/eV-spin for K$_3$C$_{60}$ and Rb$_3$C$_{60}$, respectively.

\noindent
\begin{minipage}{3.375in}
\begin{table}[t]
\caption[]{The density of states $N(0)$ (per eV and spin) for
 K$_3$C$_{60}$ and Rb$_3$C$_{60}$
as deduced from susceptibility measurements. The results have
{\sl not} been corrected for the Stoner enhancement, which
would lead to  reduced    estimates of $N(0)$.}
\begin{tabular}{ccc}
$N(0)$ (K$_3$C$_{60}$)   &   $N(0)$ (Rb$_3$C$_{60}$)/$N(0)$ (K$_3$C$_{60}$) &
Method  \\
\tableline
  14\tablenotemark[1]     &   1.36\tablenotemark[1]   &   SQUID  \\
  16\tablenotemark[2]     &         &   SQUID    \\
  11\tablenotemark[2]     &         &   EPR      \\
  15\tablenotemark[3]     &   1.40\tablenotemark[4]   &   EPR      \\
  10\tablenotemark[4]     &         &   EPR             \\
\end{tabular}
\label{tableIIIb}
\tablenotetext[1] {Ramirez {\it et al.} \ (1992b)   }
\tablenotetext[2] {Wong    {\it et al.} \ (1992)   }
\tablenotetext[3] {Tanigaki {\it et al.} \ (1995) }
\tablenotetext[4] {Wang    {\it et al.} \ (1995)   }
\end{table}

\end{minipage}

\noindent
\begin{minipage}{3.375in}
\begin{table}
\caption[]{The H$_g$ contribution to $\lambda$
  according to different theoretical calculations, using
$N(0)=7.2$ and 8.1 states per eV and spin for K$_3$C$_{60}$ \ and Rb$_3$C$_{60
}$,
respectively.}
\begin{tabular}{ccccc}
\multicolumn{1}{c}{Mode}&
\multicolumn{4}{c}{$\lambda_{\nu}$}  \\
     & Varma\tablenotemark[1] &  Schluter\tablenotemark[2] &
 Antropov\tablenotemark[3] & Faulhaber\tablenotemark[4] \\
\tableline
K$_3$C$_{60}$& 0.40& 0.37 & 0.49&  0.35 \\
Rb$_3$C$_{60}$&0.45 &0.42  &0.55  &0.40       \\
\end{tabular}
\label{tableIII}
\tablenotetext[1] {Varma {\it et al.}, 1991}
\tablenotetext[2] {Schluter {\it et al.}, 1992}
\tablenotetext[3] {Antropov {\it et al.}, 1993}
\tablenotetext[4] {Faulhaber {\it et al.}, 1993}
\end{table}

\end{minipage}

It is interesting to compare the  alkali-fullerides with intercalated
graphite, since the latter have much lower values (one to two 
orders of magnitude) of $T_c$ than the fullerides.           
 It has been reported that calculations for
intercalated graphite gives a substantially lower value (factor 
five) of $\lambda$ than for the fullerides   (Schluter {\it et al.},   1992a).
This is not ascribed
to a larger density of states in the       fullerides 
but to the curvature of the C$_{60}$ molecule
(Benning {\it et al.}, 1991; Schluter {\it et al.},  1992ab). In intercalated graphite,
to first order there is no coupling to displacements perpendicular
to the graphite planes. In C$_{60}$, on the other hand, the curvature 
mixes in $2s$ character in the primarily radial $2p$ orbital 
pointing out from the molecule and dominating the states in the $t_{1u}$
band. This allows for a coupling of these states to radial (transverse)
modes. Alternatively, we can think of the displacements of the nuclei as
setting up an approximately $p$-like perturbing potential close to the 
nuclei. Such a potential can couple a $2s$ and a $2p$ orbital but 
not two $2p$ orbitals on the same site.  
Pietronero and Str\"assler (1992) suggested as an alternative
explanation that the violation of Migdal's theorem in A$_3$C$_{60}$
may cause the large $T_c$ in these compounds. 

\subsubsection{Raman and neutron scattering}

Due to the electron-phonon coupling, a phonon can decay in an electron-hole
pair of the appropriate energy. In the metallic fullerides A$_3$C$_{60}$
this decay channel is available and contributes to the width of the
phonons. This width is therefore a measure of the electron-phonon 
coupling (Allen, 1972; Allen, 1974) and for the $\nu th$ phonon it is given by
\begin{equation}\label{eq:3.4}
\gamma_{\nu}={2\pi\omega_{\nu}^2N(0)\lambda_{\nu}\over g_{\nu}},
\end{equation}
where $\gamma_{\nu}$ is the full width at half maximum (FWHM), 
$\omega_{\nu}$ is the phonon frequency, $N(0)$ is the density of
 states per spin, $\lambda_{\nu}$ the electron-phonon coupling and
 $g_{\nu}$ is the degeneracy of
the phonon (5 for H$_g$). Due to the factor $\omega_{\nu}^2$, the width
of the high-lying phonons is often very large and those phonons are hard to
observe experimentally.

The width can be measured using Raman or neutron scattering.
 In an ordered system, Raman scattering samples the decay
of a phonon for ${\bf q} \approx 0$ and it may be misleading for the
decay in the rest of the Brillouin zone.
In A$_3$C$_{60}$ (A=K, Rb) there is, however, a strong orientational
 disorder
(Stephens {\it et al.}, 1991) and Raman scattering should be relevant for the
determination of the electron-phonon coupling integrated over ${\bf q}$
(Schluter, 1992c). The observed widths may therefore include some
dispersion effects. These are, however, expected to be small for
the intramolecular modes, due to the weak coupling between  the
molecules (Belosludov and Shpakov, 1992).

The widths of some of the H$_g$ phonons were measured very early by 
Mitch {\it et al.} \ (1992a, 1992b).
They performed Raman measurements for ultrathin films of alkali-doped
A$_x$C$_{60}$ compounds. Due to the small thickness of the film, they were
able to continuously dope the system with Na, K, Rb, Cs. 
For the H$_g$(2) mode they observed a half-width of 14, 22, 22 and 20 cm$^{-1}$
for A=Na $x$=1, A=K $x$=2.1, A=Rb $x$=1.9 and A=Cs $x$=2, respectively. They 
assigned the rather $x$-independent broadening in NaC$_{60}$ to disorder,
 and assumed
that the disorder is similar in the other compounds. The additional broadening
in these compounds was assigned to the electron-phonon coupling. Assuming
that the two mechanisms add quadratically, they deduced the widths  
17 cm$^{-1}$ for K and Rb doped C$_{60}$. 
To obtain a value for $\lambda_2$
from Eq. (\ref{eq:3.4}), one 
further has  to assume a value for $N(0)$. Since different groups 
have used different values for        $N(0)$, we have here uniformly
used $N(0)=7.2$ eV$^{-1}$ per spin, deduced from NMR 
for K$_3$C$_{60}$ (Antropov {\it et al.},  1993a), to be able 
to compare the different estimates.
The values quoted below therefore differs from the values of $\lambda$
given in the original papers. In a few cases below we have also corrected
for an incorrect prefactor in Eq. (\ref{eq:3.4}) which has appeared in the
literature. In Table \ref{tableIV} we show  different experimental 
estimates of $\lambda_{\nu}/N(0)$ for K$_3$C$_{60}$.

The phonon width has also been studied in neutron scattering
for C$_{60}$, K$_3$C$_{60}$ and Rb$_6$C$_{60}$ by Prassides
{\it et al.} \ (1992). By comparing the
broadening for the metallic  K$_3$C$_{60}$ and the insulating C$_{60}$,
the additional broadening due to the electron-phonon coupling was
deduced. This was performed for the four lowest H$_g$ modes.      
The result for H$_g$(2) is in good agreement with the result of
Mitch {\it et al.} \  (1992b).

\begin{figure}[h]
\unitlength1cm
\begin{minipage}[t]{8.5cm}
\centerline{
\rotatebox{1}{
\resizebox{3in}{!}{\includegraphics{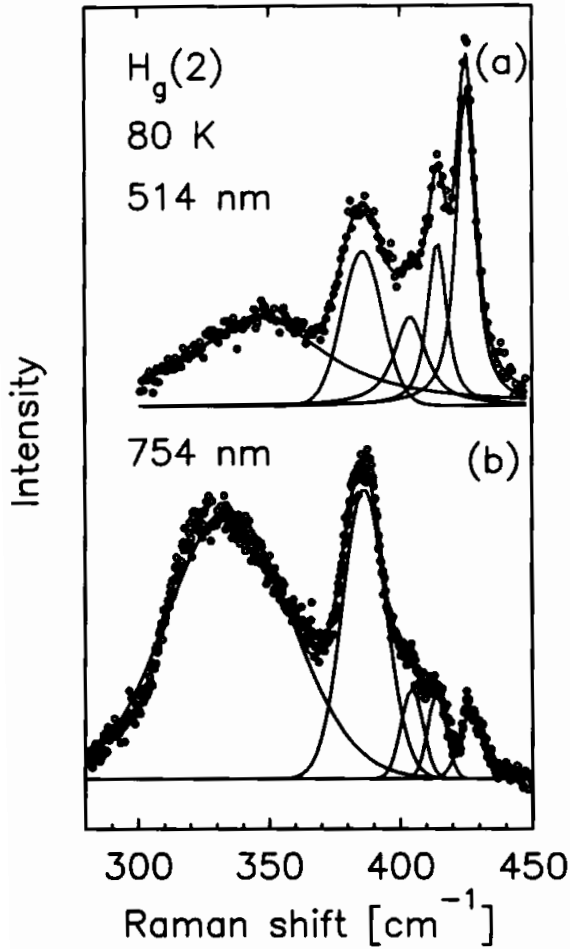}}
}}

\caption[] {\label{fig2b}Raman spectrum of the H$_g$(2) mode excited 
with a green laser
(a) and with a red laser (b). The full lines show a fit to five Voigtian
profiles (after Winter and Kuzmany, 1996).}
\end{minipage}
\end{figure}

Recently, the Raman spectrum was studied by Winter and Kuzmany (1996)
for single crystals and at low temperatures. They found that the peaks for
several of the H$_g$ phonons  were strongly asymmetric and could be fitted
with five components. This is shown in Fig. \ref{fig2b}. 

\noindent
\begin{minipage}{3.375in}
\begin{table}[t]
\caption[]{The partial electron-phonon coupling
 constants $\lambda_{\nu}/N(0)$
 [in eV] according to different experimental estimates.
 We also show the energies (in cm$^{-1}$) of
the modes for the undoped system.}
\begin{tabular}{cccccc}
\multicolumn{1}{c}{Mode}&
\multicolumn{1}{c}{Energy}&
\multicolumn{4}{c}{$\lambda_{\nu}/N(0)$}  \\
 &    & Raman\tablenotemark[1] &  Neutron\tablenotemark[2] &
Raman\tablenotemark[3] & PES\tablenotemark[4] \\
\tableline
H$_g$(8)&1575&     &      & .002&  .023 \\
H$_g$(7)&1428&     &      & .003 & .017      \\
H$_g$(6)&1250&     &      & .001 & .005      \\
H$_g$(5)&1099&     &      & .001 & .012      \\
H$_g$(4)& 774&     & .005 & .002 & .018      \\
H$_g$(3)& 710&     & .001 & .002 & .013      \\
H$_g$(2)& 437&.022 & .023 & .014 & .040      \\
H$_g$(1)& 273&     & .014 & .034 & .019      \\
\end{tabular}
\label{tableIV}
\tablenotetext[1] {Mitch {\it et al.}, 1992b}
\tablenotetext[2] {Prassides {\it et al.}, 1992}
\tablenotetext[3] {Winter   {\it et al.}, 1995}
\tablenotetext[4] {Gunnarsson {\it et al.}, 1995}
\end{table}
\end{minipage}

\noindent
It is  interesting that the component with the
largest shift relative to the position in the undoped crystal also has
the largest width. This suggests that the different components have
different electron-phonon couplings. The origin of such a difference in
the coupling, if it exists, is however not clear. It is also interesting
that the two highest H$_g$ modes seemed to have a different pattern of 
splitting. The observed line shape 
therefore poses a very interesting theoretical problem.   
If it is assumed that the formula of Allen (1972, 1974) is still applicable
to each component by itself, the results in Table \ref{tableIV} are 
obtained.

The measurement of the phonon broadening provides rather direct information
about the electron-phonon coupling. We note, however, that the interpretation
(Allen, 1972) is based on Migdal's theorem. Although this should be an
excellent approximation for most systems, it is questionable for
the fullerides since the phonon frequencies and the $t_{1u}$ band
width are comparable (Pietronero and Str\"assler, 1992;  
1994; Pietronero {\it et al.},  1995; Grimaldi {\it et al.},  1995ab). 
There are also questions if Jahn-Teller like effects studied for   
free molecules (Auerbach {\it et al.}, 1994; Manini {\it et al.}, 1994)
and going beyond Migdal's theorem,                                     
could survive 
in the solid and influence the results. Such higher order effects are
 very important
for the photoemission spectrum of a free molecule discussed below.
In particular, the
interesting line shape obtained by Winter and Kuzmany (1996) raises
questions about our understanding of how to analyze the spectra.

Reviews of Raman and neutron scattering studies of the fullerides
have been given by Kuzmany {\it et al.} \ (1994b) and by Pintschovius (1966),
respectively. 

\subsubsection{Photoemission}

An alternative way of deducing the electron-phonon coupling is
the use of photoemission data. Because of the 
relatively strong electron-phonon coupling, we
expect to see satellites due to the excitation of phonons. The weight  
of the satellites should then give information about the strength of
the coupling. This is essentially the Franck-Condon effect, except that it is
important to take the Jahn-Teller effect into account for C$_{60}$. 
The photoemmision spectra  of K$_3$C$_{60}$ and Rb$_3$C$_{60}$
have been analyzed along these lines (Knupfer {\it et al.},  1993). 
Due to the broadening effects in a solid and due to the complications
in the theoretical treatment of bands with dispersion, it was not possible,
however, to derive reliable, quantitative values for the
 electron-phonon coupling.

\begin{figure}[h]
\unitlength1cm
\begin{minipage}[t]{8.5cm}
\centerline{
\rotatebox{270}{\resizebox{2.7in}{!}{\includegraphics{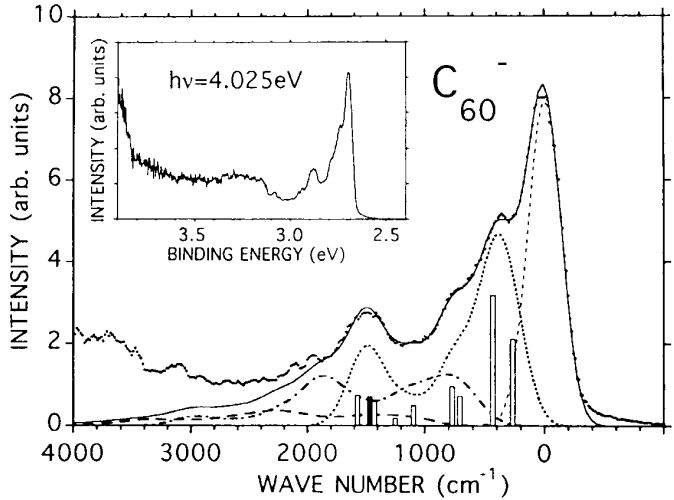}}}
}
\caption[] {\label{fig3}The experimental (dots) and theoretical (full line)
photoemission spectrum of C$_{60}^-$. The theoretical no loss (dashed),
single loss (dotted) and double loss (dashed-dotted) curves are also
shown. The contributions of the different modes to the single loss curve
are given by bars (H$_g$: open, A$_g$: solid).
The inset shows the experimental spectrum over a larger energy range
(After Gunnarsson {\it et al.}, 1995).}
\end{minipage}
\end{figure}

Photoemission experiments have, however, also been performed for free 
C$_{60}^{-}$ molecules, where the theoretical treatment is substantially
simpler (Gunnarsson {\it et al.},  1995). In these experiments
 a beam of C$_{60}^{-}$
ions was created and a photoemission experiment was performed using
a laser light source ($\hbar \omega$ =4.025 eV)  and a time of flight 
spectrometer. To analyse the results, a model was used which includes
the $t_{1u}$ level from which the electron is emitted, the two A$_g$ and
eight, five-fold degenerate H$_g$ modes and their coupling to the
electronic state. For this model the ground-state can be calculated
by numerical diagonalization to any desired accuracy. Within            
the sudden approximation (Hedin and Lundqvist, 1969),
 the photoemission spectrum can furthermore easily
be calculated. A set of coupling constants are then assumed and the 
resulting spectrum is compared with experiment. The coupling parameters are
then varied until good agreement with experiment is obtained, thereby
providing an estimate of the couplings.  The resulting 
spectrum is compared with experiment in Fig. \ref{fig3} and the
corresponding parameters are shown in 
Table \ref{tableIV}.

This approach contains mainly three uncertainties. Firstly, it is based on the 
sudden approximation. This approximation should become very accurate
as the kinetic energy of the emitted electron becomes large. Due to 
the small photon energy used, to obtain a good resolution, the
accuracy of the sudden approximation is unclear. Spectra taken at different
photon energies did not, however, show any systematic variation, suggesting
that the sudden approximation may not be a  serious problem 
here. We further observe that 
the weight of the satellites must go to zero as the photon energy is reduced
sufficiently. If this happens in a monotonous way, the use of the sudden
approximation would underestimate the couplings. As shown in Table
 \ref{tableIV}, the couplings derived from photoemission,
however, tend to be larger 
than other estimates. The second uncertainty is more technical, and is related 
to the fact that for the resolution available, it is not possible to 
distinguish between the coupling to the A$_g$ modes and the H$_g$ modes with
similar energies. The couplings to the A$_g$ modes were therefore taken  
from a calculation (Antropov {\it et al.},  1993b). With this assumption the couplings
to the H$_g$ modes can then be determined uniquely. An equally good fit
to experiment could, however, be obtained if, for instance, the coupling
to the A$_g(2)$,  $\lambda_{A_g(2)}/N(0)$, was allowed to vary between   
0.00 and 0.03 eV, provided that the total coupling to the H$_g(7)$ and 
H$_g(8)$ modes were simultaneously varied between 0.07 and 0.00 eV.
Thirdly, the coupling is determined for C$_{60}^{-}$, while we are interested
in C$_{60}^{3-}$ in a solid. Although the additional charging of C$_{60}$
is not expected to have large effects, due to the sensitivity of
the coupling to the phonon eigenvectors, the coupling could still 
be influenced.
 
\subsubsection{Resistivity}
The electric resistivity can  be caused by various scattering mechanisms.
An important mechanism is often the electron-phonon interaction, 
where electrons are scattered under the simultaneous excitation of
phonons. Another       mechanism is provided by the electron-electron
scattering. The contribution from both these mechanisms goes to zero
as $T \to 0$.  
Deviations from periodicity, e.g., the orientational disorder, on the
other hand, leads 
to an  essentially temperature independent resistivity.             
If the dominating contribution to the $T$-dependent part of the 
resistivity comes from the electron-phonon scattering, the resistivity
provides an interesting check of the values of the electron-phonon
coupling constants.

\begin{figure}[h]
\unitlength1cm
\begin{minipage}[t]{8.5cm}
\centerline{
\rotatebox{270}{\resizebox{3in}{!}{\includegraphics{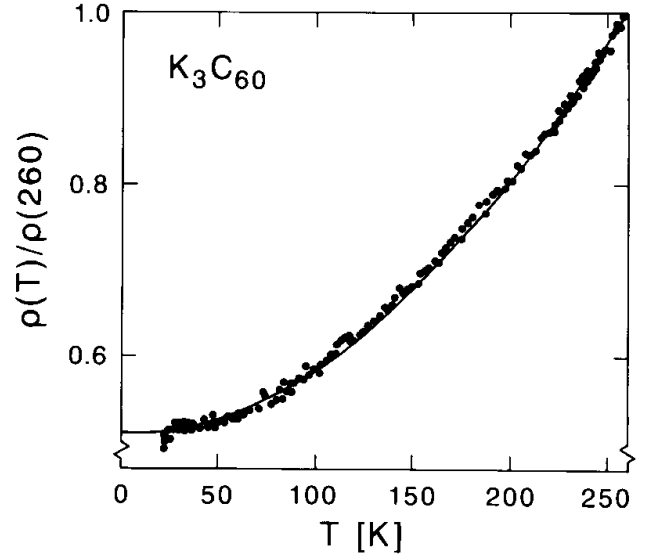}}}
}
\caption[] {\label{fig4}
 The experimental (dots)(Xiang {\it et al.}, 1992)  and theoretica
l
(full curve) $T$-dependent resistivity $\rho(T)$ normalized to
the value at $T=260$ K.}
\end{minipage}
\end{figure}

In Fig. \ref{fig4}
typical results for the temperature dependence are shown. For low $T$,
it is found that the temperature-dependent part of $\rho(T)$ at 
constant pressure is approximately quadratic in 
$T$ (Crespi, 1992). This suggests that the mechanism at these values of
$T$ is electron-electron scattering (Palstra {\it et al.},  1994). 
To obtain a reasonable magnitude, it is, however, necessary to assume
that the matrix element for electron-electron scattering is of the
order 1 eV, i.e., comparable to the $t_{1u}$ band width.
Such a large matrix element would    suggest a rather inefficient screening.
The approximate $T^2$ behavior can, however, also be obtained from
 the electron-phonon mechanism with a reasonable distributions 
of coupling constants. Therefore we focus on this mechanism below.
More recent experimental work (Vareka and Zettl, 1994) has 
 suggested that $\rho(T)$ at constant volume is linear in $T$ 
down to relatively low $T$ (100-200 K). This raises new questions,       
and the resistivity can not be considered as fully understood.

The resistivity has been studied theoretically by several groups
 (Crespi {\it et al.},  1992; Erwin and Pickett, 1992;
Antropov {\it et al.},  1993b; Lannin and Mitch, 1994), using the theory of Ziman
(Grimvall, 1981). This theory provides an approximate solution to the
Boltzmann equation for the case of the electron-phonon scattering mechanism.
\begin{eqnarray}\label{eq:3.5}
&&\rho (T)={\frac{8\pi^2 }{\omega_p^2 k_BT}} \nonumber \\
&&\label{eq:rho}\times \int_0^{\omega _{max}}d\omega {\hbar \omega
\alpha _{tr}^2F(\omega )\over {\rm cosh}(\hbar \omega /k_BT)-1}             
\end{eqnarray}
where $\omega_p$ is the plasma frequency. In the following it is assumed
that the transport coupling function $\alpha _{tr}^2F(\omega
)$ can be replaced by the electron phonon coupling appropriate for
superconductivity
\begin{equation}\label{eq:3.6}
\alpha ^2F(\omega )={\frac 12}\sum_\nu \omega _\nu \lambda _\nu
\delta (\omega -\omega _\nu ).
\end{equation}
Due to the factor cosh$(\hbar \omega /k_BT)-1$ in the denominator
of Eq. (\ref{eq:3.5}), it is clear that phonons with a frequency 
much larger than $T$ do not contribute, while they give a linear contribution
in $T$ if their frequency is  much smaller than $T$.
Using this approach, Crespi {\it et al.} \  (1992) demonstrated that
 theoretical results
for the coupling constants to  the H$_g$ modes alone of the type obtained
by Varma {\it et al.} \ (1991) or Schluter {\it et al.} \ (1992) gave a poor
 fit of experiment,
giving a much too weak temperature dependence at small $T$. The reason is that 
the H$_g$ phonons are too high in energy to give much contribution to 
$\rho(T)$ at low $T$, in particular since in these theories most of 
the coupling occurs to the higher phonons. The couplings of Jishi and
Dresselhaus (1992), emphasizing the lower H$_g$ phonons, gave better agreement.
Crespi {\it et al.} \ (1992)  
argued that one also has to consider low-lying external modes, e.g., the
intermolecular translational modes. Including such modes, they showed
that the couplings of Varma {\it et al.} \ (1991),  Schluter {\it et al.} \ (1992)
 and Jishi and Dresselhaus (1992) give  reasonable descriptions
of the experimental $T$-dependence, if the coupling to the translational
mode is chosen appropriately.

As an example of such results, we show a calculation using the 
coupling constants deduced from photoemission (Gunnarsson, 1995). 
In addition, the calculation included 
 the coupling to the librations, using a  theoretical
coupling constant (Antropov {\it et al.},  1993b), and to the intermolecular modes,
treating the coupling as an adjustable parameter, since its value is unknown.
As in the calculation of Crespi {\it et al.} \ (1992), the energy of the intermolecular
mode was set to 110 cm$^{-1}$, since neutron scattering shows many modes
at this energy. We have set $\omega_p=1.36$ eV. 
It was further assumed that some other mechanism, e.g., 
orientational disorder, provides a $T$-independent resistivity which accounts
for the resistivity just above $T_c$.
 The results are shown in Fig. \ref{fig4}. We can see that there is a good
 agreement with experiment over the whole temperature range. In particular,
the approximate $T^2$ behavior is reproduced. This results from the
contribution of many phonons at different energies, whose contributions
to the resistivity are gradually switched on one after the other as $T$
is increased. For high temperatures in Fig. \ref{fig4}, the resistivity
corresponds to a mean free path of the order of or less than the separations
between the molecules. This raises serious questions about the
 validity of the Boltzmann equation for these temperatures.

The theoretical and experimental results in Fig. \ref{fig4} are for
constant pressure. The experimental data have also been converted to 
constant volume, by measuring the pressure dependence (Vareka and Zettl,
1994). The resistivity then shows an approximately linear behavior 
for temperatures larger than about 100 to 200 K. This linear behavior
appears not to have been explained theoretically, and it is an interesting
problem.

It is of particular interest to consider the absolute value of the 
resistivity increase with $T$,  e.g., $\Delta \rho\equiv rho(T=260)-
\rho(T=0)$, since that may give information about 
the magnitude
of $\lambda$. Many estimates of the resistivity, however,  are made for low 
temperatures, thereby giving the residual resistivity and not 
$\Delta rho$.
 Since the shape of $\rho(T)$, but not its absolute
magnitude, appears to be well reproducible (Xiang {\it et al.},  1992),
we use the experimental ratio $\Delta \rho                             
 /\rho(T=0)$ two relate etimates of $\Delta rho$                   
and $\rho(T=0)$. 
Here $\rho(T=0)$ refers to the resistivity just above  the
superconductivity transition temperature.
Thus we convert our theoretical estimate of $\Delta rho$ to a theoretical
 estimate of $\rho(T=0)=$ 
0.47 m$\Omega$ cm. It is interesting that 
the resistivity from orientational disorder,                          
0.32 m$\Omega$cm (Gelfand and Lu, 1992b; 1993), is not much smaller.
Direct measurements for thin films and single crystals gave 
$\rho(T=0)\sim 1.2$ m$\Omega$cm (Palstra {\it et al.},  1994) and $\rho(T=0)=0.5$
m$\Omega$cm (Hou {\it et al.},  1993).
 Since the direct measurements involve various uncertainties
(possibly somewhat granular structure for the thin films and geometrically
very small samples for the single crystals),
it is also interesting to compare with indirect measurements of the
resistivity. Xiang {\it et al.} \ (1993) deduced $\rho(T=0)=0.12$ m$\Omega$cm
from the paraconductivity
(additional conductivity due to superconducting fluctuations just above $T_c$)
 and Crespi {\it et al.} \ (1992) deduced $\rho(T=0)=0.18$
m$\Omega$cm from the upper critical field. Degiorgi {\it et al.} \ (1992) obtained 
$\rho(T=0)=0.77$ m$\Omega$cm from a Kramer-Kronig analysis of 
the reflectivity in the normal state and Rotter {\it et al.} \  (1992)
found $\rho(T=0)=0.4$ m$\Omega$cm from the ratio between the reflectivity
in the normal and superconducting state.
These results are summarized in Table \ref{tablerho}.

 Comparison between the theoretical and experimental estimates shows
that the theoretical estimate falls in the correct range. Given the large
 range of experimental estimates it is, however, hard to assess the
accuracy of the theoretical calculation. We also observe that the theoretical
result was obtained from a set of coupling parameters which give very strong
couplings to the low-lying modes, which leads to a large resistivity, while
several other estimates of the couplings would give a substantially smaller
resistivity.

The residual resistivity is smaller than the   
Mott limit, which, due to the small
conduction electron density, is rather large for these systems,
and  has been estimated to be 1.4-2.0 m$\Omega$cm (Palstra {\it et al.},  1994).
For high temperatures, the resistivity corresponds to a mean
free path shorter than the separation between the molecules.
In such a situation the resistivity is usually expected to saturate
at a maximum value (Fisk and Webb, 1976). Such a saturation is
not observed for the doped C$_{60}$ compounds 
up to rather high temperatures ($T\lesssim 500$ K) (Hebard {\it et al.}, 1993),
but for still larger temperatures it is seen     for Rb$_3$C$_{60}$
(Hou {\it et al}, 1995). The saturation value was estimated to 
$\rho_{sat}=6\pm 3 \ {\rm m}\Omega$-cm with the corresponding mean free path
$\l_{sat}=1\pm0.5$ \AA, significantly 
smaller than the separation between the molecules.
The reason for this small mean free path appears not to be understood,
and it is an interesting question how one should conceptionally
understand such a short mean free path.

\noindent
\begin{minipage}{3.375in}
\begin{table}[t]
\caption[]{The resistivity $\rho(0)$ in m$\Omega$cm just above the 
superconductivity transition according to different theoretical 
and experimental estimates. For details see text.}      
\begin{tabular}{cl}
 $\rho(0)$               &   Method   \\                                       
\tableline
  0.47                   &   Theory. $\rho(260)-\rho(0) \to \rho(0)$ \\
  0.32\tablenotemark[1]   &  Theory. Disorder                 \\
  1.2\tablenotemark[2]   &   Direct. Film             \\
  0.5\tablenotemark[3]    &  Direct. Single crystal    \\
  0.12\tablenotemark[4]    & Paraconductivity            \\
  0.18\tablenotemark[5]    & Upper critical field        \\
  0.77\tablenotemark[6]    & Reflectivity. K-K           \\
  0.4 \tablenotemark[7]    & Reflectivity. S-N           \\
\end{tabular}
\label{tablerho}
\tablenotetext[1] {Gelfand and Lu \ (1992b, 1993)}
\tablenotetext[2] {Palstra {\it et al.} \ (1994)} 
\tablenotetext[3] {Hou {\it et al.} \ (1993)}
\tablenotetext[4] {Xiang    {\it et al.} \ (1993)}
\tablenotetext[5] {Crespi  {\it et al.} \ (1992)}
\tablenotetext[6] {Degiorgi {\it et al.} \ (1992)}
\tablenotetext[7] {Rotter    {\it et al.} \ (1992)}
\end{table}
\end{minipage}

\subsubsection{Comparison of electron-phonon coupling estimates}

We are now in the position of comparing different theoretical (Table 
\ref{tableII}) and experimental (Table \ref{tableIV}) estimates of the 
coupling constants. We first of all observe that the quantity
$\lambda/N(0)$ is obtained in the calculations and is deduced
from photoemission, while Raman and neutron scattering gives an
estimate of $\lambda N(0)$. Our assumptions about $N(0)$ (7.2 states/eV
per spin for K$_3$C$_{60}$) is therefore crucial for these comparisons.

There are substantial differences between the different calculations
of $\lambda$, 
between the different experimental estimates, and between theoretical
and experimental values. Experimentally, a strong coupling to some
of the low-lying modes is deduced. For instance, the total coupling to the
lowest two modes is experimentally several times larger than theoretically.
Thus the experimental estimates are 0.037, 0.048 and 0.059 eV, according
to neutron, Raman and photoemission experiments, respectively, while the
theoretical estimates are 0.004, 0.015, 0.009 and 0.011 eV.
For the high-lying phonons, the calculations give a rather strong 
coupling in agreement with the results from photoemission but in 
contradiction with the results from Raman. It is not clear why there is such 
a large discrepancy between the photoemission and Raman results for these
modes. 

If a larger value for $N(0)$ is assumed, the discrepancy between the
theoretical values and the Raman and neutron estimates for the 
low-lying modes would be reduced,
but the discrepancy with the photoemission values would remain. 
At the same time the difference between the theoretical and the Raman estimates
would be increased for the high-lying modes.
As discussed above, relatively small errors in the phonon eigenvectors 
could lead to a large underestimate of the theoretical coupling to the 
low-energy modes.

\subsection{Alkali phonons}

Very early it was pointed out that the alkali modes may provide
a strong electron-phonon coupling 
(Zhang {\it et al.},  1991a). Each C$_{60}$ molecule in A$_3$C$_{60}$
 is surrounded by 14 alkali atoms (6 octahedral and 8 tetrahedral).
The corresponding phonon frequencies are low   (0.013 eV
(Prassides {\it et al.},  1992); 0.014 eV (Renker {\it et al.}, 1993);
0.13 eV (Mitch and Lannin, 1993)),  implying weak force constants.
If an electron is transferred to a C$_{60}$ molecule, one may then expect 
the surrounding alkali ions to move towards the C$_{60}$ molecule,
giving an efficient screening.
It has been estimated that the alkali phonons could contribute
an attractive interaction of the order 0.9 eV (Zhang {\it et al.}, 1991a),
 which corresponds to $\lambda\sim 2$, i.e., a strong coupling.
Later experiments demonstrated that $T_c$ is approximately independent of the
alkali atom mass (see Sec. IIA and IIC), showing that the alkali modes
must contribute very little to $T_c$. This was  rationalized 
in terms of an efficient metallic screening (Gunnarsson and Zwicknagl,
1992). If an electron is transferred to a C$_{60}$ molecule, there
is a tendency to screen it by transferring charge from that molecule
to more distant molecules. The alkali ions then couple essentially
to the net charge change of the C$_{60}$ molecule. Within the RPA
this screening mechanism is very efficient, and 
when an electron is transferred to a C$_{60}$ molecule the change of its
net charge
is only a few hundredths of an electron (Gunnarsson and Zwicknagl,
1992). The coupling to the alkali modes is then reduced by one to 
two orders of magnitude. Although RPA may greatly overestimate the
efficiency of the screening, this nevertheless suggests a large reduction
of the coupling.  

The alkali phonon frequency for the tetrahedral position has been calculated
using the density functional formalism in the LD approximation
by Bohnen {\it et al.} \ (1995). They found the energy 0.015 eV, in
good agreement with experiment.

\subsection{Librations}
Early point contact measurements of the superconductivity gap
gave a large value of $2\Delta/T_c$ (Zhang {\it et al.},  1991b),
suggesting strong coupling effects and the coupling to low energy
modes. It was therefore suggested that the coupling to the librations may  
be of importance (Dolgov and Mazin, 1992; Mazin {\it et al.},  1993b).
There are several factors influencing this coupling. 
For the intramolecular phonons, the main changes in the matrix 
elements come from changes in bond lengths and bond angels, i.e., 
changes of the order of the intramolecular energy scale.
For the librations,
which involve rigid rotations of the molecules, only the hopping between 
the molecules is influenced, i.e., changes of the order of the
 intermolecular energy.
 The amplitude factor in Eq. (\ref{eq:3.2}) also
tends to be small for the librations due to the large moment of 
inertia, $I_{ii}=40MR^2$, where $M$ is the carbon mass and $R$ is the
$C_{60}$ radius. On the other hand, the very small libration frequency
($\sim 4$ meV, (Christides {\it et al.},  1992 )) tends to increase both
the amplitude factor in Eq. (\ref{eq:3.2}) and the $\lambda$ in 
Eq. (\ref{eq:3.1}). It is therefore {\sl a priori} not easy to estimate
the magnitude of the coupling to the librations.

We observe that the simplifications in the calculation of the coupling to
the intramolecular phonons (Lannoo {\it et al.},  1991) do not apply to the 
librations and the Brillouin zone sums in Eq. (\ref{eq:3.1}) should 
be performed. This has been performed within a tight-binding formalism,
which gave $\lambda_l/N(0)\sim 0.01$ (Antropov {\it et al.},  1993b). 
Comparing the splitting of the $t_{1u}$ band at the $\Gamma$ point
in the tight-binding calculation and a ASA-LMTO LDA calculation,
suggested that $\lambda_l$ should be reduced by about a factor 10,
giving $\lambda_l/N(0)\sim 0.001$ and $\lambda_l\sim 0.01$ 
(Antropov {\it et al.},  1993b). Thus the coupling to the librations is found to
be very weak.

The coupling has also been studied experimentally (Christides {\it et al.},  1992;
Reznik {\it et al.}, 1994).
They studied the librations above and below $T_c$. Since the libration
energy is smaller than the superconductivity gap $2\Delta$, the broadening
of the libration due to the decay in electron-hole pairs should 
go to zero for small temperatures (Zeyher and Zwicknagl, 1990).
Experimentally, however, no change in the width was observed when
the temperature was lowered below $T_c$. It was therefore concluded
that $\lambda_l<0.08$, since the change in width should otherwise
have been observed (Christides {\it et al.},  1992).

\subsection{Intermolecular modes}

The lattice-dynamics of K$_3$C$_{60}$ has been studied by Belosludov and
Shpakov (1992). They included the van der Waals interaction between the C$_{60}$
molecules, the electrostatic interactions between the ions,    Born-Mayer
type of repulsive interactions to describe the repulsion of the ion cores 
and a valence force model to describe the covalent interaction between
the carbon atoms inside a C$_{60}$ molecule. The parameters were 
adjusted to reproduce the experimental intramolecular $T_{1u}$
vibrations.  
This lead to acoustic modes which at larger wave vectors are mixed with
 librational modes at about 3.5 meV. At larger energies (6 meV and 16
meV) they found bands of alkali and C$_{60}$ character. This seems
to be consistent with neutron scattering experiments (Prassides {\it et al.},
1991; Christides {\it et al.}, 1992).
The translational phonons have also been studied by Zhang {\it et al.} \ (1992)
for K$_3$C$_{60}$ and by Wang {\it et al.} \ (1991) for C$_{60}$.

Pickett {\it et al.} \ (1994) have studied the electron-phonon coupling
for translational modes.  They used a tight-binding parameterization
of the hopping integrals. A distortion of the lattice corresponding to
the longitudinal zone boundary (110) intermolecular optic mode was 
introduced. 
From the displacement of the band structure energies they estimated
that $\lambda\sim 0.01$ for these modes. The intermolecular modes should  
therefore                have a small effect on $T_c$.

\section{Coulomb interaction and metallicity}
In this section we discuss the Coulomb interaction between
two electrons on a free C$_{60}$ molecule ($U_0$) and for a molecule
in the solid ($U$).
The large Coulomb interaction has played an important role in the discussion 
about superconductivity in the fullerides. This has ranged from the 
issue of why A$_3$C$_{60}$ are not Mott-Hubbard insulators
(Lof {\it et al.}, 1992), over questions
about whether or not retardation effects can reduce the Coulomb 
interaction enough to allow for superconductivity driven by the 
electron-phonon interaction (Anderson, 1991;
Gunnarsson and Zwicknagl, 1992) to the suggestion that the Coulomb interaction 
could actually be the cause of superconductivity (Chakravarty {\it et al.}, 
1991; Baskaran and Tosatti, 1991; Friedberg {\it et al.}, 1992).
\subsection{Coulomb $U_0$ for a free molecule}

A simple estimate of $U_0$ for a free molecule can be obtained
by assuming that the charge from an electron is uniformly spread out over 
a sheet with the radius $R$. The interaction between two electrons is then
$e^2/R=4.1$ eV, if $R=3.5$ \AA. This estimate neglects the relaxation of 
the charge density when an electron is added to the molecule.
To include this, one has to perform self-consistent calculations for  
different number of electrons.
This has been performed   by several groups.                          
Most of these calculations were based on the local density 
approximation of the density functional formalism (Hohenberg and Kohn,
1964; Kohn and Sham, 1965; Jones and Gunnarsson, 1989). $U_0$ is given
by the change 
of the position of the $t_{1u}$ level when charge is added to this 
level. The calculated values were 3.0 (Pederson and Quong, 1992),
2.7 (Antropov {\it et al.}, 1992), 3.0 (de Coulon {\it et al.},  
1992) and 3.1 eV (Martin and Ritchie, 1993). 

Estimates of $U_0$ can also be obtained from experimental data.
One estimate is provided by
\begin{equation}\label{eq:5.1}
U_0=I(C_{60})-A(C_{60})-E_g=3.3 \ {\rm eV},
\end{equation}
where $I(C_{60})=7.6$ eV (de Vries {\it et al.}, 1992) is the ionization energy,   
$A(C_{60})=2.7$ eV (Hettich {\it et al.}, 1991) is the affinity energy and
$E_g=1.6$ eV (Gensterblum {\it et al.}, 1991) is the band gap as measured 
by photoabsorption for a free C$_{60}$ molecule.
An alternative measure is given by
\begin{equation}\label{eq:5.2}
U_0=I(C_{60}^-)-A(C_{60}^-)\sim 2.7 \ {\rm eV},
\end{equation}
where $I(C_{60}^-)=2.7$ eV (Hettich {\it et al.}, 1991) is the ionization energy and 
and $A(C_{60}^-)$ is the affinity energy of $C_{60}^-$. $C_{60}^{2-}$
has been observed experimentally (Hettich {\it et al.}, 1991;
Limbach {\it et al.}, 1991) and if it is 
indeed stable, $A(C_{60}^-)>0$ and $U_0<2.7$ eV. It may, however,
 be metastable with
$A(C_{60}^-)$ slightly negative and $U_0$ slightly larger than 2.7 eV.
 The estimate (Eq. (\ref{eq:5.1}))
refers to the interaction between a hole in the $h_u$ level and an
electron in the $t_{1u}$ level, while
 the estimate (Eq. (\ref{eq:5.2})) refers to the interaction
between two electrons added to a C$_{60}$ molecule.
Since the electron and hole try to stay together while the two electrons
try to stay apart, it is not surprising that the latter estimate is 
smaller (Antropov {\it et al.}, 1992).

\subsection{Coulomb $U$ for the solid}
In the solid, there are additional screening effects due to the polarization
of the surrounding molecules when two electrons are added to a given molecule.
In the following, we have a Hubbard model in mind, where the on-site Coulomb 
interaction is included, but the screening due to the 
polarization of the surrounding molecules is only taken into account 
implicitely as a renormalization of  an effective $U$.
In this way the Coulomb interaction for 
a solid was found to be $U=1.27$ eV (Pederson and Quong, 1992) and
0.8-1.3 eV (Antropov {\it et al.}, 1992).
The Coulomb interaction may for many purposes be drastically reduced 
by the metallic screening. This reduction should, however, not be included
if $U$ is used in, say, a Hubbard model, since the metallic screening is then 
included explicitly in the model. To reduce the parameter $U$ by
 the metallic screening would then involve double counting.  

Experimentally, $U$ has been estimated from Auger spectroscopy. Without
any Coulomb interaction the Auger spectrum would be just a convolution 
of the one-particle spectrum, while the presence of an on-site molecular
interaction $U$ should displace the spectrum by approximately $U$.
The comparison of the Auger spectrum with the convoluted photoemission
spectrum then gives an estimate of $U=1.6$ eV (Lof {\it et al.}, 1992) and
$U=1.4$ eV (Br\"uhwiler {\it et al.}, 1992). For the highest occupied orbital
($h_u$) the estimate of $U$ is about 0.2 eV smaller (Lof {\it et al.}, 1992),
  and for the $t_{1u}$ orbital it may be still somewhat smaller.
Due to surface effects, Auger spectroscopy may also slightly 
overestimate $U$ (Antropov {\it et al.}, 1992).

\subsection{Mott-Hubbard transition in systems with a large degeneracy}

From the estimates of $U$ and $W$ it follows that the ratio
$U/W$ is substantially larger than one. It is then generally
expected that the system should be a Mott-Hubbard insulator
(Georges {\it et al.}, 1996).
It has therefore been argued that stoichiometric A$_3$C$_{60}$ must be an
insulator and that the experimental samples of A$_3$C$_{60}$
are metallic only because they are nonstoichiometric (Lof {\it et al.}, 1992). 
This  would be a similar situation as for the high $T_c$ superconductors,
and it might suggest that various properties 
of A$_3$C$_{60}$ should be understood in a   
similar way as for the high $T_c$ supersoncutors.
We therefore discuss the origin of the metallicity in A$_3$C$_{60}$ in 
more detail.

The estimates of the critical value of $U/W$ for a Mott-Hubbard 
transition have been made mostly for systems without
orbital degeneracy, while the $t_{1u}$ orbital has a threefold
degeneracy. This was considered in early work of Lu (1994), who used
the socalled Gutzwiller Ansatz and approximation, finding that
the orbital degeneracy
leads to a Mott-Hubbard transition for $U/W\sim (N+1)$, where $N$ is the
orbital degeneracy. This strong degeneracy dependence raised questions 
about its qualitative explanation and 
about why not almost all systems with partly filled bands are metals.

 To discuss  qualitatively the effect of orbital degeneracy, we consider
 a half-filled system with the orbital degeneracy $N$ and $M$ sites. 
The band gap is then
given by
\begin{equation}\label{eq:5.3} 
E_g=E(K+1)+E(K-1)-2E(K),
\end{equation}
where $K=NM$ is the number of electrons in the neutral state
and $E(K)$ is the ground-state energy for $K$ electrons.
If we consider the system with $K$ electrons in the limit of a large $U$,
 configurations with $N$ electrons per site dominate the wave function.
 The hopping of an electron costs
an energy $U$, and it is therefore strongly suppressed.              
 In the state with $K+1$ electrons, the 
extra electron can, however, hop without an extra cost in 
Coulomb energy. The extra occupancy can move through the hop of any
of $N$ electrons, which gives an extra factor $\sqrt{N}$
in the corresponding hopping matrix element compared with the
one-electron case (Gunnarsson {\it et al.}, 1996). This suggests that the
energy is lowered by about $\sqrt{N}\varepsilon_b$, where 
$\varepsilon_b$ is the bottom of the one-particle band. Similar arguments
for the $K-1$-electron state, suggest that
\begin{equation}\label{eq:5.4}
E_g=U-\sqrt{N}W.
\end{equation}
Exact diagonalization calculations and Monte Carlo calculations
 for A$_3$C$_{60}$
(Gunnarsson {\it et al.}, 1996) support these conclusions and suggest that
they can be extrapolated to intermediate values of $U$. Detailed 
calculations lead to  the conclusion that the Mott-Hubbard transition 
takes place
for $U/W\sim  2{1\over 2}$. This is at the upper limit of
both experimental and theoretical estimates of $U/W$. This 
suggests that A$_3$C$_{60}$ is on the metallic side of a Mott-Hubbard
 transition.
Although experimental samples of A$_3$C$_{60}$ may be 
nonstoichiometric, there is then no need to assume so to explain why they
are metallic. 
Since almost all physical systems have an orbital degeneracy,
the degeneracy dependence of the Mott-Hubbard transition should be of
interest for many other systems.

\section{Coulomb pseudopotential}                

The electron-phonon interaction provides an attractive interaction between 
the electrons, which may be of the order 1/10 eV. This interaction is  
counteracted by the Coulomb repulsion, and two electrons on the same      
molecule in an  undoped C$_{60}$ solid may feel a repulsion of the
order $U\sim 1-{1 \over 2}$ eV (see Sec. IVB). 
In this section we first discuss how the Coulomb repulsion may 
be reduced due to screening and retardation effects.
A dimensionless quantity 
\begin{equation}\label{eq:4.1}
\mu=UN(0)
\end{equation}
is introduced,
where $U$ is some typical (screened) interaction and $N(0)$ is the
density of states per spin. Due to retardation and other effects, $\mu$
is renormalized to $\mu^{\ast}$, the Coulomb pseudopotential.
A closely related issue is the question if the Coulomb interaction could be so 
effectively reduced that it becomes attractive, providing an
electronic mechanism for superconductivity. This goes back to calculations
for a model of a free molecule by Chakravarty {\it et al.} \ (1991),
and is discussed in the last part of this section.

\begin{figure}[h]
\unitlength1cm
\begin{minipage}[t]{8.5cm}
\vskip0.1cm
\centerline{
\rotatebox{1}{
\resizebox{0.5in}{!}{\includegraphics{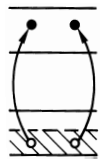}}
}}
\vskip0.2cm
\caption[]{\label{fig5}Schematic picture of two subbands and the virtual Coulomb
scattering of two electrons into the  higher subband.}
\end{minipage}
\end{figure}

\subsection{Retardation effects. Summing ladder diagrams}
\begin{figure}[h]
\unitlength1cm
\begin{minipage}[t]{8.5cm}
\centerline{
\rotatebox{270}{\resizebox{!}{3.3in}{\includegraphics{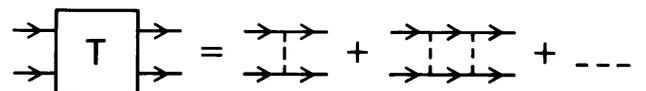}}}
}
\caption[] {\label{fig6}Ladder diagrams describing the repeated scattering of
two electrons. A   full line represents an electron and a
dashed line the (screened) Coulomb interaction.}
\end{minipage}
\end{figure}

It was very early
pointed out that the Coulomb repulsion can be reduced by retardation
effects      (Bogoliubov {\it et al.}, 1958; Morel and Anderson, 1962;
 Schrieffer, 1964;
Ginzburg and Kirzhnits, 1982). Normally, the electronic energy scale is much
larger than the energy scale of the phonons. 
By undergoing   multiple scattering  into
higher energy states, the electrons can move in a correlated way and thereby 
reduce the Coulomb interaction but still have an attractive interaction
via the phonons. This is shown schematically in Fig. \ref{fig5}
for the scattering up into a higher subband.
Mathematically, this is usually expressed by summing ladder diagrams
(see Fig. \ref{fig6}).                
 In simple models, 
e.g., assuming that all Coulomb matrix elements are equal, 
 the ladder diagrams can be
summed, giving the effective interaction
(Bogoliubov {\it et al.}, 1958; Morel and Anderson, 1962; Schrieffer, 1964;
Ginzburg and Kirzhnits, 1982)
\begin{equation}\label{eq:4.2}
\mu^{\ast}={\mu\over 1+\mu{\rm log} (B/\omega_{ph})},
\end{equation}
where $B$ is a typical electron energy (half the band width) and $\omega_{ph}$
is a typical phonon energy.   
If $B/\omega_{ph}>>1$,  $\mu^{\ast}$ 
can be strongly reduced relative to $\mu$. In the  limit $\mu{\rm ln}(B/\omega
_{ph})>>1$,
 Eq. (\ref{eq:4.2}) simplifies to 
$\mu^{\ast}\approx 1/{\rm log}(B/\omega_{ph})$, which may be of the
order 0.1-0.2. 
The importance of this reduction is illustrated by the McMillan formula
(McMillan, 1968) for the transition temperature
\begin{equation}\label{eq:4.3}
T_c={\omega_{ln} 
\over 1.2 } {\rm exp}\lbrack -{1.04 (1+\lambda)\over \lambda -\mu^{\ast}(1+0.62
\lambda)}\rbrack.
\end{equation}
In this formula, the relevant quantity is
\begin{equation}\label{eq:4.3a}
 {\lambda\over (1+\lambda)}-\mu^{\ast}{1+0.62\lambda\over 1+\lambda}
\approx  {\lambda\over (1+\lambda)}-\mu^{\ast}.
\end{equation}
It is immediately clear that $\mu^{\ast}$ is     as important as $\lambda$   
for the $T_c$.
 
Doped C$_{60}$ compounds have a number of narrow subbands with a width of
about 1/2 eV spread out over about 30 eV
(see Fig. \ref{fig1}). It has been argued that the
relevant energy scale is the total width of all the subbands (Varma, 1991).
Using Eq. (\ref{eq:4.2}) one then finds that $\mu^{\ast}\sim 0.2$ is only
slightly
larger than what is believed to be the case for conventional superconductors.
Together with current estimates of $\lambda$,  
values of $T_c$ of the right order of magnitude can then easily be obtained
 (Varma {\it et al.},  1991; Schluter {\it et al.},  1992; Mazin {\it et al.},  1992).
On the other hand, Anderson (1991) asserted that the relevant energy scale
is the width of the $t_{1u}$ subband. Since this energy is comparable
to the phonon energies, the retardation effects are then expected to be
small.
Anderson therefore argued that $\mu^{\ast}$ is large, and that the phonon 
mechanism alone cannot be enough for explaining the superconductivity.

We first discuss the retardation effects on $\mu^{\ast}$ within the formalism 
described above, where ladder diagrams
in the screened Coulomb interaction  are summed. Later we discuss effects
beyond this formalism. A crucial quantity is the Coulomb matrix element 
for scattering two electrons from the $t_{1u}$ band into higher subbands
(see Fig. \ref{fig5}). If these matrix elements are very small, the higher
subbands should play a small role in this formalism. Chakravarty {\it et al.} \ (1992b)
analyzed this in a model where they included the on-site interaction $U_C$ 
between
two electrons on the same carbon atom and the interaction $V_C$ between 
two electrons on different carbon atoms on the same C$_{60}$ molecule.
They then found that the Coulomb scattering matrix element between 
different subbands is of the order $U_C/60$, while the interaction within
the same band is $U_C/60+V_C$. Assuming that $U_C\approx 5-10$ eV and
 $V_C\approx
0.5-1.0$ eV, they found that the interband interaction is very much smaller 
than the intraband interaction.

The matrix elements of the screened Coulomb interaction 
have  been studied in the RPA (Gunnarsson and Zwicknagl, 1992;
Gunnarsson {\it et al.},  1992). The system was described in a tight-binding
approach (Satpathy, 1992; Laouini {\it et al.}, 1995) 
with one $2s$- and three $2p$-orbitals per carbon 
atom, and the screening was calculated with
local-field effects included. A long-range Coulomb interaction was included,
which involved an interaction between different atoms both on 
 the same C$_{60}$ molecule and on different molecules. 
The Coulomb interaction on a carbon atom was set to $U_C=12$ eV.
A screening calculation within the LDA approximation has been performed
by D.P. Joubert (1993), with fairly similar results.
The tight-binding results are shown in Table \ref{tableV}. 
The table shows both intraband
and interband interactions
\begin{eqnarray}\label{eq:4.4}
&&W(\alpha \beta,\gamma \delta,{\bf q})=      \\
  &&N\int d^3r d^3r^{'}
\psi^{\ast}_{\alpha {\bf k}}({\bf r})\psi_{ \beta {\bf k+q}}({\bf r})
W({\bf r},{\bf r}^{'})\psi_{ \gamma{\bf k}}({\bf r}^{'})\psi^{\ast}_{ \delta
{\bf k+q}}({\bf r}^{'}),   \nonumber
\end{eqnarray}
where
\begin{equation}\label{eq:4.5}
\psi_{\alpha {\bf k}}({\bf r})={1\over \sqrt{N}}\sum_{\nu}
e^{i{\bf k}\cdot{\bf R}
_{\nu}}\varphi_{\alpha}({\bf r}-{\bf R}_{\nu}),
\end{equation}
is a Bloch sum of molecular orbitals $\varphi_{\alpha}({\bf r})$,
$W({\bf r},{\bf r}^{'})$ is the screened interaction and $N$ is the 
number of unit cells.
The intraband integrals correspond to
the interaction of charges which integrate to one in the unit cell, i.e.,
 these integrals 
are dominated by a monopole interaction.
 On the other hand, $\phi_{\alpha}({\bf r})\phi_{\beta}
({\bf r})$ with $\alpha \ne \beta$ integrates to zero, and therefore 
the interband matrix elements $W(\alpha\beta,\alpha\beta)$ correspond to
multipole interactions.
In the solid and without screening, the intraband matrix elements are therefore 
very large for small ${\bf q}$,
 since they represent the interaction of monopoles placed
on the different molecules and with slowly varying phase factors.
When screening is introduced in the undoped solid, these integrals are reduced
 due to the polarization of the molecules. This reduction is much more 
dramatic in the doped solid due to the effective metallic screening, which 
involves a charge transfer between the molecules. The unscreened 
interband matrix elements are much smaller than the unscreened 
intraband elements, due to  
the shorter range of the multipole interaction. In the undoped solid they
are further reduced, primarily due to rearrangement of the charge density
inside the molecule. In the doped solid, the metallic screening does not    
essentially further reduce these matrix elements, since a charge transfer 
between the molecules does not contribute to the screening of multipoles.
The moderate reduction shown in Table \ref{tableV} is due to the change
in the filling of the levels and the new channels (in particular, $t_{1u}
\to t_{1g}$)  opened up for intramolecular
screening. 
It has been argued      that the intraband (metallic) screening should 
not be included in the calculation of the interband matrix elements
(Lammert {\it et al.},  1995). The calculations discussed above show, however,     
that the metallic screening of the interband matrix elements 
is negligible (Gunnarsson {\it et al.},  1992).

The Coulomb integral $W(\alpha\beta,\alpha\beta)$ with $\alpha\ne\beta$
referring to the $t_{1u}$ orbitals is also of great 
interest, since it may tend to favor singlet
states (Lammert and Rokshar, 1993). With only the short-range 
on-site carbon interaction
$U_C$, it is $U_C/100=0.12$ eV for a free molecule
(Lammert and Rokshar, 1993). Inclusion of
the long-range Coulomb interaction reduces the integral to 0.09 eV,
implying that the long range of the Coulomb interaction 
is not very important for this integral (in contrast to the monopole
integral $W(\alpha\alpha,\alpha\alpha)$).
   Allowing for screening $W(\alpha\beta,\alpha\beta)$ is reduced to 0.023 eV.
 This value is essentially unchanged if 
the screening by the surrounding molecules is included.

\noindent
\begin{minipage}{3.375in}
\begin{table}
\caption[]{The intraband Coulomb interactions (Eq. (\ref{eq:4.4}))between
two equal $W(\alpha\alpha,\alpha\alpha,
{\bf q})$ and two different $W(\alpha\alpha,\beta\beta,
{\bf q})$ $t_{1u}$ conduction band Bloch states as well as the largest
interband scattering matrix element $W^{inter}$ ($t_{1u}\to t_{1g}$).
We consider both a free molecule (Mol)
                  and a solid (Sol), where in the latter
case ${\bf q}={\bf G}/2$ and ${\bf q}=
{\bf G}/20$, where ${\bf G}=2\pi/a(1,1,1)$.
The unscreened (Unscr)  and  screened (Scr) interactions in the
undoped (Undop) and, for the solid, doped systems are shown.}
\begin{tabular}{cccccccc}
&Sys &\multicolumn{2}{c}{
  $W(\alpha\alpha,\alpha\alpha,
{\bf q})$}& \multicolumn{2}{c}{
$W(\alpha\alpha,\beta\beta,
{\bf q})$} &\multicolumn{2}{c}{$W^{inter}$}\\
\hline
 & &${\bf G}/20$&${\bf G}/2$&${\bf G}/20$&
    ${\bf G}/2$&${\bf G}/20$&${\bf G}/2$\\
\hline
Unscr      & Mol&
 \multicolumn{2}{c}{ 3.81}& \multicolumn{2}{c}{ 3.63}
&   \multicolumn{2}{c}{ 0.77}   \\
  Scr      & Mol&
 \multicolumn{2}{c}{ 3.72}& \multicolumn{2}{c}{ 3.67}
&   \multicolumn{2}{c}{ 0.14}   \\
Unscr      & Sol   &176.8 & 2.03  &176.7 & 1.86  &1.01 &  0.98      \\
Undop      & Sol &78.9 & 1.68  &78.8 & 1.63  &0.15 & 0.15      \\
Doped      & Sol &0.094 & 0.065 &0.060 & 0.031 &0.10 & 0.10     \\
\end{tabular}
\label{tableV}
\end{table}
\end{minipage}

Because of the very efficient metallic screening of the intraband
 matrix elements,
at least within RPA, these matrix elements are smaller than the largest
interband matrix elements. These results depend crucially
on the fact that these matrix elements have very different character
(monopole versus multipole interaction). This behavior
is also very different from the electron gas, 
since there is an abrupt change  
in character when going from one subband to another,
even if they are close in energy,
 due to the fact that each
subband is derived from a different molecular level with a different
spatial character. Since Eq. (\ref{eq:4.2}) is derived assuming that all matrix
elements are equal, it can not be used here. 
Approximate summing of the ladder diagrams, using the calculated matrix 
elements, shows that $\mu^{\ast}$ is close to zero in this 
formalism (Gunnarsson and Zwicknagl, 1992). 

It is interesting to ask why the intraband matrix elements are so small.
The screened interaction can be written as 
\begin{equation}\label{eq:4.6}
W=(1-vP)^{-1}v
\end{equation}
where $W$ is a matrix representing the different Coulomb matrix elements,
$v$ is the corresponding unscreened matrix elements and $P$ is the 
polarizability, which is here calculated in the RPA.
 For the intraband matrix elements, the $t_{1u}$ orbitals
dominate the screening, and we can transform to the corresponding
basis, neglecting all other orbitals. For small $|{\bf q}|$, 
the diagonal elements of $P$ are then
related to the density of states $N(0)$ per spin. Since both $N(0)$ and 
the diagonal elements of $v$ are large, we can assume that the product is 
much larger than unity. Essentially, the two factors $v$ in Eq. (\ref{eq:4.6})
then drop out, and
with appropriate assumptions one can then derive
(Gunnarsson {\it et al.},  1992) that the intraband matrix elements are
\begin{equation}\label{eq:4.7}
{1\over 3}W(\alpha\alpha,\alpha\alpha,{\bf q})+
{2\over 3}W(\alpha\alpha,\beta \beta ,{\bf q})\approx {1\over 2N(0)},
\end{equation}
for small $|{\bf q}|$.
To obtain the unrenormalized $\mu$ we multiply by $N(0)$ obtaining
$\mu\approx 0.5$. Averaging over $q$ gives $\mu \approx 0.4$.
We note that $W$ becomes very small in the RPA for large $N(0)$, since 
in the RPA
the cost of screening is purely a kinetic energy cost, and for a large $N(0)$ 
this cost is very small. For large values of $U$, we expect RPA to become 
a poor approximation, and it is therefore an interesting 
and important question how accurate RPA is in this case.

\subsection{Beyond ladder diagrams}
The considerations above were all within a framework where ladder diagrams
in the screened interaction
are summed. This raises the question about the validity of neglecting
other diagrams. For instance, Grabowski and Sham (1984) studied the lowest
order vertex corrections for the electron gas and found important
corrections. Gunnarsson {\it et al.} \ (1992) constructed other sets of diagrams, 
which to a large extent cancelled the ladder diagrams. 
This raises important questions about which diagrams to include.

To circumvent this problem, Gunnarsson and Zwicknagl (1992) and Gunnarsson
{\it et al.} \ (1992) studied a two-band model where some exact results
 can be obtained.
Thus they considered the Hamiltonian
\begin{eqnarray}\label{eq:4.8}
&&H=\sum_i\sum_{ n =1}^2\sum_{\sigma}\varepsilon_{ n }n_{i n \sigma}
+t\sum_{<ij>}\sum_{ n =1}^2\sum_{\sigma}\psi^{\dagger}_{i  n  \sigma}
\psi_{j n  \sigma} \nonumber   \\
&&+U_{11}\sum_i\sum_{ n n^{'}\sigma\sigma^{'}} {}^{'}n_{i n \sigma}
n_{i n^{'}\sigma^{'}}   \\  &&+
U_{12}\sum_{i}\lbrack \psi^{\dagger}_{i2\uparrow}\psi^{\dagger}
_{i2\downarrow} \psi_{i1\downarrow}\psi_{i1\uparrow}+h.c.\rbrack, \nonumber
\end{eqnarray}
where $i$ is a site index, $\varepsilon_{ n }$ the energies of the two
levels, $t$ is the hopping integral,
                           $U_{11}$ is the intraband interaction and
$U_{12}$ is the interband scattering.
The particularly strong interband scattering between the $t_{1u}$ and
$t_{1g}$ bands in C$_{60}$ is of a similar type,                  
 namely the excitation of  two electrons
between two bands     (see Fig. \ref{fig5}).
It was then assumed that 
\begin{equation}\label{eq:4.9}
\Delta \varepsilon\equiv \varepsilon_2-\varepsilon_1>>U_{12},t.
\end{equation}
In this limit, the upper subband can be projected out and a new effective
Hamiltonian be obtained, which describes low-energy properties correctly.
This process does not generate any new terms for the Hamiltonian (\ref{eq:4.8}),
but it renormalizes the intraband interaction to
\begin{equation}\label{eq:4.10}
U_{eff}=U_{11}-{U_{12}^2 \over 2 \Delta \varepsilon}. 
\end{equation}
If it is now assumed that $U_{12}^2/(2\Delta \varepsilon)<<U_{11}$, 
the renormalization of the intraband interaction is small. The properties
of this effective Hamiltonian then differ very little from the one-band
model, where the upper band was completely neglected. Thus the upper band
has a small influence on the low energy properties in general and 
$\mu^{\ast}$ in particular. For instance, in RPA the screened interaction
is $1/\lbrack 2 N(0)\rbrack$ in both the one-band and two-band models,
if $U_{11}N(0)>>1$.

 On the other hand,
 the ladder diagrams in the screened interaction  due to scattering
 into the upper subband can be summed.  If it is assumed that
\begin{equation}\label{eq:5.11}
{1\over N(0)} \lesssim {U_{12}^2 \over 2\Delta \varepsilon }
\end{equation}
this approach                                                          
 predicts a large renormalization of $\mu^{\ast}$.
The ladder diagrams subtract a quantity $U_{12}^2/(2\Delta \varepsilon)$,
which, although small compared with $U_{11}$, is large compared
with the screened interaction in the limit  considered.  
Thus the summation of ladder diagrams leads to a qualitatively incorrect
result in this case.

The projection method and the summation of ladder diagrams 
describe similar physics. The difference is
that in the rigorous projection method, the high energy
degrees of freedom were eliminated first, while in the ladder diagram
approach the low energy degrees of freedom were treated first (by
introducing the metallic screening) and the high energy degrees of
freedom later (by summing the ladder diagrams). 
These results should be of relevance for the retardation effects due to
the high-lying states also for conventional superconductors.

The model above contains important features of A$_3$C$_{60}$ and the
assumptions are not too unrealistic for this system.
The results therefore suggest that the summation of ladder diagrams
in the screened interaction
gives unreliable results for  A$_3$C$_{60}$ and that 
 the renormalization of $\mu^{\ast}$ due to the higher subbands
may not be large. For simplicity one may therefore assume that this 
renormalization can be neglected completely.
If in addition RPA screening is used, one arrives at $\mu^{\ast}\sim 0.4$.
The approach above has only addressed the
 renormalization due to the higher subbands but not 
due to processes within the $t_{1u}$ band itself. 
If we assume that the traditional theory is valid
for these processes, $\mu^{\ast}$ is renormalized to 0.3 according to 
formula (\ref{eq:4.2}), if we set $B=0.25$ eV and $\omega_{ph}=0.1$ eV. 
Such a value is then to be used in the McMillan formula, while the value 
0.4 applies to the Eliashberg equation including one subband.
We notice that only static screening was considered above, although
inclusion of  dynamic effects may influence the results substantially  
(Rietschel and Sham, 1983).

A very interesting work was performed by Takada (1993). He considered a model
including electrons in the  $t_{1u}$ band, phonons and plasmons, where
the plasmons screened the strong Coulomb repulsion. 
Guided by the Ward (1950) identity and work by Nambu (1960), Takada
used an approximate vertex correction. In this way he was able to 
include corrections to Migdal's theorem, vertex corrections to the 
electron-electron scattering in the $t_{1u}$ band and the dynamical screening
of the Coulomb interaction.
It was found that the corrections to Migdal's theorem enhanced the
value of $T_c$ for moderate values of the phonon frequency. Furthermore
 it was found that the vertex corrections
to the electron-electron scattering reduce $T_c$. Thus there is a partial  
cancelation of
the enhancement (Rietschel and Sham, 1983) of $T_c$ due to dynamic 
screening effects,       as had also been found earlier (Grabowski and
Sham, 1984). Takada used the value $\mu=0.4$ for the Coulomb pseudopotential
derived above, to describe static screening effects and the renormalization due
to scattering in other sub bands than the $t_{1u}$ band. Using $\lambda=0.58$
and adjusting the electron-plasmon coupling,
Takada (1993) could then reproduce the experimental $T_c$ for K$_3$C$_{60}$.
The corresponding isotope effect $\alpha=0.15$ for K$_3$C$_{60}$ was,
however, surprisingly small. To describe the lattice parameter dependence
of $T_c$, Takada had to assume that $\mu=0.4W_0/W$, where $W_0$ and
$W$ are the band widths of K$_3$C$_{60}$ and the system studied, respectively. 
This is somewhat surprising, since in RPA $\mu$ is predicted to be 
essentially independent of the lattice parameter.

\subsection{Electronic mechanisms}

Chakravarty and Kivelson (1991),    Chakravarty {\it et al.} \ (1991) 
and Baskaran and Tosatti (1991)
proposed that there may be an effective attractive interaction between 
the electrons for a C$_{60}$ molecule of a purely electronic origin.
 Chakravarty and Kivelson (1991) introduced a Hubbard model
of the $\pi$-electron system in a  free molecule with an on-site 
interaction $U_C$, only. 
Using second order perturbation theory, they calculated the energy
of C$_{60}$, C$_{60}^-$ and C$_{60}^{2-}$. This provided an estimate
of the effective repulsion for C$_{60}^{-}$. When the
singlet state of C$_{60}^{2-}$ was considered, the result
 consisted of a 
positive, linear term and a negative, quadratic term in $U_C$. 
The interaction therefore becomes attractive
for sufficiently large $U_C\sim 3t_0$, where $t_0$ is the intramolecular
hopping.  
The tendency to the formation of a singlet state results from an  
integral of the type $W(\alpha\beta,\alpha\beta)$ (see Eq. (\ref{eq:4.4}))
with $\alpha\ne\beta$ belonging to the $t_{1u}$ band 
(Lammert and Rokhsar, 1993), which may become negative 
when the higher states are projected out in second
order perturbation theory.
 The following discussion has focussed on the validity of the second
order perturbation theory and the neglect of a long-range Coulomb interaction
in this type of work.

White {\it et al.} \ (1992) addressed the question about the validity of second order
perturbation theory, by studying small model systems. In particular, they 
considered a model with twelve atoms located on a truncated tetrahedron.
For this system exact diagonalization can be performed and compared
with second order perturbation theory. White {\it et al.} \ (1992) concluded
that second order perturbation theory is qualitatively correct 
for intermediate values of $U_C$ in
the sense that exact diagonalization also gives an attractive interaction,
but that second order perturbation theory may greatly overestimate
the magnitude. 
Auerbach and Murthy (1992) and Murthy and Auerbach (1992)
 studied a model with electrons on a sphere
and with an interaction of variable range. They calculated the 
second order contribution to pairing and they also estimated the
third order terms. The second and third order contributions were 
found to be comparable at an interaction strength which was substantially
smaller than the strength giving pairing in the second order
calculation. This again raises doubts about the accuracy of the
 second order calculation.
Berdenis and Murthy (1995) performed a renormalization group
calculation for a Hubbard model of C$_{60}$ using a method of
Tokuyasu {\it et al.} \ (1993). This method remains accurate up to larger
values of the atomic $U_C$ than the second order perturbation theory,
 but its accuracy becomes questionable 
at roughly the value of $U_C$ where it predicts an attractive
interaction. Thus it seems to still be unclear whether or not
the interaction can be attractive for an on-site Hubbard model
of a free C$_{60}$ molecule.

Goff and Phillips (1992; 1993) and Auerbach and Murthy (1992) have
studied the effective interaction as a function of the range
of the interaction. They found that the use of a more long-ranged 
interaction tended to suppress the formation of an 
attractive effective interaction. 
With an on-site interaction, two electrons added to the $t_{1u}$
orbitals can almost avoid being on the same site simultaneously
and they have only a weak direct interaction $W(\alpha\alpha,\alpha\alpha)$.
 It is then not entirely
surprising that the renormalized interaction $W(\alpha\beta,\alpha\beta)$
can lead to an effective attractive interaction. If the interaction becomes
more long-ranged, the direct interaction rapidly increases, 
leading to a repulsive effective interaction.
For the free molecule, the interaction is long-range and there
is actually an ``anti-screening'' effect for electrons on the opposite
side of the molecule (Gunnarsson {\it et al.}, 1992). The reason is that
for a finite system, screening charge cannot be moved to infinity
as for an infinite system. It is therefore no surprise that the
effective interaction for the free molecule is large and 
repulsive, as discussed in Sec. IV.A. Chakravarty {\it et al.} \ (1991)
therefore from the start pointed out that the attractive electronic 
mechanism could
become operative only if the screening inside the molecule  
is much  more efficient in the metal than in  the free molecule. 

Lammert {\it et al.} \ (1995) considered a model with a long-range
Coulomb interaction. It was argued that the molecules surrounding 
a specific molecule provide a metallic environment, which was 
approximated by a metal with a spherical cavity with the radius $R_c$.
In this way the corresponding degrees of freedom were integrated out first,
although the metallic screening
corresponds to a   low   energy scale in the problem ($W\sim \omega_{pl}
\sim 0.5$ eV (Knupfer {\it et al.}, 1995)).  
The resulting screened interaction was calculated and used for
the low-frequency interactions in the problem.     
The effective singlet two-body interaction was then calculated 
diagrammatically, including the the first and second order diagrams.
An attractive interaction was obtained 
 for $R_c$ in the range 5-6 \AA \ and the on-site carbon $U_C$
in the range 8-10 eV.

An alternative way of treating the problem within this spirit, 
is to express the metallically screened interaction $U_{met.scr.}$ in terms of
the Coulomb interaction $U_0$ between two electrons on a free
C$_{60}$ molecule.
It is first assumed that in the solid the addition of a charge $q$ to a molecule
induces a potential $-qV_0$ due to the metallic screening by the
surrounding molecules, reducing $U_0$
to
\begin{equation}\label{5.1}
U_{met.scr.}=U_0-V_0
\end{equation} 
in the solid.
We here assume that due to the small hopping between the C$_{60}$
molecules, the surrounding molecules respond to a spherical symmetric
average of the charge on the central molecule.
In the model of Lammert {\it et al.} \ (1995), replacing the surrounding  
by a spherical cavity, $V_0=e^2/R_c$.
%Using $U_0 \sim 2.7$ eV (see Sec. IV.A) and $R_c=6$ \AA \ then leads to
% $U\sim 0.3$ eV, while $R_c=5$ \AA \
%leads to $U\sim -0.2$ eV. The choice of $R_c$ is therefore crucial.
To obtain an estimate of $V_0$ (or $R_c$) in  the above spirit, one can 
treat the surrounding C$_{60}$ molecules as  grounded, metallic 
 spheres with
the radius 4.46 \AA \ (which reproduces the experimental molecular
polarizability). This leads to $V_0=2.2$ eV (Krier and Gunnarsson, 1995),
corresponding to $R_c=6.6$ \AA, i.e., too large          to obtain an
attractive interaction in the work of Lammert {\it et al.} \ (1995).
If this is combined with the estimate of $U_0\sim 2.7$ eV (see Sec. IVA),
 a large and positive $U_{met.scr.}\sim 0.5$ eV ($\mu=N(0)U\sim 4$) results.
 This may suggest that it is important to also allow for  
charge transfer to the molecule where the studied electrons are  
located, as normally assumed for metallic systems.

An alternative electronic mechanism was suggested by Friedberg
{\it et al.} \ (1992). They argued that it is favorable to form a
two-electron wave-function involving one $t_{1u}$ and one $t_{1g}$ 
state. In this way a correlated state can be formed, which lowers the
Coulomb energy. Friedberg {\it et al.} \ (1992) argued that this more than outweighs
the cost in one-electron energy. This two-electron state is boson-like
and forms a Bose band. It was suggested that this may lead to      
a Bose-Einstein condensate.

\section{Beyond the Migdal-Eliashberg theory}
\subsection{Migdal's theorem}
Migdal's theorem  states  that vertex
corrections in the electron-phonon interaction can be neglected
if  the typical phonon frequencies $\omega_{ph}$
are sufficiently much smaller than the electronic energy scale, say
the Fermi energy $E_F$  
(Migdal, 1958).
 This theorem is assumed to be valid in the derivation
of the Eliashberg (1960) theory. 
For the fullerides, the highest phonon frequency is about 0.2
eV and the width of the $t_{1u}$ band is about 1/2 eV, giving $E_F\sim
0.25$ eV. Migdal's theorem may therefore be strongly violated
for the fullerides. If higher sub bands are included, one may argue   
that $E_F$ is much larger, but the conclusion that Migdal's theorem
is violated remains correct (Gunnarsson {\it et al.}, 1994b).

The lowest order vertex corrections have been studied by Grabowski
and Sham (1984), who calculated these corrections approximately,
finding that they can strongly reduce $T_c$. More recently, the 
issue has been studied extensively by Pietronero and coworkers
(Pietronero and Str\"assler, 1992; Pietronero {\it et al.}, 1995; 
Grimaldi {\it et al.}, 1995ab). They introduced a theory which is 
nonperturbative in $\lambda$ and perturbative in $m\lambda$, where
$m=\omega_{ph}/E_F$. 

\begin{figure}[h]
\unitlength1cm
\begin{minipage}[t]{8.5cm}
\vskip0.5cm
\centerline{
\rotatebox{270}{\resizebox{3in}{!}{\includegraphics{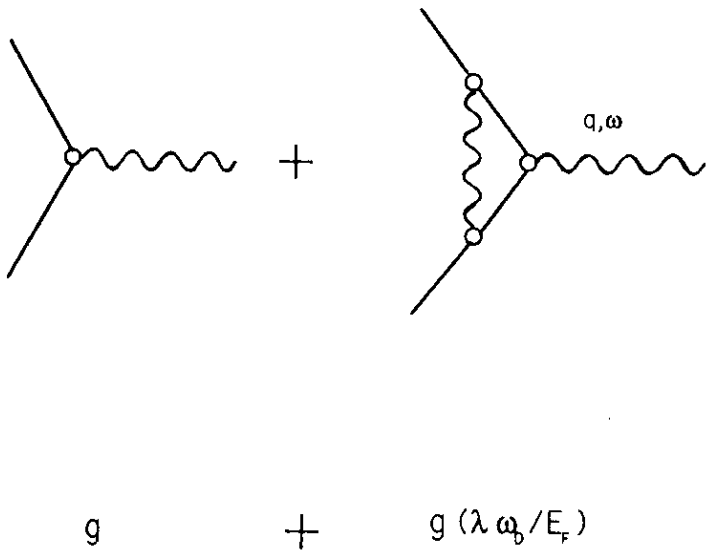}}}
}
\caption[] {\label{fig7}The zeroth order vertex and the first order vertex
 correction. (After Pietronero  {\it et al.}, 1995)}
\end{minipage}
\end{figure}

\begin{figure}[h]
\unitlength1cm
\begin{minipage}[t]{8.5cm}
\centerline{
\rotatebox{270}{\resizebox{3in}{!}{\includegraphics{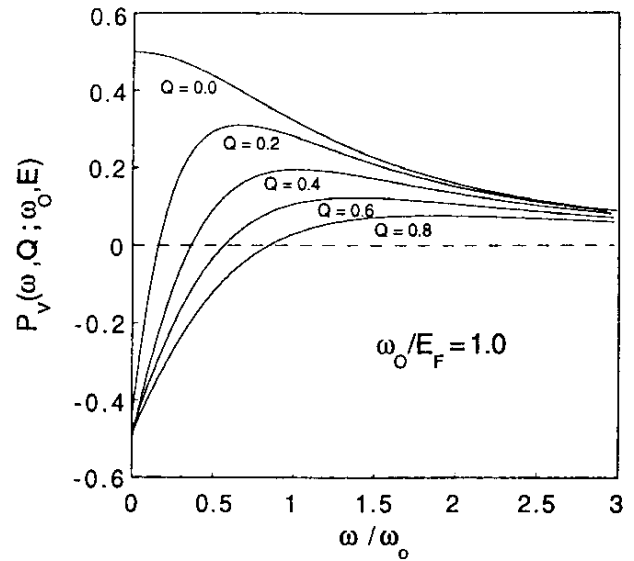}}}
}
\caption[] {\label{fig8}Behavior of the vertex correction for different values
of $Q\equiv q/(2k_F)$. (After Pietronero {\it et al.}, 1995).}
\end{minipage}
\end{figure}

\begin{figure}[h]
\unitlength1cm
\begin{minipage}[t]{8.5cm}
\centerline{
\rotatebox{0}{\resizebox{3in}{!}{\includegraphics{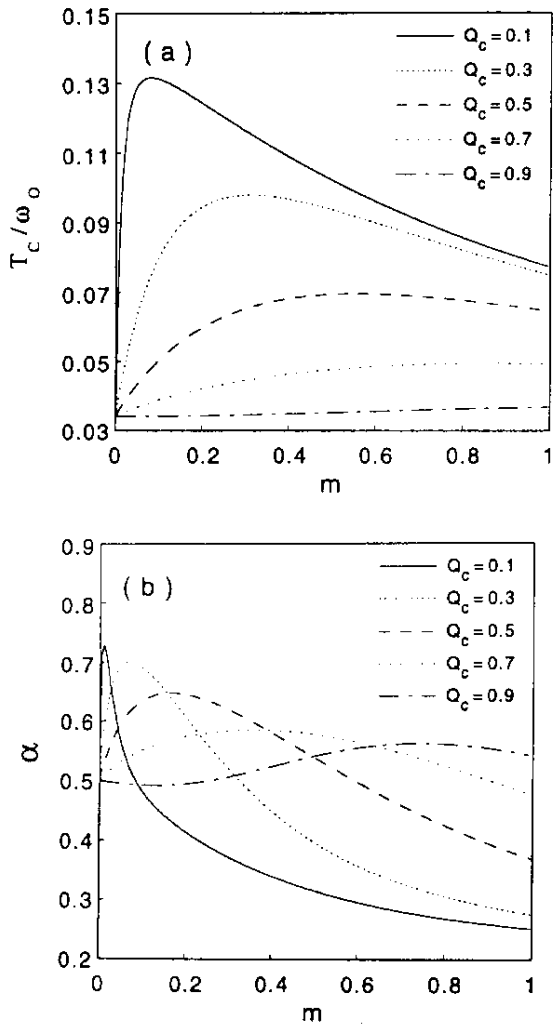}}}
}
\caption[] {\label{fig9}(a) $T_c$ as a function of $m\equiv \omega_{ph}/E_F$
and the cut off momentum $Q_c$ for $\lambda=0.5$. (b) Isotope
effect exponent $\alpha$ as a function of $m$ and $Q_c$.
(After Grimaldi {\it et al.}, 1995).}
\end{minipage}
\end{figure}

\noindent
First the lowest order vertex correction was calculated (see Fig. 
\ref{fig7}) for an electron gas like model.
 The result is shown in Fig. \ref{fig8} for $m=1$ and 
as a function  of $\omega$ and $Q=q/(2k_F)$, where $\omega$ and $q$ 
are the Matsubara frequency and wave vector of the phonon line and $k_F$ is
the Fermi wave vector. The energy of the incoming electron was
set to zero, and $\omega_0=\omega_{ph}$ is the energy of the Einstein phonon.
 This should be compared with the zeroth order contribution,
which is unity. The vertex correction has a very strong $Q$ dependence.
For small $Q\ne 0$ and small $\omega$ it is large and negative,
while it is positive elsewhere. This suggests that the influence
on $T_c$ may depend strongly on the system. An Eliashberg-like 
theory was then developed, including the lowest order vertex correction
and two ``crossed'' phonon lines in the scattering between two electrons
(Grimaldi {\it et al.}. 1995ab).
Fig. \ref{fig9} shows their results for $T_c$ and for the 
isotope effect
\begin{equation}\label{eq:6.1}
\alpha=-{d {\rm ln} T_c \over d {\rm ln} M},
\end{equation}
where $M$ is the carbon mass.
These results were calculated assuming that there is a cut off
in $q$-space at some critical $Q_c$. It was argued 
(Grimaldi {\it et al.}, 1995ab)
that this cut off may simulate
specific one-particle effects  or correlation
effects (Kulic and Zheyer, 1994). Depending on the value of $Q_c$,
the theory may predict a large enhancement of $T_c$ or no enhancement
at all. The value of the isotope effect can also vary substantially,
and even be larger than 0.5. In the limit where Migdal's theorem
is valid, this theory gives $\alpha=0.5$, since no Coulomb effects are
 included ($\mu^{\ast}=0$).
 
As described in Sec. Vi. Takada (1993) also considered
corrections to Migdal's theorem, and found that these enhance $T_c$
for moderate values of the phonon frequencies.
Ikeda {\it et al.} \ (1992) showed that Migdal's theorem becomes valid again 
when the phonon energies are much larger than the electronic energies.
Kostur and Mitrovic (1993) studied
the vertex corrections with the emphasis on two-dimensional systems
and Krishnamurthy {\it et al.} \ (1994) and  Cappelluti and
Pietronero (1996) studied 
systems with a van Hove singularity.
Kostur and Mitrovic (1994) studied the corrections to $T_c$ from 
vertex corrections for several model interactions. For a BCS-type of 
interaction and an isotropic electron-phonon interaction, they
found that vertex corrections reduce $T_c$, while for a highly 
anisotropic interaction, inspired by the cuprate superconductors,
they found that $T_c$ can also be enhanced by the vertex corrections.
Asai and Kawaguchi (1992) argued that nonadiabatic effect could enhance
the coupling by 10-30 $\%$.

The violation of Migdal's theorem is also expected to play an
important role for the photoemission spectrum. 
According to Migdal's theorem, the spectrum should just show 
single phonon satellites. Due to various broadening effects,
this cannot be checked experimentally for A$_3$C$_{60}$.  
Theoretical calculations suggest, however, that multiple satellites
develop (Gunnarsson {\it et al.}, 1994ab). This involves a substantial
transfer of spectral weight to higher binding energies, and the coupling
to phonons and plasmons may explain (Knupfer {\it et al.}, 1993) the large
observed width of the $t_{1u}$ band in photoemission (Chen {\it et al.}, 1991).
.
The work above illustrates that the violation of Migdal's theorem 
could have important effects for the fullerides, but the magnitude,
and perhaps even the sign, of the corrections seem to be not fully 
established.

\subsection{Jahn-Teller effects beyond Migdal-Eliashberg treatment}

The three-fold degeneracy of the $t_{1u}$-level is lifted when a
vibration of H$_g$ symmetry is excited. This makes C$_{60}$ molecules
with a partly filled $t_{1u}$ level unstable to a Jahn-Teller 
distortion (Jahn and Teller, 1937).
It has been shown (Ihm, 1994; Auerbach {\it et al.}, 1994; Manini {\it et al.}, 1994)
that this Jahn-Teller system has a Berry phase (Berry, 1984), which imposes
selection rules on the allowed transitions. 
These Jahn-Teller modes are included in the Migdal-Eliashberg theory
under the assumption that the band width is much larger than the
phonon frequency. Below we discuss how in the opposite limit,
 the molecular limit,
some new interesting effects appear. These effects may possibly be relevant 
also for doped C$_{60}$ compounds, although the band width may be too 
large for these systems to allow the survival of the molecular properties.

It is interesting to study the energy of C$_{60}^{n-}$ as a function of the 
electron-phonon coupling strength. A Jahn-Teller energy is defined as
\begin{equation}\label{eq:6.2}
E_{JT}=\sum_{\nu=1}^8 {g_{\nu}^2\over \omega_{\nu}},
\end{equation}
where the sum is over the eight H$_g$ modes and $g_{\nu}$ and $\omega_{\nu}$
are the coupling strength and energy, respectively, of the $\nu$th mode.
In the strong-coupling limit, the energy can then be written as
(Lannoo {\it et al.}, 1991)
\begin{equation}\label{eq:6.3}
E^{SC}(n)=C(n)E_{JT},
\end{equation}
where $C(n)=1$ $(n=1,5)$, 4 $(n=2,4)$ or 3 $(n=3)$. It was emphasized by 
Yabana and Bertsch (1992),
Auerbach {\it et al.} \ (1994) and Manini {\it et al.} \ (1994) that in the weak-coupling
limit the prefactor in the energy expression is a factor 5/2 larger
\begin{equation}\label{eq:6.4}
E^{WC}(n)={5\over 2}C(n)E_{JT},
\end{equation}
due to the Jahn-Teller effect, while for a mode of A$_g$ symmetry
the two expressions (\ref{eq:6.3}) and (\ref{eq:6.4}) are identical.
This result is valid for a free molecule or a solid with zero band
width, while the Migdal-Eliashberg theory is valid in the opposite
limit of a large band width. Auerbach {\it et al.} \ (1994) and Manini {\it et al.} \
(1994) observed that in the zero band width, weak-coupling limit,
 the attractive 
interaction between two electrons due the the electron-phonon interaction
for C$_{60}^{3-}$, 
\begin{equation}\label{eq:6.5}
U_{ph}^{WC}(3)=-5E_{JT},
\end{equation}
is a factor of three larger than in the Eliashberg theory. This could
potentially have important implications for the superconductivity.
We notice, however, that for realistic estimates of the coupling
constants, the system is not in the weak-coupling limit and may actually
be closer to the strong-coupling limit (Gunnarsson, 1995). The prefactor
 in Eq. (\ref{eq:6.4}, \ref{eq:6.5})
is then strongly reduced already in the zero band width limit.

\section{Calculated properties}
\subsection{Transition temperature}
It was very early pointed out that the estimated values 
of the electron-phonon coupling $\lambda$ to the intramolecular modes 
are of the right
order of magnitude to explain the transition temperature $T_c$
(Varma {\it et al.}, 1991; Schluter {\it et al.}, 1992; Mazin {\it et al.}, 1992).
Since the calculated values of 
$\lambda$ are only intermediate, it is important  for the explanation 
of the rather high values of $T_c$ that the
intramolecular phonon frequency are quite high.				
For values of $\lambda$ in the upper range of the estimates, 
the rather large values of $T_c$ can be understood even if 
it is assumed that the inefficient renormalization
of $\mu^{\ast}$ by retardation effects in the fullerides leads to
substantially larger $\mu^{\ast}$   than for conventional superconductors. 
For instance, Gunnarsson {\it et al.} \ (1995) solved the Eliashberg 
equation, using the coupling constants deduced from photoemission
for a free C$_{60}^-$ molecules (see Sec. III.A.4),
  a $t_{1u}$ band with the width
1/2 eV and the density of states at the Fermi energy  
deduced from NMR (see, Sec. III.A.4). It was found that to reproduce 
the values of $T_c$ for K$_3$C$_{60}$ and Rb$_3$C$_{60}$, $\mu^{\ast}$ had to be assumed 
to be 0.6, i.e., even larger than the estimate in Sec. V.B.  We note that 
the value 0.6 used in the Eliashberg equation corresponds to the 
value 0.4 in the McMillan formula, due to renormalization effects
inside the $t_{1u}$ band. 
From the discussion in the previous sections it is, however,
clear that the uncertainties in the estimates of $\lambda$ and $\mu^{\ast}$
are substantial, and the use of the Migdal-Eliashberg theory introduces
further uncertainties. 

Zheng and Bennemann (1992) studied effects of the phonon energies being 
comparable to the band width. They used the Eliashberg theory, i.e., 
Migdal's theorem was implicitly assumed to be valid, but they did not
assume the band width to be large in the solution of the Eliashberg
theory. As a result they found that the quasi-particle weight was
less renormalized than is normally assumed (the weight was larger than 
$1/(1+\lambda)$). As a result their calculated value of $T_c$ was larger than 
according to, e.g., the McMillan (1968) formula. 

\subsubsection{Lattice parameter dependence}
\begin{figure}[h]
\unitlength1cm
\begin{minipage}[t]{8.5cm}
\centerline{
\resizebox{3in}{!}{\includegraphics{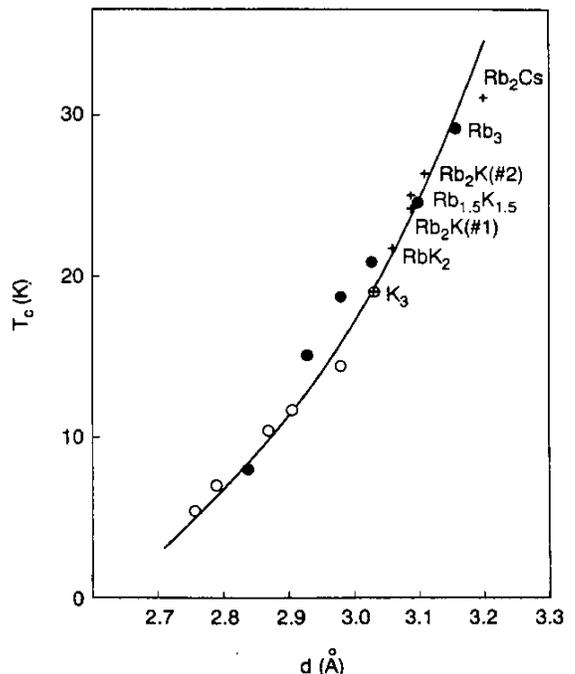}}
}
\caption[]{\label{fig10}Experimental (circles and crosses) and theoretical
(full line) variation of $T_c$ with the shortest distance $d$
between two closest carbon atoms on neighboring C$_{60}$
molecules. The density of states was assumed to have a power
law dependence on $d$, $N(0)\sim d^{-n}$, with $n=2.7$
adjusted to experiment. (After Schluter {\it et al.}, (1992b))}

\end{minipage}
\end{figure}

As discussed in Sec. II.A, for the A$_3$C$_{60}$ superconductors, except
Cs$_3$C$_{60}$, $T_c$ increases with the lattice parameters. 
The explanation of this behavior was provided very early (Fleming {\it et al.}, 1991;
Zhou {\it et al.}, 1992;
Varma {\it et al.}, 1991; Schluter {\it et al.}, 1992; Mazin {\it et al.}, 1992).
As discussed in Sec. III.A.2, for the coupling to the intramolecular
phonons $\lambda=VN(0)$ can to a very good approximation be factorized
in an intramolecular part $V$ and the density of states $N(0)$ (Lannoo
{\it et al.}, 1991). The intramolecular part $V$ can be expected to be
essentially independent of the lattice parameter $a$, while $N(0)$
grows with $a$. This follows, since as $a$ increases, the overlap
 between the molecules 
is reduced, leading to a reduced band width and an increased density
of states.  Fig.  \ref{fig10} shows a calculation of $T_c$ based
on a tight-binding calculation of the $N(0)$, assuming a  power law 
reduction of the hopping matrix elements with the separation between 
the closest atoms on neighboring molecules
(Schluter {\it et al.}, 1992b). The exponent of the power law was adjusted
to experiment.                                    
 
The very strong variation of $T_c$ with the lattice parameter $a$
in Na$_2$AC$_{60}$ (see Fig. \ref{fig2}) 
suggests a strong variation of $N(0)$ with $a$
in this system (see, e.g., Yildirim {\it et al.}, 1995).
Indeed, Maniwa {\it et al.} \ (1995) deduced $N(0)$ from the NMR relaxation
rate for  Na$_2$AC$_{60}$ (A=K, Rb, Cs) at ambient pressure
and found a much stronger dependence than for the Fm$\bar 3$m
structure, although possibly not quite strong enough to fully  
explain the strong variation of $T_c$. 
Simple tight-binding calculations including 60 ``radial'' $p$-orbital
per C$_{60}$ molecule did, however,  not give a very different
$a$-dependence for the density of states for the two structures
(Gunnarsson, 1993).           
It has instead been suggested that Na$_2$KC$_{60}$ and Na$_2$RbC$_{60}$ 
may have structural distortions,  
and that these distortions possibly could reduce $N(0)$ much more than the 
change in the lattice parameter (Maniwa, {\it et al.}, 1995). Since the Fermi
energy falls fairly close to a maximum in the density of states
in the undistorted Pa${\bar 3}$ structure (Satpathy {\it et al.}, 1992), such a 
reduction is possible. 
%This may be consistent with EPR measurements of the susceptibility
%as a function of temperature, which were interpreted in terms of changes
%in the density of states as the lattice expands (Tanigaki and Prassides, 
%1995). It was found that the deduced density of states changed in a similar
%way for the Pa3 and Fm${\bar 3}$m structures.
 An intermediate $a$-dependence was     
 observed in pressure experiments for Na$_2$CsC$_{60}$
(Mizuki, 1994), possibly consistent with the observation that
the structure may also distort under pressure (Zhu, 1995; Fischer, 1996).       
Whether $N(0)$ in the ideal structure has a strong dependence on $a$
or the strong dependence could be explained by other factors       
remains an interesting issue. For instance,  
it has also been proposed that paramagnetic impurities may suppress
$T_c$ in Na$_2$KC$_{60}$ and
Na$_2$RbC$_{60}$ (Tanigaki and Prassides, 1995).

\subsubsection{Doping dependence}

As discussed in Sec. II.A, Yildirim {\it et al.} \ (1995, 1996ab) have found a
strong doping ($n$) dependence for $T_c$, 
 by studying Na$_2$Cs$_x$C$_{60}$ ($n \le 3$) and Rb$_{3-x}$Ba$_x$C$_{60}$
($n \ge 3$). They found a maximum for $T_c$ close to half-filling
($n=3$) and rapid decrease as $n$ was reduced or increased.
The reduction for $n>3$ may be explained by considering the density   
of states for a system with merohedral disorder (Yildirim {\it et al.}, 1995).
The density of states shows a steady reduction with energy,
except at the bottom of the band (Gelfand and Lu, 1992). As the
filling $n$ is increased, the density of states at the Fermi energy $N(0)$
then drops, and $\lambda$ is reduced, which may explain the drop in $T_c$.
It appears, however, that there is no such explanation for $n<3$.
A tight-binding calculation for the appropriate structure (Pa${\bar 3}$)
shows an increase in the density of states as the energy is lowered
(Satpathy {\it et al.}, 1992). The situation is changed slightly if the
potential from the alkali atoms is taken into account (Gunnarsson, 1996).
 The nearest alkali neighbors 
of a given C$_{60}$ molecule set up a potential with $l=4$, which
cannot be efficiently screened by the C$_{60}$ molecule, and which
modifies the density of states. If it is taken into account, however,
that the density of states is sampled over a finite energy, due to the 
rather large phonon energies, the basic tendency of the density of states
to grow when the energy is reduced remains. One might have expected 
the potential of the vacancies to strongly influence the density of states.
The potential from the vacancies has, however, mainly monopole or dipole
character relative to a neighboring C$_{60}$ molecules, and it is 
therefore efficiently screened. Thus the drop in $T_c$ as $n$ is reduced
appears not to be understood.

\subsection{Isotope effect}
 
As discussed in Sec. II.C different experiment give widely different
results for the isotope effect ($T_c\sim M^{-\alpha}$), with 
$\alpha$ even being larger than unity. In the case of almost complete
substitution $\alpha=0.30$ was found for K$_3$C$_{60}$ and Rb$_3$C$_{60}$ (Chen and Lieber,
1992). 

It was pointed out early that a reasonable value of $\alpha\sim 0.3$ can 
be obtained if the mechanism for superconductivity is assumed to be
the interaction with the intramolecular phonons (Schluter, 1992b).
For instance, from the McMillan formula (\ref{eq:1.1}), the prefactor
$\omega_{ph}$ gives $\alpha=0.5$ and the renormalization of $\mu^{\ast}$
as defined in Eq. (\ref{eq:4.2}) gives a reduction of $\alpha$. Thus 
$\alpha$ is reduced to
about 0.3 if it is assumed that $\mu^{\ast}\sim 0.2$ and $\omega_{ph}
\sim 1000$ cm$^{-1}$. We note, however, that the McMillan formula
is not very accurate for calculating the isotope effect. The isotope
effect was also calculated in the model discussed in the previous (VII.A)
section using the Eliashberg theory and the coupling constants derived from 
photoemission (Gunnarsson {\it et al.}, 1995). The values of $\alpha$ were
 then found to be 0.32 and
0.37 for K$_3$C$_{60}$ and Rb$_3$C$_{60}$, respectively.

Deaven and Rokhsar (1993) addressed the large spread in the experimentally
observed isotope effect $\alpha$ for different isotope
substitions. They considered a model with harmonic phonons and a
linear electron phonon coupling. Within the Migdal-Eliashberg framework,
they showed that the anomolous isotope effect cannot be explained.
In particular, they showed that $\lambda$ is independent of the 
isotope distribution.
Anharmonic effects have been considered by, e.g., Crespi {\it et al.}
 (1991) and Inada and Nasu (1992). These effects could in principle lead to
$\alpha > 0.5$.

The isotope effect has traditionally been considered as a sign of 
an electron-phonon mechanism. It has, however, been pointed out
that even for an electronic mechanism, there is an isotope effect
due to the change of the electronic structure by zero-point
vibrations (Chakravarty {\it et al.}, 1992a, Ashcroft and Cyrot, 1993). 
Ashcroft and Cyrot (1993) emphasized the opposing effects of the translational
phonons and the librations on the electronic structure. They argued that
an increase in the carbon mass would tend to reduce the hopping 
integral due to the translational vibrations but increase it due to
the librations. They found that depending on the precise assumptions
about the parameters, $\alpha$ could take a broad range of values,
including values larger than 0.5. In view of this it is not clear
how much is proven by the reasonable values for $\alpha$ found from 
the intramolecular phonons alone. 

\subsection{Reduced gap}

The superconductivity gap $\Delta$ is of interest, since it may give
 indications
about strong-coupling effects. Gunnarsson {\it et al.}, \ (1995) calculated the gap   
using the Eliashberg equation in the model discussed above, using the
couplings deduced from photoemission. They obtained the values
$2\Delta/T_c=3.59$ and 3.66 for K$_3$C$_{60}$ and Rb$_3$C$_{60}$, respectively. 
This is close to the BCS value 3.52 and in agreement with most 
recent measurements. 

The much larger values found in point contact tunneling measurements
have stimulated a substantial amount of work.  
Mazin {\it et al.} \ (1993b) studied a model with coupling to both intramolecular
phonons (with frequencies around 1000 cm$^{-1}$) and to low-lying
modes with frequencies of the order 40 cm$^{-1}$. The latter modes could
be librations, intermolecular or alkali modes. Due to the huge difference
between the frequencies, this model shows very interesting properties. 
The low-lying modes hardly influence $T_c$ but have a strong influence
on the reduced gap $2\Delta/T_c$. Thus Mazin
{\it et al.} (1993b) could use typical couplings 
to the intramolecular modes ($\lambda_1\sim 0.5$) to produce the measured
values of $T_c$, but at the same time use strong coupling to the
low-lying modes ($\lambda_2\sim 2.7$) to produce a large reduced gap 
($2\Delta/T_c\sim 5$). More recent work has suggested that the coupling
to the low-lying modes is much weaker (see Sec. III.B, C and D) and that
the reduced gap may be much smaller than found in point contact tunneling. 
The model of Mazin {\it et al.} \ (1993b) may therefore not be applicable to
A$_3$C$_{60}$, but it is still a nice illustration of the unconventional
properties that can be obtained if bosons of very different energies
couple to the electrons.

Mele and Erwin (1993) considered the detailed Fermi surface for a
system with {\sl ordered} molecules. The studied an axial, nodeless 
order parameter with $d$-symmetry. This led to the ratio $2\Delta/T_c\sim4.6$.
The temperature dependence  and the quasi-particle 
spectrum  were also non-BCS. Since the solution, however, depends on 
specific features of the Fermi surface, it is not clear if it 
survives for the disordered system.

Mele {\it et al.} \ (1994) studied the effects of orientational disorder. 
They found that this introduces some gap anisotropy.
Furthermore they obtained some reduction of both  $T_c$ and $\Delta$ 
in such a way that the reduced gap $2\Delta/T_c$ was increased 
to values of the order 3.8-4.0. They therefore concluded that 
disorder can lead to an increased gap ratio even within a weak-coupling theory.

\subsection{Other properties}

Erwin and Pickett (1991) calculated the Fermi velocity  
$\nu_F$ for a system with
ordered C$_{60}$ molecules, using the local density approximation.
From $\nu_F=1.8$ $10^7$ cm s$^{-1}$ they deduced the Drude plasma energy
$\hbar \Omega=1.2$ eV and the clean limit London penetration depth
$\lambda=1600$ \AA. Considering that A$_3$C$_{60}$ is in the dirty 
limit, they corrected this result by a factor $(1+\xi/l)^{1/2}$,
where $\xi$ is the coherence length and $l$ is the mean free path.
Using $\xi=26$ \AA \ and $l=10$ \AA, they obtained $\lambda \sim 3000$
\AA, in reasonable agreement with experiment (the scattering between 
different experiments is, however, large).

\section{Concluding remarks}
It is widely believed that the A$_3$C$_{60}$ (A=K, Rb) 
are  $s$-wave BCS superconductors, driven by the 
coupling to the intramolecular H$_g$ phonons.
 The work     reviewed here is consistent with and supports 
such a picture, although there is  no conclusive evidence that this
picture is correct or that an electronic mechanism is excluded.  
Accurate and detailed tunneling data would be very interesting in this 
context.

It appears to be understood why A$_3$C$_{60}$ are metals and not 
Mott-Hubbard insulators, in spite of the large Coulomb interaction.
 Experimental and theoretical estimates of the electron-phonon
coupling $\lambda$ give values of about 0.5-1. 
The spread in           these estimates indicate that
the coupling strength is still not very accurately  known. For the other 
important parameter determining $T_c$, the Coulomb pseudopotential
$\mu^{\ast}$, there are        arguments
that $\mu^{\ast}$ is not strongly  renormalized by retardation effects
due to scattering of the electrons into sub bands other than the $t_{1u}$ band. 
This,  however, leaves interesting questions about the screening and 
retardation effects on $\mu^{\ast}$ within the $t_{1u}$ band. 
Theoretical calculations of $T_c$ based on estimates of $\lambda$ and 
$\mu^{\ast}$ are consistent with experiment and support an electron-phonon
picture. 

The lattice parameter dependence of $T_c$ seems to be understood
 for  A$_3$C$_{60}$ in the Fm${\bar 3}$m structure
but not for  Na$_2$AC$_{60}$ 
in the Pa${\bar 3}$ structure. The doping dependence is also not well 
understood. The BCS like superconducting gap seen in most experiment,
follows from the picture of the superconductivity being driven
by the H$_g$ intramolecular phonons, while the isotope effect still gives
rise to some interesting experimental and theoretical questions.

We have emphasized that because several energy scales are comparable,
the theoretical treatment of the fullerides is particularly demanding
but also unusually interesting. Thus                                         
one may expect Migdal's theorem to be violated and many-body effects
due to the Coulomb interaction to be important.
The electron-phonon interaction should also be important for
 many properties. On the other hand, the molecular solid character
leads to certain simplifications, for instance  
the electronic properties are expected to be dominated by the three-fold
degenerate $t_{1u}$ orbital. Due to the unusual parameter range,
 many interesting
general issues are raised in a particular clear-cut way for the
 doped fullerides,
and their solutions should have implications in many other fields.

\vskip0.7cm
{\bf ACKNOWLEDGMENTS}

\vskip0.4cm
We would like to thank I.I. Mazin for many helpful suggestions.

\parindent -10pt
\vskip0.4cm
{\bf REFERENCES}
\vskip0.4cm 

Adams, G.B., J.B. Page, O.F. Sankey, K. Sinha, J. Menendez, 
and D.R. Hoffman, 1991,
Phys. Rev. B {\bf 44}, 4052.

Aldersey-Williams, H., 1995, {\it The most beautiful molecule}, Aurum Press,
London.

Allemand, P.-M., K.C. Khemani, A. Koch, F. Wudl, K. Holczer, S. Donovan,
G. Gr\"uner, and J.D. Thompson, 1991, Science {\bf 253}, 301.

Allen, P.B., 1972, Phys. Rev. B {\bf 6}, 2577. Observe a misprint, which
was corrected by Allen (1974).

Allen, P.B., 1974, Solid State Commun. {\bf 14}, 937.

Allen, P.B., and B. Mitrovic, 1982, 
in {\it Solid State Physics}, edited by
H. Ehrenreich, D. Turnball, and F. Seitz (Academic, New York), Vol. 37,
p. 1.
 
Anderson, P.W., 1991, ``Theories of Fullerene $T_c$'s which will not work'',
preprint.

Antropov, V.P., 1992, unpublished.
 
Antropov, V.P., O. Gunnarsson, and O. Jepsen, 1992, Phys. Rev. B 
{\bf 46}, 13647.

Antropov, V.P., I.I. Mazin, O.K. Andersen, A.I. Liechtenstein,
and O. Jepsen, 1993a, Phys. Rev. B {\bf 47}, 12373. 

Antropov, V.P., O. Gunnarsson, and A.I. Liechtenstein, 1993b, Phys. Rev.
B {\bf 48}, 7651.
    
Aryasetiawan, F., O. Gunnarsson, E. Koch, and R.M. Martin, 1996, (to be
published).

Asai, Y., and Y. Kawaguchi, 1992, Phys. Rev. B {\bf 46}, 1265.

Aschroft, N.W., and M. Cyrot, 1993, Europhys. Lett. {\bf 23}, 605.

Auban-Senzier, P., G. Quirion, D. Jerome, P. Bernier, S. Della-Negra, 
C. Fabre, and A. Rassat, 1993, Synthetic  Metals {\bf 56}, 3027.

Auerbach, A., and G.N. Murthy, 1992, Europhys. Lett. {\bf 19}, 103.

Auerbach, A., N. Manini, and E. Tosatti, 1994, Phys. Rev. B \ {\bf 49}, 12998.

Baenitz, M., M. Heinze, K. L\"uders, H. Werner, R. Schl\"ogl, M. Weiden,
G. Sparn, and F. Steglich, 1995, Solid State Commun. {\bf 96}, 539.

Baggot, J., 1994, {\it Perfect Symmetry}, Oxford Press, Oxford.

Baskaran, G. and E. Tosatti, 1991, Current Science {\bf 61}, 33.

Belosludov, V.R., and V.P. Shpakov, 1992, Mod. Phys. Lett. B {\bf 6}, 1209.

Benning, P.J.,  J.L. Martins, J.H. Weaver,
L.P.F. Chibante, and R.E. Smalley, 1991 
Science {\bf 252} 1417.

Benning, P.J., F. Stepniak, D.M. Poirier, J.L. Martins, J.H. Weaver,
L.P.F. Chibante, and R.E. Smalley, 
 1993, Phys. Rev. B {\bf 47}, 13843 (1993).

Berdenis, N., and G. Murthy, 1995, Phys. Rev. B {\bf 52}, 3083.

%Bergmann, G., and D. Rainer, 1973, Z. Phys. {\bf 263}, 59.

Berry, M.V., 1984, Proc. R. Soc. London A {\bf 392}, 45.

Bethune, D.S., G. Meijer, W.C. Tang, H.J. Rosen, W.G. Golden,
H. Seki, C.A. Brown, and M.S. de Vries, 1991, Chem. Phys. Lett. {\bf 179}, 
181.

Bethune, D.S., 1992, as quoted by Deaven and Rokhsar (1993).

Billups, W.E., and M.A. Ciufolini, 1993, Editors {\it Buckminsterfullerenes},
VCH Publishers, New York.

Boebinger, G.S.,  T.T.M. Palstra, A. Passner, M.J. Rosseinsky,
D.W. Murphy, and I.I. Mazin, 1992, Phys. Rev. B {\bf 46}, 5876.

Bogoliubov, N.N., V.V. Tolmachev, and D.V. Shirkov, 1958, 
{\it A new method  in Superconductivity}, Consultants Bureau, New York.

Bohnen, K.-P., R. Heid, K.-M. Ho, and C.T. Chan, 1995, Phys. Rev. B
{\bf 51}, 5805.

Br\"uhwiler, P.A., A.J. Maxwell, A. Nilsson, N. M\aa rtensson, and  
O. Gunnarsson, 1992, Phys. Rev. B {\bf 48}, 18296.   

Buntar, V., and H.W. Weber, 1996, Superconductor Science and Technology
(in press).

Burk, B., V.H. Crespi, M.S. Fuhrer, A. Zettl, and M.L. Cohen, 
1994a, Physica C {\bf 235-240}, 2493.

Burk, B., V.H. Crespi, A. Zettl, and M.L. Cohen, 1994b, Phys Rev. Lett. {\bf 72},
3706. 

Burkhart, G.J., and C. Meingast, 1996, (to be published).

Cappelluti, E., and L. Pietronero, 1996, Phys. Rev. B {\bf 53}, 932.

Carbotte, J.P., 1990, Rev. Mod. Phys. {\bf 62}, 1027.

Chakravarty, S. and S. Kivelson, 1991, Europhys. Lett. {\bf 16}, 751.

Chakravarty, S., M.P. Gelfand and S. Kivelson, 1991, Science {\bf 254}, 970.

Chakravarty, S., S.A. Kivelson, M.I. Salkola, and S. Tewari, 1992a,
Science {\bf 256}, 1306.

Chakravarty, S., S. Khlebnikov, and S. Kivelson, 1992b, Phys. Rev. Lett.
{\bf 69}, 212.

Chauvet, O., G. Oszlanyi, L. Forro, P.W. Stephens, M. Tegze, G. Faigel,
and A. Janossy, 1994, Phys. Rev. Lett. {\bf 72}, 2721.

Chen, C.-C, and C.M. Lieber, 1992, J. Am. Chem. Soc. {\bf 114}, 3141.

Chen, C.-C, and C.M. Lieber, 1993, Science {\bf 259}, 655.

Chen, C.T., L.H. Tjeng, P. Rudolf, G. Meigs, J.E. Rowe, J. Chen, 
J.P. McCauley Jr, A.B. Smith III, A.R. McGhie, W.J. Romanow, 
and E.W. Plummer, 1991, Nature {\bf 352}, 603.

Chen, G., and W.A. Goddard III, 1993, Proc. Natl. Acad. Sci. USA
{\bf 90}, 1350.

Chen, X.H., and G. Roth, 1995, Phys. Rev. B \ {\bf 52}, 15534.        

Chen, Y., D.M. Poirier, M.B. Jost, C. Gu, T.R. Ohno, J.L. Martins,
J.H. Weaver, L.P.F. Chibante, and R.E. Smalley, 1992, Phys. Rev. B 
{\bf 46}, 7961.

Christides, C., D.A. Neunman, K. Prassides, J.R.D. Copley, J.J. Rush,
M.J. Rosseinsky, D.W. Murphy, and R.C. Haddon, 1992, Phys. Rev. B     
{\bf 46}, 12088.

Crespi, V.H., M.L. Cohen, and D.R. Penn, 1991, Phys. Rev. B
{\bf 43}, 12921.

 Crespi, V.H., J.G. Hou, X.-D. Xiang, M.L. Cohen,
and A. Zettl, 1992, Phys. Rev. B {\bf 46}, 12064.

Deaven, D.M., and D.S. Rokhsar, 1993, Phys. Rev. B {\bf 48}, 4114.

de Coulon, V., J.L. Martins, and F. Reuse, 1992, Phys. Rev. B 
{\bf 45}, 13671.

Degiorgi, L.,  P. Wachter, G. Gr\"uner, S.-M. Huang,
J. Wiley, and R.B. Kaner, 1992, Phys. Rev. Lett. {\bf 69}, 2987.

Degiorgi, L., G. Briceno, M.S. Fuhrer, A. Zettl, and P. Wachter, 1994,
Nature {\bf 369}, 541.

Degiorgi, L., 1995, Mod. Phys. Lett. B {\bf 9}, 445.

De Seta, M., and F. Evangelisti, 1995, Phys. Rev. B {\bf 51}, 1096.

de Vries, J., H. Steger, B. Kamke, C. Menzel, B. Weisser, W. Kamke,
and I.V. Hertel, 1992, Chem. Phys. Lett. {\bf 188}, 159.

Diederichs, J., J.S. Schilling, K.W. Herwig, and W.B. Yelon, 1996, J. 
Phys. Chem. Solids  

Dolgov, O.V., and I.I. Mazin, 1992, Solid State Commun. {\bf 81}, 935.
 
Duclos, S.J., R.C. Haddon, S. Glarum, A.F. Hebard, and K.B. Lyons, 1991,
Science {\bf 254}, 1625.

Ebbesen, T.W., J.S. Tsai, K. Tanigaki, J. Tabuchi, Y. Shimakawa,
Y. Kubo, I. Hirosawa, and J. Mizuki, 1992a, Nature {\bf 355}, 620.

Ebbesen, T.W., J.S. Tsai, K. Tanigaki, H. Hiura,   Y. Shimakawa,
Y. Kubo, I. Hirosawa, and J. Mizuki, 1992b, Physica C {\bf 203}, 163.

Eliashberg, G.M., 1960, J. Expt. Theor. Phys. (U.S.S.R) {\bf 38}, 966        
[Sov. Phys. JETP {\bf 11}, 696 (1960)].

Erwin, S.C., and M.R. Pederson, 1991, Phys. Rev. Lett. {\bf 67}, 1610.

Erwin, S.C., and E.J. Mele, 1994, Phys. Rev. B {\bf 50} 5689.

Erwin, S.C. and W.E. Pickett, 1991, Science {\bf 254}, 842.

Erwin, S.C. and W.E. Pickett, 1992, Phys. Rev. B {\bf 46}, 14257.

Erwin, S.C., 1993, in {\sl Buckminsterfullerenes}, Eds. W.E.
Billups and M.A. Ciufolini (VCH Publishers, New York), p. 217.

Erwin, S.C., and C. Bruder, 1994, Physica B {\bf 199$\&$200}, 600.

Erwin, S.C. and M.R. Pederson, 1993, Phys. Rev. B {\bf 47}, 14657.

Erwin, S.C., G.V. Krishna and E.J. Mele, 1995, Phys. Rev. B {\bf 51}, 7345.

Fabrizio, M, and E. Tosatti, 1996 (preprint).

Faulhaber, J.C.R., D.Y.K. Ko, and P.R. Briddon, 1993, Phys. Rev. B
{\bf 48}, 661.

Fischer, J.E., and D.E. Cox, 1992, Eds.,
{\it Physics and Chemistry of Fullerene-Based Solids}, J. Phys.
Chem. Solids {\bf 53}, No. 12.

Fischer, J.E., Y. Maruyama, D.E. Cox, and R.N. Shelton, 1993, Eds.,
{\it Physics and Chemistry of Fullerene-Based Solids - II}, J. Phys.
Chem. Solids {\bf 54}, No. 12.

Fischer, J.E., 1996,  NATO ASI series X, Kluwer.

Fisk, Z., and G.W. Webb, 1976, Phys Rev. Lett. \ {\bf 36}, 1084.

Fleming, R.M., A.P. Ramirez, M.J. Rosseinsky, D.W. Murphy, R.C. Haddon,
S.M. Zahurak, and A.V. Makhija, 1991, Nature {\bf 352}, 787.

Friedberg, R., T.D. Lee and H.C. Ren, 1992, Phys. Rev. B \ {\bf 46}, 14150.

%Geilikman, B.T., and V.Z. Kresin, 1972, Phys. Lett. A {\bf 40}, 123.

Gelfand, M.P.,  and J.P. Lu, 1992a, Phys. Rev. Lett. {\bf 68}, 1050.

Gelfand, M.P.,  and J.P. Lu, 1992b, Phys. Rev. B {\bf 46}, 4367.

Gelfand, M.P.,  and J.P. Lu, 1993, Phys. Rev. B {\bf 47}, 4149.

Gelfand, M.P., 1994, Superconductivity Review {\bf 1}, 103.    

Gensterblum, G., J.J. Pireaux, P.A. Thiry, R. Caudano, J.P. Vigneron,
Ph. Lambin, A.A. Lucas, and W. Kr\"atschmer, 1991, Phys Rev. Lett. {\bf 67}, 2171.

Georges, A., G. Kotliar, W. Krauth, and M.J. Rozenberg, 1996,
Rev. Mod. Phys. {\bf 68}, 13 
have given a review of the results for large dimensions.

Ginzburg, V.L., and D.A. Kirzhnits, 
1982, {\sl High-Temperature superconductivity},
Consultants Bureau, New York.

Goff, W.E., and P. Phillips, 1992, Phys. Rev. B {\bf 46}, 603. 

Goff, W.E., and P. Phillips, 1993, Phys. Rev. B {\bf 48}, 3491.

Grabowski, M, and L.J. Sham, 1984, Phys. Rev. B {\bf 29}, 6132.

Grimaldi, C., L. Pietronero, and S. Str\"assler, 1995, Phys. Rev. B  
{\bf 52}, 10530.

Grimaldi, C., L. Pietronero, and S. Str\"assler, 1995b, Phys. Rev. Lett.
{\bf 75}, 1158.

Grimvall, G., 1981, {\sl The Electron-Phonon Interaction in Metals},
(North-Holland, Amsterdam), pp. 210-223. 

Gu, C., B.W. Veal, R. Liu, A.P. Paulikas, P. Kostic, H. Ding, K. Gofron,
J.C. Campuzano, J.A. Schlueter, H.H. Wang, U. Geiser, and J.M. Williams,
1994, Phys. Rev. B {\bf 50}, 16566.

Gunnarsson, O., S. Satpathy, O. Jepsen, and O.K. Andersen, 1991, Phys. Rev. 
Lett. {\bf 67}, 3002.

Gunnarsson, O., and G. Zwicknagl, 1992, Phys. Rev. Lett. {\bf 69}, 957. 

Gunnarsson, O., D. Rainer, and G. Zwicknagl, 1992, Int. J. Mod. Phys. B {\bf 
 6}, 3993.

Gunnarsson, O., 1993 (unpublished).

Gunnarsson, O., V. Meden, and K. Sch\"onhammer, 1994a, Phys. Rev. B 
{\bf 50}, 10462.

Gunnarsson, O., V. Meden, and K. Sch\"onhammer, 1994b
 {\it Progress in Fullerne Research} (World Scientific, Singapore).
Kuzmany, H., J. Fink, M. Mehring, and S. Roth,  Eds., p. 317.

Gunnarsson, O., H. Handschuh, P.S. Bechthold, B. Kessler, G. Gantef\"or,
and W. Eberhardt, 1995, Phys. Rev. Lett. {\bf 74}, 1875.

Gunnarsson, O., 1995, Phys. Rev. B {\bf 51}, 3493.

Gunnarsson, O.,  V. Eyert, M. Knupfer, J. Fink, J.F. Armbruster, 1996,
J. Phys.: Condens. Matter {\bf 8}, 2557.

Gunnarsson, O., E. Koch and R.M. Martin, 1996 (to be published).

Gunnarsson, O., 1996 (unpublished).

Haddon, R.C., A.F. Hebard, M.J. Rosseinsky, D.W. Murphy, S.J. Duclos,
K.B. Lyons, B. Miller, J.M. Rosamilia, R.M. Fleming, A.R. Kortan, 
S.H. Glarum, A.V. Makhija, A.J. Muller, R.H. Eick, S.M. Zahurak, 
R. Tycko, G. Dabbagh, and F.A. Thiel, 1991, Nature {\bf 350}, 320.

Hammond, G.S., and V.J. Kuck, 1992, Editors, American Chemical Company,
Washington, DC. 

Hebard, A.F., M.J. Rosseinsky, R.C. Haddon, D.W. Murphy, S.H. Glarum, 
T.T.M. Palstra, A.P. Ramirez, and A.R. Kortan, 1991, Nature
{\bf 350}, 600.
 
Hebard, A.F., 1992, Physics Today, {\bf 45}, No. 11, p. 26.

Hebard, A.F., T.T.M. Palstra, R.C. Haddon, and R.M. Fleming,
1993, Phys. Rev. B \ {\bf 48}, 9945.

Hedin, L., and S. Lundqvis, 1969, in {\it Solid State Physics}, edited by
H. Ehrenreich, D. Turnball, and F. Seitz (Academic, New York), Vol. 23,
p. 1.

Hettich, R.L., R.N. Compton, and R.H. Ritchie, 1991, Phys Rev. Lett. {\bf 67}, 1242.

Hirosawa, I., K. Prassides, J. Mizuki, K. Tanigaki, M. Gevaert,
A. Lappas, and J.K. Crockcroft, 1994, Science, {\bf 264}, 1294.

Hohenberg, P., and W. Kohn, 1964, Phys. Rev. {\bf 136}, B864.

Holczer, K., O. Klein, S.-M. Huang, R.B. Kaner, K.-J. Fu, R.L. Whetten, 
and F. Diederich, 1991a, Science {\bf 252}, 1154.

Holczer, K., O. Klein, G. Gr\"uner, J.D. Thompson, F. Diederich, and    
R.L. Whetten, 1991b, Phys. Rev. Lett. {\bf 67}, 271. 

Holczer, K. and R.L. Whetten, 1992, Carbon {\bf 30}, 1261.   

Hou, J.G., V.H. Crespi, X.-D. Xiang, W.A. Vareka, G. Briceno, A. Zettl, 
and M.L. Cohen, 1993, Solid State Commun. {\bf 86}, 643.

Hou, J.G., L. Lu, V.H. Crespi, X.-D. Xiang, A. Zettl, 
and M.L. Cohen, 1995, Solid State Commun. {\bf 93}, 973.

Huang, M.-Z., Y.-N. Xu and W.Y. Ching, 1992, Phys. Rev. B {\bf 46}, 6572.
 
Ihm, J., 1994, Phys. Rev. B \ {\bf 49}, 10726.

Ikeda, M.A., A. Ogasawara and M. Sugihara, 1992, Phys. Lett. A {\bf 170}, 319.

Inada, Y., and K. Nasu, 1992, J. Phys. Soc. Japan {\bf 61}, 4511.
 
Jahn, H.A., and E. Teller, 1937, Proc. R. Soc. London A {\bf 161}, 220.

Jess, P., S. Behler, M. Bernasconi, V. Thommen-Geiser, H.P. Lang,
M. Baenitz, K. L\"uders, 
and H.-J. G\"untherodt, 1994, Physica C {\bf 235-240}, 2499..

Jess, P., U. Hubler, S. Behler, V. Thommen-Geiser, H.P. Lang,
and H.-J. G\"untherodt, 1996, Synthetic Metals {\bf 77}, 201.

Jishi, R.A., and M.S. Dresselhaus, 1992, Phys. Rev. B {\bf 45}, 2597.

Jishi, R.A., R.M. Miere, and M.S. Dresselhaus, 1992, Phys. Rev. B {\bf 45}, 
13685.

Jones, R.O., and O. Gunnarsson, 1989, Rev. Mod. Phys. {\bf 61}, 689.

Jones, R., C.D. Latham, M.I. Heggie, V.J.B. Torres, S. \"Oberg,
and S.K. Estreicher, 1992, Phis. Mag. Lett. {\bf 65}, 291.

Joubert, D.P., 1993, J. Phys.: Condens. Matter {\bf 5}, 8047.

Kiefl, R.F., T.L. Duty, J.W. Schneider, A. MacFarlane, K. Chow, J.W. Elzey,
P. Mendels, G.D. Morris, J.H. Brewer, E.J. Ansaldo, C Niedermayer, 
D.R. Noakes, C.E. Stronach, B. Hitti, and J.E. Fischer,
  1992, Phys. Rev. Lett. {\bf 69}, 2005.

Kiefl, R.F., W.A. MacFarlane, K.H. Chow, S. Dunsiger, T.L. Duty,
T.M.S. Johnston, J.W. Schneider, J. Sonier, L. Brard, R.M. Strongin,
J.E. Fischer, and A.B. Smith III, 1993, Phys Rev. Lett. {\bf 70}, 3987.

Kniaz, K, J.E. Fischer, Q. Zhu, M.J. Rosseinsky, O. Zhou, and D.W. Murphy,
1993, Solid State Commun {\bf 88}, 47.
 
Knupfer, M., M. Merkel, M.S. Golden, J. Fink, O. Gunnarsson, V.P.
Antropov, 1993, Phys. Rev. B {\bf 47}, 13944.

Knupfer, M., F. Stepniak, and J.H. Weaver, 1994, Phys. Rev. B {\bf 49}, 7620.

Knupfer, M., J. Fink, J.F. Armbruster, and H.A. Romberg, 1995, Z. Phys. 
{\bf 98}, 9.

Kohanoff, J., W. Andreoni, and M. Parrinello, 1992, Phys. Rev. B 
{\bf 46}, 4371.

Kohn, W., and L.J. Sham, 1965, Phys. Rev. {\bf 140}, A1133.

Kortan, A.R., N. Kopylov, S. Glarum, E.M. Gyorgy, A.P. Ramirez, 
R.M. Fleming, F.A. Thiel, and R.C. Haddon, 1992a, Nature {\bf 355}, 529.

Kortan, A.R., N. Kopylov, S. Glarum, E.M. Gyorgy, A.P. Ramirez, 
R.M. Fleming, O. Zhou,
 F.A. Thiel, P.L. Trevor, and R.C. Haddon, 1992b, Nature {\bf 360}, 566.

Kortan, A.R., N. Kopylov, E. \"Ozdas, A.P. Ramirez, R.M. Fleming,
and R.C. Haddon, 1994, Chem. Phys. Lett. {\bf 223}, 501.

Kostur, V.N., and B. Mitrovic, 1993, Phys. Rev. B {\bf 48}, 16388.

Kostur, V.N., and B. Mitrovic, 1994, Phys. Rev. B {\bf 50}, 12774.
 
Kr\"atschmer, W., L.D. Lamb, K. Fostiropoulos, and D.R. Huffman, 1990,
Nature {\bf 347}, 354.

Krier, G., and O. Gunnarsson, 1995 (unpublished).

Krishnamurthy, H.R., D.M. Newns, P.C. Pattnaik, C.C. Tsuei,
and C.C. Chi, 1994, Phys. Rev. B {\bf 49}, 3520.

Kroto, H.W., J.R. Heath, S.C. O'Brien, R.F. Curl, and R.E. Smalley, 1985 
Nature {\bf 318}, 162.

Kroto, H.W., A. W. Allaf, and S.P. Balm, 1991, Chem. Rev. {\bf 91}, 1213.

Kulic, M.L., and R. Zheyer, 1994, Phys. Rev. B \ {\bf 49}, 4395.

Kumar, V., T.P. Martin, and E. Tosatti, 1992, Editors,
{\it Clusters and fullerenes},  Int. J. Mod. Phys. B
{\bf 6}, No. 23$\&$24.

Kuzmany, H., J. Fink, M. Mehring, and S. Roth, 1993, Eds., {\it
Electronic properties of fullerenes} (Springer, Berlin).

Kuzmany, H., J. Fink, M. Mehring, and S. Roth, 1994a, Eds., {\it
Progress in Fullerne Research} (World Scientific, Singapore).

Kuzmany, H., J. Fink, M. Mehring, and S. Roth, 1995a, Eds., {\it
Physics and chemistry of fullerenes and derivatives}
 (World Scientific, Singapore).

Kuzmany, H., M. Matus, B. Burger, and J. Winter, 1994b, 
Adv.   Mater.    {\bf 6}, 731. 

Kuzmany, H., R. Winkler, and T. Pichler, 1995b, J. Phys.: Cond. Matter
{\bf 7}, 6601.

Lammert, P.E., and D.S. Rokhsar, 1993, Phys. Rev. B \ {\bf 48}, 4103. 

Lammert, P.E., D.S. Rokhsar, S. Chakravarty, S. Kivelson, and
M.I. Salkola, 1995, Phys. Rev. Lett. {\bf 74}, 996.

Lannin, J.S., and M.G. Mitch, 1994, Phys. Rev. B {\bf 50}, 6497.

Lannoo, M., G.A. Baraff, M. Schluter, and D. Tomanek, 1991, Phys. Rev. B \
{\bf 44}, 12106.

Laouini, N.,  O.K. Andersen, O. Gunnarsson, 1995,      
 Phys. Rev. B {\bf 51}, 17446.

Lieber, C.M., and Z. Zhang, 1994, Solid State Physics, Edited by 
H. Ehrenreich and  F. Spaepen, Academic Press, New York, p. 349.

Liechtenstein, A.I., O. Gunnarsson, M. Knupfer, J. Fink, and J.F. Armbruster,
1996, J. Phys.: Condens. Matter {\bf 8}, 4001. 

Limbach, P.A., L. Schweikhard, K.A. Cowen, M.T. McDermott,
A.G. Marshall, and J.V. Coe, 1991, J. Am. Chem. Soc. {\bf 113}, 6795.

Lin, C.L., T. Mihalisin, N. Bykovetz, Q. Zhu, and J.E. Fischer, 1994,
Phys. Rev. B {\bf 49}, 4285.

Lof, R.W., M.A. van Veenendaal, B. Koopmans, H.T. Jonkman, and G.A.
Sawatzky, 1992, Phys. Rev. Lett. {\bf 68}, 3924.

Loktev, V.M., 1992, Fiz. Nizk. Temp. {\bf 18}, 217 [Sov. J. Low Temp. Phys.
{\bf 18}, 149].

Lu, J.P., 1994, Phys. Rev. B {\bf 49}, 5687. 

Manini, N., E. Tosatti, and A. Auerbach, 1994, Phys. Rev. B \ {\bf 49}, 13008.
 
Maniwa, Y., T. Saito, K. Kume, K. Kikuchi, I. Ikemoto, S. Suzuki,
Y. Achiba, I. Hirozawa, M. Kosaka, and K. Tanigaki, 1995, Phys. Rev. B \
{\bf 52}, R7054.

Martin, R.L., and J.P. Ritchie, 1993, Phys. Rev. B {\bf 48}, 4845. 

Martins, J.L., and N. Troullier, 1992, Phys. Rev. B {\bf 46}, 1766.

Mazin, I.I., S.N. Rashkeev, V.P. Antropov, O. Jepsen, A.I. Liechtenstein,
and O.K. Andersen, 1992, Phys. Rev. B {\bf 45}, 5114.

Mazin, I.I., A.I. Liechtenstein, O. Gunnarsson, O.K. Andersen, V.P. Antropov,
and S.E. Burkov, 1993a, Phys. Rev. Lett. {\bf 70}, 4142.

Mazin, I.I., O.V. Dolgov, A. Golubov, and S.V. Shulga, 1993b, Phys. Rev. B \
{\bf 47}, 538.

McMillan, W.L., 1968, Phys. Rev. {\bf 167}, 331.

Mele, E.J., and S.C. Erwin, 1993, Phys. Rev. B \ {\bf 47}, 2948.

Mele, E.J., and S.C. Erwin, 1994, Phys. Rev. B \ {\bf 50}, 2150.

Mele, E.J., S.C. Erwin, and M.S. Deshpande, 1994, Synthetic Metals
{\bf 65}, 255.

Mitch, M.G., S.J. Chase, and J.S. Lannin, 1992a, Phys. Rev. Lett. {\bf 68},
883.

Mitch, M.G., S.J. Chase, and J.S. Lannin, 1992b, Phys. Rev. B {\bf 46}, 3696. 

Mitch, M.G., and J.S. Lannin, 1993, Phys. Rev. B {\bf 48}, 16192.

Mitrovic, B., H.G. Zarate, and J.P. Carbotte, 1984, Phys. Rev. B {\bf 29},
184.

Migdal, A.B., 1958, J. Expt. Theor. Phys. (U.S.S.R.) {\bf 34}, 1438
 [Sov. Phys. JETP {\bf 7}, 996 (1958)].

Mizuki, J., M. Takai, H. Takahashi, N. Mori, K. Tanigaki, 
I. Hirosawa, and K. Prassides, 1994, Phys. Rev. B \ {\bf 50}, 3466.

Morel, P., and P.W. Anderson, 1992, Phys. Rev.  {\bf 125}, 1263.

Murphy, D.W., M.J. Rosseinsky, R.M. Fleming, R. Tycko, A.P. Ramirez,
R.C. Haddon, T. Siegrist, G. Dabbagh, J.C. Tully, and R.E. Walstedt, 
 1992, J. Phys. Chem. Solids {\bf 53}, 1321.

Murthy, G.N., and A. Auerbach, 1992, Phys. Rev. B {\bf 46}, 331.
 
Nambu, Y., 1960, Phys. Rev. {\bf 117}, 648.
 
Negri, F., G. Orlandi, and F. Zerbetto, 1988, Chem. Phys. Lett.
{\bf 144}, 31.

Novikov, D.L., V.A. Gubanov and A.J. Freeman, 1992, Physica C {\bf 191}, 399.

Onida, G., and G. Benedek, 1992, Europhys. Lett. {\bf 18}, 403.

\"Ozdas, E., A.R. Kortan, N. Kopylov, A.P. Ramirez, T. Siegrist,
K.M. Rabe, H.E. Bair, S. Schuppler, and P.H. Citrin, 1995, Nature,
{\bf 375}, 126.

Palstra, T.T.M., A.F. Hebard, R.C. Haddon, and P.B. Littlewood,
1994, Phys. Rev. B {\bf 50}, 3462. 

Palstra, T.T.M., O. Zhou, Y. Iwasa, P.E. Sulewski, R.M. Fleming,  
and B.R. Zegarski, 1995, Solid State Commun. {\bf 93}, 327.

Parks, R.D., 1969, {\it Superconductivity} (Dekker, New York).

Pederson, M.R., and A.A. Quong, 1992, Phys. Rev. B {\bf 46}, 13584.

Pichler, T., M. Matus, J. K\"urti, and H. Kuzmany, 1992, Phys. Rev. B \
{\bf 45}, 13841.

Pickett, W.E., 1994, Solid State Physics, Edited by H. Ehrenreich and  
F. Spaepen, Academic Press, New York, p. 226. 

Pickett, W.E., D.A. Papaconstantopoulos, M.R. Pederson, and S.C. Erwin,
1994, J. Superconductivity, {\bf 7}, 651. The values of $\lambda$ quoted
here differ from the ones given by Pickett {\it et al.}, due to a different
convention used by them.

Pietronero, L., and S. Str\"assler, 1992, Europhys. Lett. {\bf 18}, 627.

Pietronero, L., S. Str\"assler, and C. Grimaldi, 1995, Phys. Rev. B \ {\bf 52},
10516. 

Pintschovius, L., 1996, Rep. Prog. Phys. {\bf 59}, 473.

Prassides, K., J. Tomkinson, C. Christides, M.J. Rosseinsky, D.W. Murphy,
and R.C. Haddon, 1991, Nature {\bf 354}, 462.

Prassides, K., C. Christides, M.J. Rosseinsky, J. Tomkinson,
D.W. Murphy, and R.C. Haddon, 1992, Europhys. Lett. {\bf 19}, 629.

Prassides, K., C. Christides, I.M. Thomas, J. Mizuki, K. Tanigaki, 
I. Hirosawa, and T.W. Ebbesen, 1994, Science {\bf 263}, 950.

Prassides, K., 1994, Editor, {\it Physics and Chemistry of the fullerenes},
NATO ASI series,
Kluwer Academic Publishers, Dordrect.

Quong, A.A., M.R. Pederson, and J.L. Feldman, 
1993, Solid State Commun. {\bf 87}, 535.

Rainer, D., 1986, Prog. Low Temp. Phys. {\bf 10}, 371.

Ramirez, A.P., A.R. Kortan, M.J. Rosseinsky, S.J. Duclos, A.M. Mujsce, 
R.C. Haddon, D.W. Murphy, A.V. Makhija, S.M. Zahurak, and K.B. Lyons,
1992a, Phys Rev. Lett. {\bf 68}, 1058.

Ramirez, A.P., M.J. Rosseinsky, D.W. Murphy, and R.C. Haddon, 
1992b, Phys Rev. Lett. {\bf 69}, 1687.
  
Ramirez, A.P., 1994, Superconductivity Review {\bf 1}, 1.
 
Ramirez, A.P., M.J. Rosseinsky, and D.W. Murphy, 1994, as quoted in
Ramirez (1994).

Renker, B., F. Gompf, H. Schober, P. Adelmann, H.J. Bornemann,
and R. Heid, 1993, Z. Physik. B {\bf 92}, 451.

Reznik, D., W.A. Kamitakahara, D.A. Neumann, J.R.D. Copley,
J.E. Fischer, R.M. Strongin, M.A. Cichy, and A.B. Smith III, 1994, Phys. Rev.
B {\bf 49}, 1005.
 
Rietschel, H., and L.J. Sham, 1983, Phys. Rev. B \ {\bf 28}, 5100.

Rosseinsky, M.J., A.P. Ramirez, S.H. Glarum, D.W. Murphy, R.C., Haddon, 
A.F. Hebard, T.T.M. Palstra, A.R. Kortan, S.M. Zahurak,  and
A.V. Makhija, 1991, Phys. Rev. Lett. {\bf 66}, 2830.

Rosseinsky, M.J., D.W. Murphy, R.M. Fleming, and O. Zhou, 1993,
Nature {\bf 364}, 425.

Rotter, L.D.,  Z. Schlesinger, J.P. McCauley Jr,
N. Coustel, J.E. Fischer, and A.B. Smith III, 1992, Nature {\bf 355},
532.

Saito, S., and A. Oshiyama, 1991a, Phys. Rev. Lett. {\bf 66}, 2637. 

Saito, S., and A. Oshiyama, 1991b, Phys. Rev. {\bf 44}, 11536.

Saito, S., and A. Oshiyama, 1993, Phys. Rev. Lett. {\bf 71}, 121.  

Sasaki, S., A. Matsuda, and C.W. Chu, 1994, J. Phys. Soc. Japan {\bf 63}, 
1670.

Satpathy, S., V.P. Antropov, O.K. Andersen, O. Jepsen, O. Gunnarsson,
and A.I. Liechtenstein, 1992, Phys. Rev. B {\bf 46}, 1773.

Sawatzky, G.A., 1995, in 
{\it Physics and chemistry of fullerenes and derivatives}
Kuzmany, H., J. Fink, M. Mehring, and S. Roth,  Eds., 
 (World Scientific, Singapore) p. 373.

Scherrer, H., and G. Stollhoff, 1993, Phys. Rev. B \ {\bf 47}, 16570.

Schirber, J.E., D.L. Overmyer, W.R. Bayless, M.J. Rosseinsky.
D.W. Murphy, Q. Zhu, O. Zhou, K. Kniaz, and J.E. Fischer, 1993, 
J. Phys. Chem. Solids {\bf 54}, 1427. 

Schluter, M., M. Lannoo, M. Needels, G.A. Baraff, and D. Tomanek, 1992a,
Phys. Rev. Lett. {\bf 68}, 526.

Schluter, M., M. Lannoo, M. Needels, G.A. Baraff, and D. Tomanek, 1992b,
J. Phys. Chem. Solids {\bf 53}, 1473.

Schluter, M.A., M. Lannoo, M.F. Needels, G.A. Baraff and D. Tomanek, 1992c,
Phys Rev. Lett. {\bf 69}, 213. 

Schober, H., B. Renker, F. Gompf, and P. Adelmann, 1994, Physica
{\bf 235-240}C, 2487.

Schrieffer, J.R., {\sl Theory of superconductivity}, 1964, Benjamin, New York.

Sparn, G., J.D. Thompson, S.-M. Huang, R.B. Kaner, 
F. Diederich, R.L. Wetten, G. Gr\"uner, and K. Holczer, 1991, Science  
{\bf 252}, 1829.

Sparn, G., J.D. Thompson, R.L. Whetten, S.-M. Huang, R.B. Kaner, 
F. Diederich, G. Gr\"uner, and K. Holczer, 1992, Phys. Rev. Lett.
{\bf 68}, 1228.

Stenger, V.A., C.H. Pennington, D.R. Buffinger, and R.P. Ziebarth, 1995,
Phys. Rev. Lett. {\bf 74}, 1649.

Stephens, P.W., L. Mihaly, P.L. Lee, R.L. Whetten, S.-M. Huang,
R. Kaner, F. Deiderich, and K. Holczer, 1991, Nature {\bf 351}, 632.

Stephens, P.W., G. Bortel, G. Falgel, M. Tegze, A. Janossy,
S. Pekker, G. Oszlanyl, and L. Forro, 1994, Nature {\bf 370}, 636.

Stollhoff, G., 1991, Phys. Rev. B \ {\bf 44}, 10998.

Takada, Y., 1993, J. Phys. Chem. Solids {\bf 54}, 1779.

Tanigaki, K., T.W. Ebbesen, S. Saito, J. Mizuki, J.S. Tsai, Y. Kubo,
and S. Kuroshima, 1991, Nature {\bf 352}, 222.

Tanigaki, K., I. Hirosawa, T.W. Ebbesen, J. Mizuki, Y. Shimakawa, 
Y. Kubo, J.S. Tsai, and S. Kuroshima, 1992, Nature {\bf 356}, 419.

Tanigaki, K., I. Hirosawa, T.W. Ebbesen, J.-I. Mizuki, and J.-S. Tsai,
1993, J. Phys. Chem. Solids {\bf 54}, 1645.

Tanigaki, K., M. Kosaka, T. Manako, Y. Kubo, I. Hirosawa, K. Uchida,
and K. Prassides, 1995, Chem. Phys. Lett. {\bf 240}, 627.

Tanigaki, K., and K. Prassides, 1995, J. Mater. Chem. {\bf 5}, 1515.

Teslic, S., T. Egami, and J.E. Fischer, 1995, Phys. Rev. B {\bf 51}, 5973.

Tinkham, M., 1975, {\sl Introduction to superconductivity},
(McGraw-Hill, New-York).

Tokuyasu, T., M. Kamal and G. Murthy, 1993,  Phys Rev. Lett. {\bf 71}, 4202.

Troullier, N. and J.L. Martins, 1992, Phys. Rev. B {\bf 46}, 1754.   

Tycko, R., G. Dabbagh, M.J. Rosseinsky, D.W. Murphy, A.P. Ramirez, 
and R.M. Fleming, 1992, Phys Rev. Lett. {\bf 68}, 1912.

Tycko, R., G. Dabbagh, D.W. Murphy, Q. Zhu and J.E. Fischer, 
 1993, Phys. Rev. B {\bf 48}, 9097. 

Uemura, Y.J., A. Keren, L.P. Le, G.M. Luke, B.J. Sternlieb, 
W.D. Wu, J.H. Brewer, R.L. Whetten, S.M. Huang, S. Lin, R.B. Kaner,
F. Diederich, S. Donovan, G. Gr\"uner, and K. Holczer, 1991, Nature
{\bf 352}, 605.

Uemura, Y.J., L.P. Lee, and G.M. Luke, 1992, Synthetic Metals {\bf 57}, 2845.
 
Vareka, W.A., and A. Zettl, 1994, Phys. Rev. Lett. {\bf 72}, 4121.

Varma, C.M., J. Zaanen, K. Raghavachari, 1991, Science {\bf 254}, 989.

Walstedt, R.E., D.W. Murphy and M. Rosseinsky, 1993, Nature {\bf 362}, 611.

Wang, D., 1995, Solid State Commun. {\bf 94}, 767.

Wang, Y., D. Tomanek, and G.F. Bertsch, 1991, Phys. Rev. B \ {\bf 44}, 6562.

Ward, J.C., 1950, Phys. Rev. {\bf 78}, 182.

Weltner Jr., W., and R.J. van Zee, 1989, Chem. Rev. {\bf 89}, 1713.

Wertheim, G.K., D.N.E. Buchanan, and J.E. Rowe, 1992, Science {\bf 258}, 1638.

Wertheimer, N.R., E. Helfand,  and P.C. Hohenberg,  1966, Phys. Rev.
{\bf 147}, 295.

White, S.R., S. Chakravarty, M.P. Gelfand, and S. A. Kivelson, 1992, 
Phys. Rev. B {\bf 45}, 5062.

Wilson, M.A., L.S. Pang, G.D. Willett, K.J Fisher and I.G. Dance,
1992, Carbon {\bf 30}, 675.

Winter, J. and H. Kuzmany, 1996, Phys. Rev. B {\bf 53}, 655. 

Wong, W.H., M.E. Hanson, W.G. Clark, G. Gr\"uner, J.D. Thompson, 
R.L. Whetten, S.-M. Huang, R.B. Kaner, F. Diederich, P. Petit, 
J.-J. Andre and K. Holczer, 1992, Europhys. Lett. {\bf 18}, 79.

Xiang, X.-D.,  J.G. Hou, G. Briceno, W.A. Vareka,
R. Mostovoy, A. Zettl, V.H. Crespi, and M.L. Cohen, 1992, Science {\bf 256},
1190.        

Xiang, X.-D.,  J.G. Hou, V.H. Crespi, A. Zettl,
and M.L. Cohen, 1993, Nature {\bf 361}, 54.

Yabana, K., and G. Bertsch, 1992, Phys. Rev. B {\bf 46}, 14263.

Yildirim, T., J.E. Fischer, R. Dinnebier, P.W. Stephens, and C.L. Lin,
1994, 

Yildirim, T., J.E. Fischer, R. Dinnebier, P.W. Stephens, and C.L. Lin,
1995, Solid State Commun. {\bf 93}, 269.

Yildirim, T., L. Barbedette, K. Kniaz, J.E. Fischer, C.L. Lin, 
N. Bykovetz, P.W. Stephens, P.E. Sulewski, and S.C. Erwin, 1995,
in {\it Science and technology of fullerene materials}, Eds.
P. Bernier, T.W. Ebbesen, D.S. Bethune, R.M. Metzger, L.Y. Chiang, and
J.W. Mintmire (Mat. Res. Soc. Proc. {\bf 359}, Pittsburg, 1995) p. 273.

Yildirim, T., L. Barbedette, J.E. Fischer, C.L. Lin, J. Robert, 
P. Petit, and T.T. M. Palstra, 1996a, Phys. Rev. Lett. {\bf 77}, 167. 

Yildirim, T., L. Barbedette, J.E. Fischer, G. Bendele, P.W. Stephens,
C.L. Lin, C. Goze, F. Rachdi, J. Robert, P. Petit, and T.T.M. Palstra,
1996b, Phys. Rev. B  (submitted).

Zakhidov, A.A., K. Imaeda, D.M. Petty, K. Yakushi, H. Inokuchi,
K. Kikuchi, I. Ikemoto, S. Suzuki, and Y. Achiba, 1992, Phys. Lett. A
{\bf 164}, 355.

Zhang, F.C., M. Ogata, and T.M. Rice, 1991a, Phys. Rev. Lett.
{\bf 67}, 3452.

Zhang, W., H. Zheng, and K.H. Bennemann, 1992, Solid State Commun.
{\bf 82}, 679.

Zhang, Z., C.-C. Chen, S.P. Kelty, H. Dai, and C.M. Lieber, 1991b,
Nature {\bf 353}, 333.

Zheng, H. and K.H. Bennemann, 1992, Phys. Rev. B {\bf 46}, 11993.

Zheyer, R., and G. Zwicknagl, 1990, Z. Phys. B {\bf 78}, 175.

Zhou, O., G.B.M. Vaughan, Q. Zhu, J.E. Fischer, P.A. Heiney, N. Coustel,
J.P. McCauley, Jr., and A.B. Smith III, 1992, Science {\bf 255}, 833. 

Zhou, O., R.M. Fleming, D.W. Murphy, M.J. Rosseinsky, A.P. Ramirez,
R.B. van Dover, and R.C. Haddon, 1993, Nature {\bf 362}, 433.

Zhou, O., T.T.M. Palstra, Y. Iwasa, R.M. Fleming, A.F. Hebard, P.E. Sulewski,
D.W. Murphy, and B.R. Zegarski, 1995 Phys. Rev. B \ {\bf 52}, 483.

Zhu, Q., Phys. Rev. B {\bf 52}, R723 (1995).
 
Zimmer, G., M. Helmle, M. Mehring, F. Rachdi, J. Reichenbach, 
L. Firlej, and P. Bernier, 1993, Europhys. Lett. {\bf 24}, 59.          

\vfill\eject

\end{multicols}
\end{document}